\documentclass[sts,preprint]{imsart}

\RequirePackage{amsthm,amsmath,amsfonts,amssymb}
\RequirePackage[authoryear]{natbib}
\RequirePackage[colorlinks,citecolor=blue,urlcolor=blue]{hyperref}
\RequirePackage{graphicx}

\startlocaldefs
\usepackage{booktabs}
\usepackage{paralist}
\usepackage{siunitx}

\theoremstyle{plain}

\theoremstyle{definition}

\allowdisplaybreaks

\newcommand{\cf}{{\cal F}}

\newcommand{\cx}{{\cal X}}

\newcommand{\bsa}{\boldsymbol{a}}
\newcommand{\bsb}{\boldsymbol{b}}

\newcommand{\bsh}{\boldsymbol{h}}

\newcommand{\bsu}{\boldsymbol{u}}
\newcommand{\bsv}{\boldsymbol{v}}
\newcommand{\bsw}{\boldsymbol{w}}
\newcommand{\bsx}{\boldsymbol{x}}

\newcommand{\bsz}{\boldsymbol{z}}

\newcommand{\natu}{\mathbb{N}}
\newcommand{\real}{\mathbb{R}}
\newcommand{\ints}{\mathbb{Z}}

\newcommand{\tran}{\mathsf{T}} 

\newcommand{\tmod}{\ \mathsf{mod}\ }

\newcommand{\rd}{\mathrm{\, d}}

\newcommand{\var}{{\mathrm{Var}}}

\newcommand{\vol}{{\mathbf{vol}}}
\newcommand{\wh}{\widehat}

\renewcommand{\emptyset}{\varnothing}
\renewcommand{\ge}{\geqslant}
\renewcommand{\le}{\leqslant}

\newcommand{\dnorm}{\mathcal{N}}

\newcommand{\dunif}{\mathbf{U}} 

\newcommand{\e}{{\mathbb{E}}} 

\newcommand{\hk}{{\mathrm{HK}}}

\newcommand{\simiid}{\stackrel{\mathrm{iid}}{\sim}}
\newcommand{\bszero}{\boldsymbol{0}}
\newcommand{\bvhk}{\mathrm{BVHK}}
\newcommand{\bsk}{\boldsymbol{k}}
\newcommand{\bbf}{\mathbb{F}}
\newcommand{\bsgamma}{\boldsymbol{\gamma}}
\newcommand{\wce}{\mathrm{wce}}
\newcommand{\ine}{\mathrm{ine}}

\endlocaldefs

\begin{document}

\begin{frontmatter}
\title{Randomized quasi-Monte Carlo integration}
\runtitle{RQMC integration}

\begin{aug}
  \author[A]{\fnms{Art B.}~\snm{Owen} 
    \ead[label=e1]{owen@stanford.edu}}


\address[A]{Art B.\ Owen, Stanford University\printead[presep={\ }]{e1}.}

\end{aug}

\begin{abstract}
Quasi-Monte Carlo sampling is a numerical integration method
that uses points with a space-filling property in $[0,1]^s$
designed to give better estimates than plain Monte Carlo methods do.
For integrands of bounded variation in the sense of Hardy and Krause,
errors of $O(n^{-1+\epsilon})$ for any $\epsilon>0$
are obtained from $n$ sample points.
Randomized quasi-Monte Carlo (RQMC) points are individually uniformly
distributed but collectively space-filling and then independent
replications provide variance estimates.  For smooth enough
integrands the randomization can give a root mean squared
error of $O(n^{-3/2+\epsilon})$.
This article explains RQMC for a statistical readership recounting
some history and presenting some current directions.

\smallskip\par\noindent
This article is dedicated to the memory of Henri Faure and Ilya M.\ Sobol'.
\end{abstract}


\end{frontmatter}

\section{Introduction}

This paper is about randomized quasi-Monte Carlo (RQMC)
for numerical integration especially where it might be
useful for problems in statistics and machine learning.
The emphasis here is on an approach to the problems
of high dimensional integration making use of
statistical methods.
That emphasis combined with limited space means
that many other important results are either omitted
or given only brief coverage.

We may write our quantity of interest  as an integral but when we need
a numerical evaluation then the computation might not be straightforward.
Classical integration methods become less effective in high dimensions.
Then Monte Carlo (MC) sampling becomes competitive.  Quasi-Monte Carlo (QMC)
sampling creates deterministic point sets that have better
regularity than MC points do and this can yield better
accuracy.  However deterministic sampling removes the
uncertainty quantifications that we can get from randomization.
RQMC lets us have it both ways. We get QMC accuracy with
uncertainty quantification based on independent random replicates.  

Randomized QMC points can even have a better convergence
rate than the original unrandomized QMC points do.  In favorable settings 
RQMC can be much more accurate than MC. In other
settings it can have roughly the same accuracy despite
having a better asymptotic mean squared error (MSE).

We can write the problem as solving $\int_{\real^d}f(\bsz)\rd\bsz$ for a given integrand
$f$ on $\real^d$.  Very often we want to estimate an expectation
\begin{align}\label{eq:muasmean}
\mu =\e_p( g(\bsz))=\int_{\real^d}g(\bsz)p(\bsz)\rd\bsz
\end{align}
for a probability density function $p$.  Often, though not always,  we may
use transformations such as those in \cite{devr:1986} to find a function $\phi:[0,1]^s\to\real^d$
such that $\bsz=\phi(\bsx)\sim p$ when $\bsx\sim\dunif[0,1]^s$. 
Our problem is in dimension $d$, the solution comes from dimension
$s$ and we do not always have $s=d$.

Normalizing flows \citep{reze:moha:2015} provide
numerically obtained transformations $\psi$ from a base
distribution such as $\dnorm(0,I_s)$ to
$p$ in settings where we might not find a solution
in \cite{devr:1986}.  
Then we may take $\phi = \psi\circ \Phi^{-1}$ where $\Phi^{-1}$ is the $\dnorm(0,1)$
inverse cumulative distribution function applied componentwise.
Some gains from employing RQMC for normalizing flows
were reported in \cite{andr:2026}. \cite{liu:2026} devises normalizing
flows customized to RQMC.

Both of those approaches reduce to evaluating
\begin{align}\label{eq:muasintegral}
\mu=\mu(f)=\int_{[0,1]^s}f(\bsx)\rd\bsx 
\end{align}
where $f(\bsx) = g(\phi(\bsx))$.
Evaluating~\eqref{eq:muasintegral} under various conditions on $f$
is the problem we will consider in depth. 
In Section~\ref{sec:nondevroye} we remark
on extensions to include  acceptance-rejection sampling, Markov
chain Monte Carlo (MCMC) and particle methods.

Standard MC, QMC and RQMC algorithms all estimate $\mu$ by
\begin{align}\label{eq:muhat}
\hat\mu = \hat\mu(f) = \frac1n \sum_{i=0}^{n-1} f(\bsx_i)
\end{align}
for points $\bsx_0,\dots,\bsx_{n-1}\in[0,1]^s$.  Where they differ
is in how the sample points are chosen.
We will write $\hat\mu_n$ and $\hat\mu_n(f)$ when it is necessary
to specify $n$.


The treatment here takes a statistical view of the numerical problem.
In a high dimensional setting we can only get a sparse sampling of
the inputs while the quantity of interest depends on the function value
at places we did not sample.  Choosing where to sample the integrand
is very similar to problems of experimental design.  The QMC sampling methods
have been developed using number theory and the algebra of finite fields, reminiscent of
classical experimental design work by \cite{bose:1938} and \cite{kemp:1947} and many others with a more recent account in \cite{heda:sloa:stuf:1999}.

Once we have an estimate of $\mu$ we will want to judge its accuracy
using the data from that experiment.  
Injecting randomness into the computations helps to make the sampled
points representative of the whole set of points and that supports
statistical methods to estimate variance and compute approximate
confidence intervals. Inference based on randomization goes back
at least to \cite{fish:1926}.  For more recent coverage see \cite{athe:imbe:2017}.

This (frequentist) 
statistical approach is not the only way to approach these problems.
Bayesian numerical analysis (see \cite{cock:oate:sull:giro:2019} and \cite{diac:1988} who traces the ideas back
to Poincar\'e) is similarly based on statistical
thinking, largely Gaussian process models for the integrand $f$.  
Non-statistical methods such as
information-based complexity \citep{nova:wozn:2010} are more
studied than statistical ones. They work by formulating a class
of integrands and a class of algorithms and finding upper and lower
bounds for the errors at given cost or for the cost with a given
error budget.  It is common to study worst case errors
over functions in some class but there are also approaches 
based on averages over a function class and there is a lot of
attention paid to more general information than function evaluations,
such as linear functionals of $f$.

\begin{figure}
\centering
\includegraphics[width=1.0\hsize]{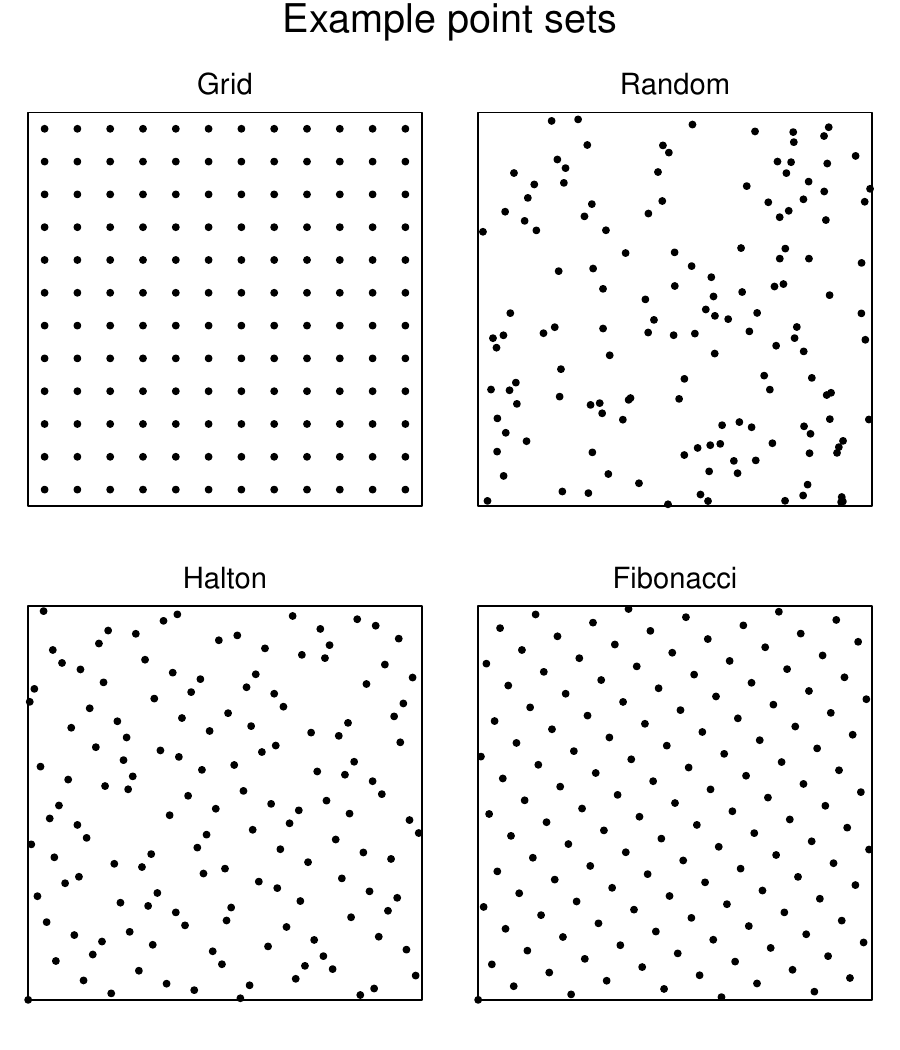}
\caption{\label{fig:examplepoints}
Each panel shows $144$ points in the unit square $[0,1]^2$:
a grid, random points, Halton points and a Fibonacci lattice.
}
\end{figure}

Our first task is to select $n$ points in  $[0,1]^s$.
Figure~\ref{fig:examplepoints} shows some ways
we might do that.  The upper left has a plain grid which
can be very effective with unequal point weighting in low
dimensions but is infeasible to use in high dimensions.
The upper right shows plain MC points.  They have randomly
located clumps and voids that we seek to even out with
QMC sampling.  The lower left shows Halton points
that we describe in Section~\ref{sec:halton} and the lower
right has a Fibonacci lattice. Lattices are described in Section~\ref{sec:lattices}.

RQMC sampling has some differences from MC sampling that
can trip up new users.  
Many RQMC rules are designed for specific sample sizes
like powers of $2$ or large prime numbers. Then using $n=1000$
could be much worse than $n=1024$ and even worse than $n=512$.
Skipping the first few points (like an MCMC burn-in) can be
very harmful. So can thinning by taking every $k$'th point for
some $k>1$. 
In some cases, skipping even one point can harm the rate of convergence,
not just the constant.
These things don't always cause problems but if one does not
understand how the points were constructed it is better to use them
as directed than to fiddle with them.  

Another difference between (R)QMC and MC arises when we
compare empirical results to theoretical ones. Plain MC estimates
have variance $\var(f(\bsx))/n$ so the finite sample variance
exactly matches the asymptotic one.  While there can be exceptions,
such as rare event calculations, sample variances in MC for
different sample sizes commonly follow an $O(1/n)$ pattern.
For (R)QMC the empirical results are often a surprise.
The available theory describes worst case bounds that
are never seen in applications.  The
most optimistic results give errors or mean squared
errors (MSEs)  decreasing at a rate like $O(n^{-\alpha+\epsilon})$
for an integer $\alpha\ge1$ and any $\epsilon>0$.
Empirical results commonly show non-integer rates.
The best results seem to require a larger value of $n$
to be seen when $s$ is large.


\section{Numerical examples}\label{sec:examples}
We begin with some numerical examples to show RQMC in action.
We will see below how the RQMC estimates $\hat\mu_n$
in~\eqref{eq:muhat} are unbiased estimates of $\mu$ and that
we can take $R$ independent replicates of them.

There are tradeoffs among the various (R)QMC
methods.  A reasonable default is to use the
digital nets of \cite{sobo:1967:tran} selecting the `direction numbers'
from \cite{joe:kuo:2008} with the scramble of \cite{mato:1998:2}
for $n=2^m$ and $R=10$ independent replications of the RQMC estimates. 
The results can be compared to using $nR$ plain Monte Carlo points,
or to some other number of MC points to match wall clock time.
Those scrambled Sobol' points have several implementations described below. 
It is possible to construct integrands $f$ where the default works badly, 
but that would be true of other defaults and the worst case integrands 
are pretty contrived.

We can try the default on some example functions from \cite{surj:bing:2013}.
They list six functions from emulators.  We use uniform distributions
over the ranges quoted, except for the Borehole where we use one
Gaussian variable and one log-normal variable as indicated in  \cite{surj:bing:2013}.
These functions have motivations
from science and engineering. The motivations are about approximation,
not integration but these functions at least are not purely synthetic
combinations of polynomials, sinusoids and exponentials.
One of them just has tabular data, so we
look at the other five.  For each of them we compute the $R=10$
replicates of $n\in\{2^{10},2^{13},2^{16},2^{19}\}$ integral estimates using
scrambled Sobol' points. For the RQMC estimates we 
divide the sample variance $S^2$ of the $R$ replicated estimates by $R$.
For MC we divide the sample variance $s^2$ of the $nR$ function
evaluations by $nR$.  The results are in Table~\ref{tab:variance-all}.
The whole computation took under 5 seconds on an M4 MacBook Air as described in the  Materials and Methods section. 

\begin{table}[t]
\centering
\footnotesize
\begin{tabular}{l S[table-format=1.2e-2] S[table-format=1.2e-2] S[table-format=1.2e-2]}
\toprule
{$n=2^{10}$} & {RQMC $S^2/R$} & {MC $s^2/(nR)$} & {Ratio} \\
\midrule
Borehole & 7.50e-06 & 7.93e-02 & 1.06e+04 \\
OTL Circuit & 6.06e-11 & 1.26e-04 & 2.08e+06 \\
Piston & 1.63e-10 & 1.92e-06 & 1.18e+04 \\
Robot Arm & 7.99e-06 & 2.72e-05 & 3.40e+00 \\
Wing Weight & 8.18e-07 & 2.29e-01 & 2.80e+05 \\
\midrule
{$n=2^{13}$} & {RQMC $S^2/R$} & {MC $s^2/(nR)$} & {Ratio} \\
\midrule
Borehole & 3.85e-07 & 1.00e-02 & 2.60e+04 \\
OTL Circuit & 3.42e-13 & 1.61e-05 & 4.70e+07 \\
Piston & 4.37e-12 & 2.38e-07 & 5.44e+04 \\
Robot Arm & 1.23e-07 & 3.41e-06 & 2.78e+01 \\
Wing Weight & 1.33e-08 & 2.83e-02 & 2.13e+06 \\
\midrule
{$n=2^{16}$} & {RQMC $S^2/R$} & {MC $s^2/(nR)$} & {Ratio} \\
\midrule
Borehole & 1.06e-09 & 1.25e-03 & 1.18e+06 \\
OTL Circuit & 6.38e-18 & 2.00e-06 & 3.13e+11 \\
Piston & 9.27e-15 & 2.96e-08 & 3.20e+06 \\
Robot Arm & 3.49e-09 & 4.28e-07 & 1.23e+02 \\
Wing Weight & 1.70e-12 & 3.52e-03 & 2.07e+09 \\
\midrule
{$n=2^{19}$} & {RQMC $S^2/R$} & {MC $s^2/(nR)$} & {Ratio} \\
\midrule
Borehole & 1.95e-11 & 1.56e-04 & 7.99e+06 \\
OTL Circuit & 9.19e-22 & 2.49e-07 & 2.72e+14 \\
Piston & 4.21e-17 & 3.71e-09 & 8.81e+07 \\
Robot Arm & 9.15e-11 & 5.35e-08 & 5.85e+02 \\
Wing Weight & 2.18e-16 & 4.41e-04 & 2.02e+12 \\
\bottomrule
\end{tabular}
\caption{Variance of the pooled RQMC estimate ($R=10$ replicates) versus i.i.d.\ Monte Carlo variance, both using the same total budget $nR$.\label{tab:variance-all}}
\end{table}


In hindsight the large variance reductions here 
can be explained in part by noting that the dimensions
range from 6 to 10, most of them are bounded functions and
they have relatively simple closed form expressions.  The
borehole function as given has a tiny chance of taking
the logarithm of a negative value. That could be eliminated
by replacing a Gaussian by a truncated one,
but it was not necessary here.

It is common
in RQMC that explanations happen in hindsight connecting results to
theorems instead of using the theorems prospectively.  One
reason is that RQMC is most effective on integrands with
an ANOVA decomposition dominated by main effects and
lower dimensional interactions.  As \cite{diac:1988} notes
when asking what it means to understand a function,
we cannot necessarily know such things just by looking at a formula.

With only $R=10$ replicates a user could not be very confident
of the RQMC variance or the variance ratio.
The evidence in the table does however make it clear that
RQMC is much more accurate than MC for these problems.
Reasonable next steps include looking for confidence
intervals on the estimate or trying several other RQMC
methods.

\section{Integration and discrepancy}\label{sec:intanddiscrep}

We begin this section with some notation.  
Some further notation is introduced closer to where it is used.

We use
$\ints$ for the integers,
$\natu = \{1,2,\dots\}$ for the natural numbers, 
and $\natu_0=\{0\}\cup\natu$.
For integers $b\ge2$ the set $\ints_b=\{0,1,\dots,b-1\}$
is used to denote the base $b$ digits of real numbers.

There are several different useful versions of the unit cube.
Some QMC constructions are easier to present for $[0,1)^s$
because it partitions easily into congruent subcubes.
Riemann integration and multivariate bounded variation
are best presented for $[0,1]^s$.  Some singular integrands
are only finite within $(0,1)^s$. As a result different versions
of the unit cube are used in different places. 
RQMC uses randomized points and there are no differences among
the $\dunif(0,1)^s$, $\dunif[0,1]^s$ and $\dunif[0,1)^s$ distributions,
so RQMC sampling does not depend on which unit cube we choose.

The point $\bsx_i\in[0,1]^s$ has components $x_{ij}$.
Sometimes we use a comma, as in $x_{n-1,j}$ for typographical
clarity, but there is no difference in meaning between $x_{ij}$ and $x_{i,j}$.

We use $1{:}s$ to denote the set $\{1,2,\dots,s\}$ of components.
The cardinality of $u\subseteq1{:}s$ is $|u|$ and the largest
index $j\in u$ is $\lceil u\rceil$ with $\lceil\emptyset\rceil=0$
by convention.

For $f:[0,1]^s\to\real$ and 
$u\subseteq1{:}s$ we use  $\partial^uf$ to denote
the partial derivative of $f$ taken once with respect to $x_j$
for each $j\in u$ with $\partial^\emptyset f=f$.

For $x\in\real$ we use $\lfloor x\rfloor$ for the greatest
integer no larger than $x$. We also use $\{x\}\in[0,1)$ for the fractional
part $x-\lfloor x\rfloor$ of $x$.  The distinction from the set
containing $x$ only will be clear from context.

Summations are commonly over the first variable in them, 
with a range given by context.  For example $\sum_{|u|\ge1}$
denotes $\sum_{u\subseteq1:s,|u|\ge1}$.




\subsection{Classical quadrature/cubature}
Computing the integral in \eqref{eq:muasintegral} is a classical problem
addressed by approaches such as Simpson's method, Newton-Cotes rules and Gauss rules.  See \cite{davrab}
for detailed accounts.  For $s=1$ and $f$ with $r$ continuous derivatives on $[0,1]$
we can get an error of $O(n^{-r})$ using $n$ function evaluations.  The implied constant 
in that rate is typically proportional to $\Vert f^{(r)}\Vert_\infty$. For $s\ge2$,
the convergence rate in iterated integration falls to $O(n^{-r/s})$.

That rate cannot be improved as was shown
by a lower bound due to \cite{bakh:1959}. Let $\cf(s,r,M)$ be the set of functions on
 $[0,1]^s$ with all partial derivatives of order up to $r$ bounded by $M\in(0,\infty)$.
Then there exists $k>0$ such that for any $\bsx_0,\dots,\bsx_{n-1}\in[0,1]^s$ there
is $f\in \cf(s,r,M)$ with $|\hat\mu-\mu|>kn^{-r/s}$.  

In one dimensional problems, some of the best methods sample
disproportionately many points near the edges of the interval $[0,1]$ 
and apply unequal weights to the resulting function values.
Bakhvalov's  `curse of dimension' result  also applies to
estimates of the form $\sum_{i=0}^{n-1}w_if(\bsx_i)$ for $w_i\in\real$.
Iterated integration in particular, using tensor products of univariate
integration weights, cannot attain a good convergence rate and the resulting
product of univariate weights becomes extremely unequal for large $s$.

It follows that deterministic multivariate quadrature (also called cubature)  is not
very effective for functions of  high dimension or limited smoothness.
If $\sigma^2=\var(f(\bsx))<\infty$ under $\bsx\sim\dunif[0,1]^s$ 
then MC sampling has a root mean squared error (RMSE)
of $\e( (\hat\mu-\mu)^2)^{1/2} = \sigma/\sqrt{n}$. 
By this measure MC is not harmed by high dimension or
non-smoothness. Nor does it benefit from low dimension
or high smoothness.  This RMSE for MC does not contradict Bakhvalov's theorem.
That theorem allows for an adversarial integrand $f$ to be chosen
after the $\bsx_i$ are revealed while for MC we assume that the 
sample points are generated randomly after $f$ has been specified.

In MC sampling we can get an unbiased estimate of $\sigma^2$, namely
$s^2 = \sum_{i=0}^{n-1}(f(\bsx_i)-\hat\mu)^2/(n-1)$ and from that
construct asymptotic confidence intervals for $\mu$ based on the
central limit theorem.  This is a great practical advantage of MC.
For deterministic rules we may be able to bound $|\hat\mu-\mu|$
in terms of some norm of $f$ but that norm can be harder
to estimate than $\mu$. Then we may have to fall back on
using $|\hat\mu_{2n}-\hat\mu_n|$ as an estimate of the
error in $\hat\mu_n$. 
See Chapter 4.9 of \cite{davrab} for a discussion.

\subsection{Discrepancy and star discrepancy}

Our goal is to estimate $\mu$, the expected value of $f(\bsx)$
for $\bsx$ uniformly distributed on $[0,1]^d$.  The estimate
$\hat\mu$ is the expected value of $f(\bsx_i)$ for
$i$ uniformly distributed on $\{0,1,2,\dots,n-1\}$. When the $\bsx_i$
are distinct as we assume from here on, $\hat\mu$ is the
expected value of $f(\bsx)$ for $\bsx\sim\dunif\{\bsx_0,\bsx_1,\dots,\bsx_{n-1}\}$.
We want to choose points $\bsx_i$ so that the discrete uniform
distribution on our $n$ points is close to the continuous uniform
distribution on $[0,1]^s$.  This is a common goal.  We often
specify our problems over a continuum but have to resort to
a discrete set of inputs for computations. 
We will see below how
matching these two uniform distributions gives accurate estimation.

We use discrepancies to quantify the distance between 
our discrete and continuous uniform distributions.  There are many
discrepancies.  For  $A\subset[0,1]^s$ let $\vol(A)$
be its Lebesgue measure and $\wh\vol(A) = (1/n)\sum_{i=0}^{n-1}1\{\bsx_i\in A\}$.
Then the signed discrepancy of $A$ is $\delta(A)=\wh\vol(A)-\vol(A)$.  We would
like this to be near zero for a rich collection of sets $A$.

For an axis parallel box $[\bsa,\bsb)\subset[0,1]^s$ let
$\delta(\bsa,\bsb) = \delta([\bsa,\bsb))$.
One of the most natural discrepancies to consider is
\begin{align}\label{eq:extremediscrep}
D_n=D_n(\bsx_0,\dots,\bsx_{n-1}) = \sup_{[\bsa,\bsb)\subset[0,1]^s}\,
  \bigl|\delta(\bsa,\bsb)\bigr|.
\end{align}
An infinite sequence of points $\bsx_0,\bsx_1,\bsx_2,\dots$ is called uniformly
distributed if $D_n(\bsx_0,\dots,\bsx_{n-1})\to0$ as $n\to\infty$.
If that sequence is uniformly distributed then $\hat\mu_n(f)\to\mu(f)$ holds
for any Riemann integrable $f$. Conversely if $\hat\mu_n(f)$ converges
to a limit as $n\to\infty$ for any uniformly distributed sequence then
$f$ is Riemann integrable \citep{nied:1978}.
This discrepancy result is analogous to the law of large numbers
for random $\bsx_i$. It does not give a bound or a rate.

A grid of $n=r^s$ points in $[0,1]^s$ will have very bad discrepancy for large $s$.
For any positive $\epsilon$, we can find a set $[\bsa,\bsb)$ of volume less than $\epsilon$
that has $n/r=r^{s-1}$ of the points. It follows that $D_n \ge1/r=n^{-1/s}$
for the grid.

The most commonly studied discrepancy is taken over `anchored boxes'
of the form $[\bszero,\bsa)$ for $\bsa\in[0,1]^s$.
We define the local discrepancy $\delta(\bsa)=\delta([\bszero,\bsa))$
and let
\begin{align*}
D_n^*=D^*_n(\bsx_0,\dots,\bsx_{n-1}) = \sup_{\bsa\in[0,1]^s}|\delta(\bsa)|.
\end{align*}
For $s=1$, this is the Kolmogorov-Smirnov distance between $\dunif\{x_0,\dots,x_{n-1}\}$ and $\dunif[0,1]$.

Boxes anchored at the origin are not necessarily more important
than others, and the origin is not necessarily the most important vertex
of $[0,1]^s$.  Instead, the star discrepancy is a convenient
shortcut.  It satisfies $D_n^* \le D_n\le 2^sD^*_n$
where the upper bound  comes from an inclusion-exclusion argument
over $2^s$ anchored boxes.
Then to show that some points are asymptotically uniform we only need
to show that $D_n^*\to0$.  The star discrepancy also appears in
the Koksma-Hlawka inequality
\begin{align}\label{eq:kh}
|\hat\mu_n-\mu| \le D_n^*(\bsx_0,\dots,\bsx_{n-1})V_{\hk}(f)
\end{align}
from \cite{koks:1942} and \cite{hlaw:1961} 
where $V_{\hk}$ denotes total variation in the sense of Hardy and Krause.
Our integration error is bounded by a product of two terms, $D_n^*$
that quantifies how non-uniform our points are, and $V_{\hk}$ that
quantifies how non-constant our integrand is.  For $s=1$, $V_{\hk}$
is the usual total variation. For general $s$, the definition is 
more involved.  See \cite{variation}.
When $V_{\hk}(f)<\infty$, we say that $f\in\bvhk$.

Equation~\eqref{eq:kh} is like Chebyshev's inequality in probability.
Like Chebyshev's inequality it is tight.
For given points $\bsx_i$, an adversary can construct
an integrand $f=f_\epsilon$ with $|\hat\mu-\mu|>V_{\hk}(f)(D_n^*-\epsilon)$
for any $\epsilon>0$.  
It is also loose in the sense that the right hand side can be far larger
than the error it bounds, just as Chebyshev's 
bound gives the loose bound $\Pr( |\dnorm(0,1)|>10)\le 0.01$.

There are known constructions of $n$ points $\bsx_i\in[0,1]^s$
for which $D_n^* = O( \log(n)^{s-1}/n)$. For extensible constructions taken
as the first $n$ points of an infinite sequence it is possible to have
$D_n^* = O( \log(n)^s/n)$.   This shows that we can potentially do much better
than MC asymptotically, when $f\in\bvhk$.

These rates are commonly written as $O(n^{-1+\epsilon})$
for any $\epsilon>0$.  For moderately large $s$,
$\log(n)^{s-1}/n$ is by no means small for computationally relevant $n$.
Bounds of that magnitude are unrealistically pessimistic.
One never sees errors so large in applications.  For a lengthy
discussion see \cite{thelogs}.
In addition to the Koksma-Hlawka inequality using
the worst case over all $f\in\bvhk$, no points can yield $D_n^*>1$
despite the bounds being proportional to $\log(n)^{s-1}/n$.
\cite{faur:lemi:2017} study some of the discrepancy bounds
and show that for large $s$ it requires an enormous $n$
before the bound is below 1.  Some RQMC methods 
in Section~\ref{sec:rqmc} control the powers of $\log(n)$
so they are not relevant until they are negligible (if then).

The Koksma-Hlawka inequality shows that the interval
$\hat\mu\pm D_n^*V_{\hk}$ always contains $\mu$.  We cannot use
it as a 100\% confidence interval because both factors are typically unknown.
For modestly large $s$, 
the star discrepancy is very expensive to compute and $V_{\hk}$ is
ordinarily far harder to compute than $\mu$ is.

\subsection{Other discrepancies}
Discrepancy theory is a rich area of mathematics going back at
least to \cite{weyl:1916}.
For an introduction to discrepancy theory see the collection \cite{chen:sriv:trav:2014}.
Here we mention a few of the results.
A famous result by \cite{roth:1954},
shows that no sequence of points in $[0,1]^s$ can have $D_n^* = o((\log n)^{(s-1)/2}/n)$ and 
it is still not known what the best attainable rate is for general $s$.
Discrepancies can be defined by taking the supremum over other collections
of sets besides axis parallel boxes.  Triangles, disks, boxes not parallel
to the axes and general convex sets have all been considered.  The rates
for those other sets are much less favorable than for axis parallel boxes.
Fortunately, the rates for axis parallel boxes are what we need for most Koksma-Hlawka
type results.
For fixed $s$ the best known way to compute
the star discrepancy costs $O(n^{s/2+1})$ \citep{dobk:epps:mitc:1996},
and computing  $D_n^*$ is NP-hard as $s=n\to\infty$
\citep{gnew:sriv:winz:2009}.

For $s=1$, the Koksma-Hlawka inequality arises by writing $\hat\mu-\mu$
as an inner product of $\delta$ and another functions and bounding it by $\Vert \delta\Vert_\infty$ times the $L_1$ norm of that other 
function.  For $s\ge1$ we sum $2^s-1$ such inner products \citep{hlaw:1961}.
There are $L_2$ versions based on Cauchy-Schwarz.  
The $L_2$ version can be computed in $O(sn^2)$
time using a formula from \cite{warn:1972}.

Just as discrepancies have been extended from axis parallel
boxes to less favorable regions, QMC sampling has also
been extended from $[0,1]^s$ to less favorable sets.
For some entry points to that literature see
\cite{brau:saff:sloa:wome:2014} for spheres,
\cite{dong:etal:2024} and references therein for
triangles,  and \cite{basu:owen:2015} for some more general spaces.

\subsection{Optimization}

Most constructions of point sets with low discrepancy are based
on the algebra of finite fields, or on Fourier analysis.
A very direct approach to point set construction is to seek a numerical
optimization of the discrepancy $D_n$ or more commonly the star
discrepancy $D_n^*$.  So far that has been limited to relatively small
values of $n$ and $s$.  
For some recent work, see  \cite{rusc:kirk:bron:lemu:rus:2024}
who use deep neural networks
and \cite{clem:doer:klam:paqu:2025} who give provably optimal
star discrepancy values for some small $n$ and~$s$.

\section{Digital QMC constructions}\label{sec:digital}
In this section we look at some digital constructions of
QMC points.  The point $\bsx_i$ has $j$'th component
$x_{ij}$ which has digits $x_{ijk}\in\ints_b$ in base $b$.
These constructions specify the digits $x_{ijk}$. 
Base $b=2$ is the most important choice. 
Section~\ref{sec:lattices} 
has brief remarks on lattice rules for QMC.

\subsection{Van der Corput sequence}
We begin with the sequence of \citep{vand:1935:I}.
The integer $i\ge0$ can be written in base $2$ as
$$
i = \sum_{k=1}^\infty a_{ik}2^{k-1}\quad a_{ik}\in\{0,1\}
$$
where only finitely many $a_{ik}$ are nonzero.
Then we let
$$
x_i = \sum_{k=1}^\infty a_{ik}2^{-k}\quad a_{ik}\in\{0,1\}.
$$
We generate $x_i\in[0,1)$ using the same bits as $i\in\natu_0$
but in the opposite order.

\begin{table}
\centering
\begin{tabular}{lrll}
\toprule
$i$ & $i_{(2)}$ & $x_{i(2)}$ & $x_i$\\
\midrule
$0$ & $0$ & $0.0$ & $0$\\
$1$ & $1$ & $0.1$ & $0.5$\\
$2$ & $10$ & $0.01$ & $0.25$\\
$3$ & $11$ & $0.11$ & $0.75$\\
$4$ & $100$ & $0.001$ & $0.125$\\
$5$ & $101$ & $0.101$ & $0.625$\\
$6$ & $110$ & $0.011$ & $0.375$\\
$7$ & $111$ & $0.111$ & $0.875$\\
\bottomrule
\end{tabular}
\caption{\label{tab:vdc} Eight
van der Corput points. The middle two
columns are in base $2$.}
\end{table}

Table~\ref{tab:vdc} shows the first few van der Corput points.
As integers alternate between even and odd, the points $x_i$
alternately belong to $[0,1/2)$ and $[1/2,1)$.  As a result
the first $n$ points have nearly equal numbers of points
in the left and right halves of $[0,1)$.  More generally any
consecutive $2^r$ points for $r\in\natu$ have exactly
one point in each of $2^r$ subintervals of the form $[a/2^r,(a+1)/2^r)$
for $a\in\ints_{2^r}$.
If $n=2^m$ then
$D_n^*=1/n$ and generally $D_n^* = O(\log(n)/n)$.

\subsection{Halton sequences}\label{sec:halton}
Van der Corput's digit reversal 
in base $b\ge2$ is
\begin{align}\label{eq:vdcbaseb}
x_i = \phi_b(i) :=\sum_{k=1}^\infty a_{ik}b^{-k}\quad\text{for\quad $i=\sum_{k=1}^\infty a_{ik}b^{k-1}$}
\end{align}
for digits $a_{ik}\in\ints_b$.
For the Halton sequence \citep{halt:1960} we generate $\bsx_i\in[0,1)^s$
by taking $x_{ij}$ from van der Corput's method in base $b_j$.
It is typical to take $b_j$ equal to the $j$'th prime. As long as the bases
$b_j$ are relatively prime, the Halton sequence has 
$D_n^* = O( (\log n)^s/n)$ as $n\to\infty$ where the implied constant
grows rapidly with $s$.  

\begin{figure}
\centering
\includegraphics[width=1.0\hsize]{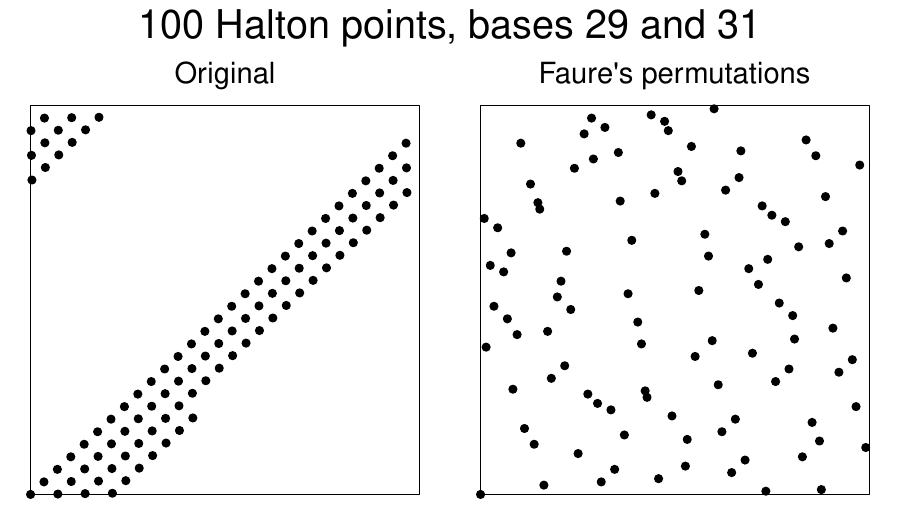}
\caption{\label{fig:haltonscram}
The first 100 Halton points.
Left: $x_{i,11}$ versus $x_{i,10}$.
Right: generalized Halton using permutations from \cite{faur:1992}.
}
\end{figure}

Figure~\ref{fig:haltonscram} shows 
the 10th and 11th components of Halton points
using bases $29$ and $31$ respectively.  The first 100 points 
are not very uniformly distributed especially in
comparison to their first and second components (bases 2 and 3)
shown in Figure~\ref{fig:examplepoints}.
The Halton points can be improved by permuting
their digits, taking $x_{ij} = \sum_{k=1}^\infty \pi_{b_j}(a_{ik})b_j^{-k}$
where $\pi_{b_j}$ is a permutation of $\ints_{b_j}$.
Such generalized Halton sequences preserve the 
convergence rate for $D_n^*$. If $\pi_b(0)=0$ then
the summation for $\phi_b(i)$ in~\eqref{eq:vdcbaseb}
is finite.  \cite{braa:well:1979} made a computer
search for effective permutations. \cite{faur:1992} gave a simple
recursive choice of permutations. \cite{faur:lemi:2009} survey many
deterministic proposals.

Like many QMC constructions, the first indices of these
points have better equidistribution than later ones.
When we know which of the $s$ inputs of $f$
are most important then it pays to use the first
components of $\bsx_i$ for those inputs.

For $s=2$, we can split the unit square into $72$
rectangular cells of size $1/8\times 1/9$. Then any
consecutive $72$ points of the Halton sequence for $s=2$
place exactly one point in each of those cells.  This
follows by the Chinese remainder theorem. More generally
if $n$ is a multiple of $2^m3^\ell$ then every cell in
a grid  of size $2^{-m}\times 3^{-\ell}$ gets exactly $n/(2^m3^\ell)$ 
of  the points.
The projections of the second and third variables have a comparable
equidistribution property when $n$ is a multiple of $3^m5^\ell$ and
generally to get such exact equidistribution over a collection of variables
requires $n$ to be a multiple of a corresponding product of powers 
of prime numbers.

\subsection{Digital nets}
Digital nets  allow us to use the same base $b$ for
every component of $\bsx$. Then sample sizes that are powers of $b$
can be good for many coordinate projections of the $\bsx_i$.
We begin by defining the cells mentioned above for the case where
the same base is used in all dimensions.  For a vector $\bsk=(k_1,\dots,k_s)\in\natu_0^s$ and integers $0\le a_j<b^{k_j}$ the
set
\begin{align}\label{eq:elemint}
E(\bsk,\bsa) = \prod_{j=1}^s \Bigl[ \frac{a_j}{b^{k_j}}, \frac{a_j+1}{b^{k_j}}\Bigr)
\end{align}
is an `elementary interval' in base $b$.
For integers $m\ge t\ge0$ and  $n=b^m$, the points
 $\bsx_0,\dots,\bsx_{n-1}\in[0,1]^s$ are a $(t,m,s)$-net
in base $b$ if $\wh\vol(E(\bsk,\bsa))=b^t/n$ whenever
$\vol(E(\bsk,\bsa))=b^t/n$. Every elementary interval that `deserves'
$b^t$ points gets exactly $b^t$ points.  We say that those
elementary intervals have been balanced by the net.
Smaller $t$ are better and $t=0$ is the best.  However,
nets with $t>0$ allow smaller values of $b$. They balance
elementary intervals of volume $b^{t-m}$ which can be smaller, hence
better, than we could get with $t=0$ and larger $b$.

\begin{figure}
\centering
\includegraphics[width=1.0\hsize]{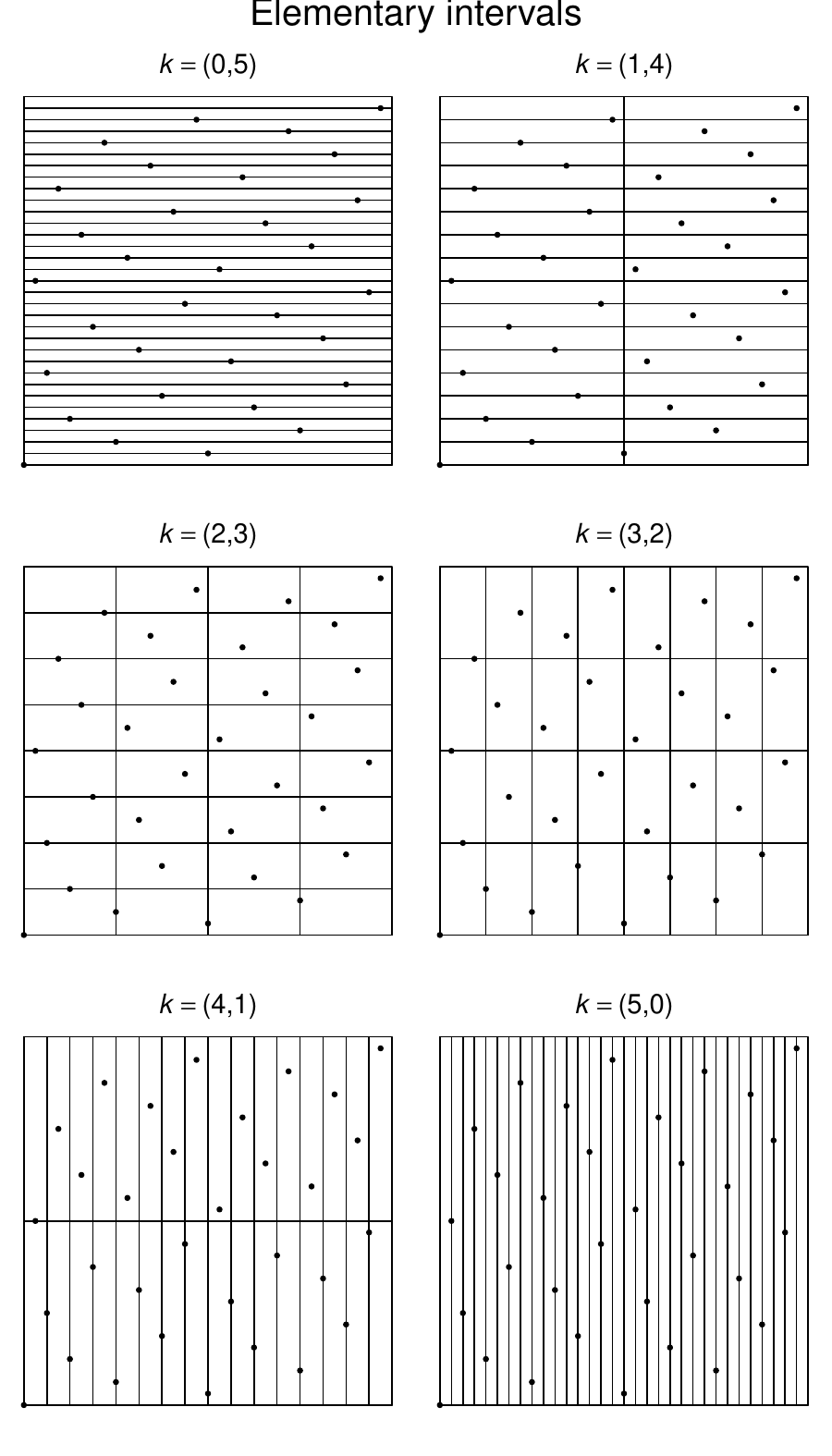}
\caption{\label{fig:element2}
  Some elementary intervals in base $2$ in $[0,1]^2$
  along with 32 Hammersley points in base $2$.
  Here 32 points are simultaneously stratified
  over $192$ strata.
}
\end{figure}

Figure~\ref{fig:element2} shows some two dimensional elementary intervals in base $2$
along with $32$ points of a $(0,5,2)$-net in base $2$.  Those $32$ points balance
$6\times 32$ elementary intervals.  For large $s$ and increasing $n$ the number of
balanced intervals can be much larger than $n$.  Many of the points are on the
boundaries of their elementary intervals.  Random scrambling of that
$(0,5,2)$-net puts one point
uniformly distributed within each elementary interval. That holds simultaneously for all
of those partitions.

\cite{faur:1982} devised some $(0,m,s)$-nets in base $b$ where
$b$ is a prime number and $b\ge s$.  We must take $n\ge s^m$
and for large $s$ this sharply limits the size of $m$ that we can use.
\cite{sobo:1967:tran} developed $(t,m,s)$-nets in base $b=2$. The value
of $t$ grows with $s$.  \cite{schu:schm:2009} track the best known
values of $t$ given $s$ and $b$ and $m$ along with lower bounds for them.
The published values of $t$ apply to the points $\bsx_i\in[0,1]^s$.
The lower dimensional coordinate projections of those points can
have much smaller values of $t$.  For Sobol' points the univariate
margins $x_{0,j},\dots,x_{2^m-1,j}$ have $t=0$.

Commonly used digital nets are taken from digital sequences.
For $t\ge0$ a $(t,s)$-sequence in base $b$ has points $\bsx_i$ for $i\ge0$
where $\bsx_{rb^m},\dots,\bsx_{(r+1)b^m-1}$ 
form a $(t,m,s)$-net in base $b$
for all integers $r\ge0$ and $m\ge t$.
We get an infinite sequence of $(t,m,s)$-nets.  We can
group those into $b$-tuples of nets and get an infinite sequence of $(t,m+1,s)$-nets
which can be grouped into $(t,m+2,s)$-nets and so on ad infinitum.
Faure's $(0,m,s)$-nets are derived from $(0,s)$-sequences in prime bases $b\ge s$.
Sobol's $(t,m,s)$-nets in base $2$ are derived from $(t,s)$-sequences in base $2$.
The value $t$ for Sobol's sequences holds for the whole infinite sequence and
smaller values of $t$ may hold for nets derived from them \citep{schm:1999}.

\subsection{Digital net construction}

The construction of digital  nets is described in depth in the texts by \cite{nied:1992}
and \cite{dick:pill:2010}. Here we give an explanation that serves as the basis
for RQMC in Section~\ref{sec:rqmc}.  We will generate the first $E$ bits of 
$2^m$ points $\bsx_i\in[0,1)^s$.  To do this, we use $s$ generator matrices
$C_j\in\{0,1\}^{E\times m}$ with $E\ge m$.

Let $\vec i$ have the first $m$ bits of the integer $i$ using the same base $2$
expansion that we used for the van der Corput points.
For $x\in[0,1)$ let $\vec x$ be the first $E$ bits in the binary expansion of $x$.
When $x$ has two binary expansions we choose the one that terminates in $0$s,
for instance, preferring $0.01\bar{0}$ over  $0.00\bar{1}$ for $x=1/4$. The only
number in $\natu_0\cap[0,1)$ is $0$ and $\vec 0$ is all zeros either way.

Digital nets, like Sobol's, in base $2$ are generated as follows.
For $i=0,\dots,n-1$ and $j=1,\dots,s$ let
\begin{align}\label{eq:netmatrix}
\vec x_{ij} = C_j \vec i
\end{align}
using arithmetic modulo $2$. Then $\bsx_i = (x_{i1},\dots,x_{is})$
where $x_{ij}$ is the point in $[0,1)$ with the given value of $\vec x_{ij}$.
In the above description each $x_{ij}$ will be an integer multiple of $2^{-E}$.
The binary matrices $C_j$ are usually constructed using formulas
to give infinitely many rows, though only finitely many of those are
relevant for floating point computations.

The quality parameter $t$ of the digital net can be obtained from the
matrices $C_j$.  Suppose that we take the first $r_j$ rows of $C_j$
for $j=1,\dots,s$ and make one binary matrix out of all those rows. 
Let $r=\sum_{j=1}^sr_j$.
If the matrix we make always has full rank over $\ints_2$ when $r\le m-t$, then the digital net has parameter $t$.  

The first $b^m$ van der Corput points are a digital net with  $C_1$ equal to the $m\times m$ identity matrix in base $b$.
The Hammersley points in $[0,1]^2$ use that $C_1$ along with $C_2$ an $m\times m$
matrix of zeros except for ones on the other diagonal.  Taking $r_1$ rows of $C_1$
and $r_2=m-r_1$ rows of $C_2$ gives a permutation matrix which is of full rank.
More generally \cite{hamm:1960} prepends $i/n$ to each of the first $n$ Halton points
to increase their dimension by one.

These digital constructions are available in any prime number base $b$
using arithmetic modulo $b$.
They are also available in any prime power base $b=p^k$ where $p$ is
prime and $k\in\natu$. In that case the matrices $C_j$ must
have elements from the finite field  $\bbf_{p^k}$.
Then we can take  $\vec{x}_{ij} = 
\phi_1( C_j \phi_0(\vec i))$ where $\phi_0$ is a bijection mapping
$\ints_{p^k}$ onto $\bbf_{p^k}$ and $\phi_1$ is a bijection
mapping $\bbf_{p^k}$ onto $\ints_{p^k}$.

\section{RQMC}\label{sec:rqmc}

RQMC points are constructed by applying some randomizations
to QMC points.  They satisfy $\bsx_i\sim\dunif[0,1]^s$ individually
while having low discrepancy collectively:
$\Pr( D_n^* \le A \log(n)^s/n)=1$ for some $A<\infty$.

With RQMC we can get $R\ge2$ IID unbiased estimates $\hat\mu_{n,r}$ 
of $\mu$.
Then  $\hat\mu_n = (1/R)\sum_{r=1}^R\hat\mu_{n,r}$ is unbiased for $\mu$
and
$$
\wh\var(\hat\mu_n) = \frac{S^2}R\quad\text{for}\quad S^2=\frac1{R-1}\sum_{r=1}^R(\hat\mu_{n,r}-\hat\mu_n)^2
$$
as an unbiased estimate of $\var(\hat\mu)$.  This is much easier than attempting to compute
$V_{\hk}(f)$.  In special circumstances described below, some randomizations can improve
the QMC error $O(n^{-1+\epsilon})$ to an RMSE of $O(n^{-3/2+\epsilon})$.

\subsection{Rotations and digital shifts}
The simplest way to get RQMC points is to apply
a random shift.  Given QMC points $\bsa_i\in[0,1]^s$ for $0\le i<n$, 
take $\bsu\sim\dunif[0,1]^s$ and return $\bsx_i = \{\bsa_i + \bsu\}$.
This approach, due to \cite{cran:patt:1976}, is called a Cranley-Patterson
rotation.  Another simple randomization is to write
$a_{ij} = \sum_{k=1}^\infty a_{ijk}/2^k$ for bits $a_{ijk}\in\{0,1\}$
and let $x_{ijk} = a_{ijk} + u_{jk} \tmod 2$
where $u_{jk}\simiid \dunif\{0,1\}$. This randomization, now known as a `digital shift'
appears in \cite{lecu:lemi:1999} who credit R.\ Couture for it.
It produces $\bsx_i\sim\dunif[0,1]^s$.  If the points $\bsa_i$ are
a digital net then so are the points $\bsx_i$.

Those two randomizations both bring $\bsx_i\sim\dunif[0,1]^s$
making $\hat\mu$ a random variable with $\e(\hat\mu)=\mu$.
Then if $f\in\bvhk$, $\var(\hat\mu) = O(n^{-2+\epsilon})$.
The most studied randomizations for digital nets
are  nested uniform scrambling and linear matrix scrambling, that
we describe next.

\subsection{Nested uniform scrambling}\label{sec:nus}
Let $\pi$ be a uniform random permutation of $\ints_b$.
Then taking $x_{ijk} = \pi(a_{ijk})$ makes $x_{ijk}\sim\dunif(\ints_b)$,
or even take $x_{ijk} = \pi_j(a_{ijk})$ 
where $\pi_1,\dots,\pi_s$
are IID uniformly distributed permutations. 
We could also take $x_{ijk} = \pi_{jk}(a_{ijk})$.
The nested uniform scramble (NUS) of \cite{rtms}
takes this further where the permutation applied
to $a_{ijk}$ depends on $j$ and $k$ and $a_{ij\ell}$
for all $\ell <k$ and all the permutations are independent.

Nested uniform scrambling of a  $(t,m,s)$-net in base $b$
has the following properties, where limits are as $n=b^m\to\infty$
and $\epsilon>0$:
\begin{compactenum}[\ \ \bf1)]
\item $\e(\hat\mu)=\mu$.
\item $f\in L_2[0,1]^s$ $\implies$ $\var(\hat\mu)=o(1/n)$.
\item $f\in L_2[0,1]^s$ $\implies$ $\var(\hat\mu)\le \Gamma\sigma^2/n$, some $\Gamma<\infty$.
\item $f\in L_{1+\epsilon}[0,1]^s$ $\implies$ $\Pr( \lim_{n\to\infty}\hat\mu_n=\mu)=1$.
\item $f$ `smooth' $\implies$ $\var(\hat\mu) = O( \log(n)^{s-1}/n^3)$.
  \end{compactenum}
  Property 1 follows because $\bsx_i\sim\dunif[0,1]^s$.  Property 2 is
  from \cite{snetvar}. Property 3 is in \cite{snetvar} for $t=0$ and in \cite{snxs} for all $t$.
The strong law in property 4 is from \cite{owen:rudo:2020}.
For property 5, smoothness means that $\partial^u f\in L_2[0,1]^s$ for all $u\subseteq1{:}s$.
That result is from \cite{smoovar} with a correction in \cite{localanti}. The weakest
conditions for it are in \cite{yue:mao:1999}.
\cite{loh:2003} proves a central limit theorem for NUS sampling under smoothness
conditions when $t=0$.

Properties 1 through 4 show that scrambled nets can handle
integrands that are singular or otherwise fail to be in BVHK
so long as they have finite variance.  Functions $f:[0,1]^s\to\{0,1\}$
with $f(\bsx) = 1\{\bsx\in S\}$
are an important special case because for $s\ge2$ they are generally
not in BVHK unless $S$ has axis parallel boundaries.

Property 3 shows that scrambling protects against powers of $\log(n)$.
Those powers of $\log(n)$ cannot make the RMSE
larger than $\sqrt{\Gamma}\sigma/\sqrt{n}$.

The variance for these scrambled nets has a contribution from
each term in a  Walsh function expansion of $f$ \citep{dick:pill:2010}. 
Those terms contribute uncorrelated errors whose variance is a computable multiple
of their MC variance.
The `coarsest' Walsh functions (constant within large elementary intervals) contribute no variance, the finest
contribute the same as they would under MC, and the ones in between
contribute at most $\Gamma$ times their MC variance.  As $n$ increases
more terms move to the zero variance category, providing the $o(1/n)$
variance rate.   The variance of $\hat\mu_n(f)$  depends on
details of how $f$'s Walsh coefficients decay.
The $O(n^{-3+\epsilon})$
rate for smooth functions comes from a bound on Walsh function coefficients
in terms of derivatives of $f$.
If $f\in L_2$ is not smooth the variance may still appear to decay
as some power of $n$ but that power may not be predictable
before we compute $\hat\mu$.

To store the permutations for a nested uniform
scramble (NUS) takes space proportional to $ns$.
One only needs to store the first $m$
permutations because digits $k>m$ of $x_{ij}$
can be taken IID $\dunif(\ints_b)$.
After generating $m$ digits of $x_{ij}$ we can
add $u_{ij}/b^m$ to them where $u_{ij}\simiid \dunif[0,1]$.
The NUS storage is smaller for $b=2$ because a permutation of $\ints_2$
only needs one bit. 
For $n$ points in $s$ dimensions with $b=2$ and $m\ge2$ we only need 
$$
s\times\sum_{k=1}^{m}2^{k-1}=s(2^{m}-1)=s(n-1)
$$
bits.
This storage  is less of a burden now than 
in the 1990s.  

Higher order digital nets can attain even better convergence rates.
To integrate over $[0,1]^s$ they use a $(t,m,sd)$-net
of points $\bsu_i$ for a positive integer $d$.  Each integration variable is
constructed by interlacing the digits of $d$ variables in the net.
For example with $s=1$ and $d=2$, $x_{i1}$ has base $b$ digits
$u_{i11}u_{i21}u_{i12}u_{i22}u_{i13}u_{i23}\cdots$.
When $f$ has square integrable mixed partial derivatives of order
$\alpha\ge1$ in each variable then \cite{dick:2011} shows
that scrambling higher order nets yields 
$\var(\hat\mu) = O( n^{-2\min(d,\alpha)-1+\epsilon})$.
Such rates are seldom seen in difficult problems.  Section \ref{sec:bestrates}
explores that point for this and other algorithms.

\subsection{Linear matrix scrambling}
The linear matrix scramble (LMS) of \cite{mato:1998:2}
is now used more than NUS because it uses less memory.
It has the same variance as NUS. In particular it has the
same near $n^{-3}$ variance on smooth integrands
and the same bound $\Gamma$ for finite $n$ as NUS.

LMS is simplest to present for nets in base $b=2$. There it takes the form 
\begin{align}\label{eq:lms}
\vec{x}_{ij} = M_j C_j \vec{i} + \vec{u}_j \ \tmod 2
\end{align}
for generator matrices $C_j\in\{0,1\}^{E\times m}$,
lower triangular random matrices $M_j\in\{0,1\}^{E\times E}$
and vectors $\vec{u}_j\in\{0,1\}^E$.  The components
of $\vec{u}_j$ are IID $\dunif\{0,1\}$ as are the components
of $M_j$ below the diagonal.  The diagonal components of $M_j$
are all equal to $1$, so $M_jC_j$ has full rank $m\le E$,
when as usual $C_j$ has rank $m$.
Adding $\vec{u}_j$ ensures that $\bsx_i\sim\dunif[0,1]^s$.

The values $\hat\mu$ from LMS do not have the same
distribution as those from NUS. For nets with $t=0$ and smooth
enough integrands, \cite{loh:2003} proves a central
limit theorem for NUS.  On the other hand \cite{superpolyone} show
that some such smooth integrands lead to $\hat\mu$ with a kurtosis
that diverges to $\infty$ as $n\to\infty$ using LMS on Sobol'
sequences. So it is clear that the
fourth moments differ between these two forms of scrambling.

\subsection{Other scrambles}
In addition to his matrix scramble, \cite{mato:1998:2} describes 
how to do NUS by caching random seeds and
regenerating permutations as needed while \cite{frie:kell:2002}  generate permutations
on the fly as needed. 
\cite{burl:2020}  and \cite{phar:etal:2023} use hashing schemes to generate 
permutations for $x_{ijk}$ that depend on $j$ and $(a_{i,j,1},\dots,a_{i,j,k-1})$.
Because the scrambles for different components of $\bsx_i$ are dependent, the points are not 
uniformly distributed over $[0,1]^s$, though the hashing may mean that only very strange integrands 
show significant bias.  Exact uniformity may be less important
than speed in graphics applications and quality is judged in part on how the
resulting images look.
Their implementations use fixed seeds instead of generating them.

\subsection{Gain coefficients}

The bounds $\Gamma$  in Section~\ref{sec:nus} are called `gain coefficients'.
They quantify the worst case ratio between an RQMC variance and an MC
variance taken over all non-trivial $f\in L_2$.
For Sobol' sequences they are a power of $2$ that can
grow exponentially with $s$ \citep{nonzerogain} but cannot exceed $n$.
For scrambled Faure points they cannot exceed $\exp(1)\doteq 2.718$.
Empirically the scrambled Sobol' points usually provide more accurate
estimates than Faure points, which are quite difficult to use for large $s$.
A nested uniform or linear matrix scramble of Halton points provides
a gain coefficient that is $O(\log(s))$ \citep{scrambledhalton}, but it 
does not provide an $O(n^{-3+\epsilon})$ variance for smooth
integrands.  The coarse scrambling of \cite{suzu:2026} scrambles component
$j$ of a base $b$ net using a base $b^{e_j}$ scramble where $e_j$
is the degree of the polynomial over $\bbf_b$ used to construct $C_j$.
It can attain variance $O(n^{-3+\epsilon})$ for smooth integrands 
and it has gain coefficients that are $O(\log(s))$. 
It is well suited to integrands of low truncation dimension
as opposed to low superposition dimension where NUS and LMS
do well; those notions are defined in Section~\ref{sec:anova}.

\subsection{Confidence intervals}

We generally prefer confidence intervals to variance estimates.
In special circumstances, such as bounded integrands, it is 
possible to obtain (conservative) confidence intervals for RQMC.
See \cite{jain:hick:owen:soro:2026} who adapt the betting
confidence intervals of \cite{waud:ramd:2024} to RQMC.

Commonly used confidence intervals only attain their
coverage asymptotically.
Letting $R\to\infty$ for fixed $n$ allows standard
confidence intervals to be based on the central limit theorem.
See \cite{naka:tuff:2024} for details.  For a fixed budget of $nR$
function evaluations we can get better accuracy with large $n$
and small $R$ because we only get the Monte Carlo rate in $R$
but we get a better rate in~$n$. This makes large $R$ unattractive.

\cite{ci4rqmc} report on simulations over 
6 families of integrands in  4 different dimensions
using 4 values of $R$ for 5 RQMC algorithms at 5 sample sizes.
Those 2400 test cases were used to make 1000 95\% approximate confidence
intervals via these algorithms: Student's $t$,
the percentile bootstrap, and the bootstrap $t$.  
A 95\% interval was deemed to be a failure if it would cover below 94\%.
If the observed coverage was below 927 of 1000 trials that was taken
to be evidence of failure. The percentile bootstrap was seen to have 1689
failures, the bootstrap $t$ had 81 and Student's $t$ had 3.
Those failures of Student's $t$ were all for $R=5$.  There were none for $R\in\{10,20,30\}$.
This shows that it may not be necessary to use very large $R$
to form reasonable approximate confidence intervals.

Some theoretical understanding of that result is developing.
\cite{skewrqmc} 
show that the skewness of $\hat\mu$
from LMS scrambled Sobol' points is $O(n^\epsilon)$. 
Also if the generator matrices are drawn using independent
random bits, the skewness is $O(n^{-1/2+\epsilon})$.
From a result in \cite{superpolyone}, the sample average of a smooth main
effect will have a kurtosis growing at the rate $O(n^{1-\epsilon})$ as $n\to\infty$.
The bootstrap $t$ is ordinarily a very good method to get
confidence intervals for the mean
of non-Gaussian data \citep{hall:1988}  but Student's $t$ is very good on nearly
symmetric random variables and high kurtosis also gives it high coverage.
Student's $t$ can have high coverage in settings where the kurtosis
is so high we do not get a good estimate of variance \citep{efro:1969}.

When the LMS estimates have high kurtosis we expect
to have $\Pr( S^2/R < \var(\hat\mu_n))>1/2$, even though
$\e( S^2)=\var(\hat\mu_n)$.
Then the variance reductions with LMS from small $R$ 
will typically be overestimated.  

If we want a good estimate of $\var(\hat\mu_n)$ as opposed
to a good estimate $\hat\mu_n$, then we do need a
larger value of $R$.  We might also switch from LMS to
NUS.  Large kurtoses are not as big a problem for NUS
as in LMS.  The data from \cite{ci4rqmc}  are available at \cite{lecu:rqmc:wsc23:data}.
The sample value of the excess kurtosis of $\hat\mu$
using NUS was below 1.4 for all 5 sample sizes, all 4 dimensions
and 5 of the 6 integrands.  The exceptional integrand comes
from Johnson's $S_u$ distribution which gives $f(\bsx)$ very heavy tails
similar to a lognormal.

\subsection{Median of means}

Near symmetry and high kurtosis of the LMS estimate $\hat\mu$
give it a variance dominated by outliers. That 
motivates taking the median of replicated
scrambled Sobol' estimates $\hat\mu_1,\dots,\hat\mu_R$
instead of using their average.  \cite{superpolymulti} show that
the median has an RMSE of $O(n^{-c\log(n)/s})$, for any $c<3\log(2)/\pi^2\approx0.21$
for analytic integrands on $[0,1]^s$, as long as $R/m^2$ is bounded away from
$0$ as $m=\log_2(n)\to\infty$.
\cite{pan:2026:automatic} shows that the median of 
scrambled Sobol' point estimates $\hat\mu_r$
still converges to $\mu$ with RMSE $O(n^{-\alpha-1/2+\epsilon})$
when $f$ has a finite generalized Hardy-Krause variation of order $\alpha\in\natu$
and there are $R\ge m$ replicates. 

Taking the median can remove other difficulties.
\cite{goda:lecu:2022} study the median of $\hat\mu_r$
when using rank one lattices (described in Section~\ref{sec:lattices}).
They select the lattice rules randomly with a
distribution  tuned for one set of interaction  weights
(as described in Section~\ref{sec:weights}). The integrand might
be much better suited to very different weights.  The
median estimate is robust to such misspecification.

\subsection{Variance reduction methods}

RQMC can be combined with the sort of variance reductions one
does in MC but there are some differences.  
First, there isn't much to gain from antithetic sampling as the
RQMC points are already almost antithetic in their
low dimensional projections.

Similarly there may be little gain from
control variates.  See \cite{hick:lemi:owen:2005}.
In MC a good control variate
correlates well with the integrand.  For RQMC
we  want the RQMC errors in integrating
the integrand to correlate with the RQMC errors
in integrating the control variate.  
Often the control variates are integrated much
more accurately than the integrand and can
then offer little extra benefit.
The optimal control variate coefficient is different
for RQMC than for MC and it depends on $n$.

Conditional Monte Carlo is known as Rao-Blackwell\-ization
in MCMC and it is called
pre-integration in QMC \citep{grie:kuo:leov:sloa:2018}.
There we might integrate out one of the $s$ variables
in closed form or by using a classical quadrature rule
of negligible error.  It may seem surprising that replacing a 100
dimensional problem by one of 99 dimensions could
be helpful. It can in fact help because the $s-1$
dimensional integral can be much smoother than the
original.  When we have to integrate with respect
to $\dnorm(0,I)$,  \cite{liu:preintegration} 
show how to select a linear combination of the Gaussians
to integrate out.

RQMC can be combined with importance sampling.
The interaction between those two approaches is subtle.
An early reference is \cite{chel:1976}.
For some recent work in this area see
\cite{ouya:wang:he:2024}
who use it to obtain an RMSE of $O(n^{-3/2+\epsilon})$
for some unbounded integrands.

Multilevel Monte Carlo (MLMC) 
writes a complicated integral as a telescoping sum of simpler
integrals and estimates that sum term by term. 
\cite{hein:1998,hein:2001} introduces it for
integral equations and parametric integration. \cite{gile:2008} develops it for
stochastic differential equations.
It can bring a big efficiency gain when the terms have
quite unequal costs and sampling variances.
\cite{gile:wate:2009} merge RQMC
into MLMC to value some financial options
defined via stochastic differential equations.
They report a large improvement
over  MLMC alone  from using randomly shifted lattice rules.

\section{ANOVA and effective dimension}\label{sec:anova}

Mindful of how large $\log(n)^{s-1}/n$ can be,
\cite{brat:fox:nied:1992} restricted their implementation of
$(t,m,s)$-nets in base $b$ to $s\le 12$.
Shortly thereafter \cite{pask:trau:1995} reported 
some very successful uses of QMC for some financial
valuation problems with $s=360$.
The specific integrands in that work were proprietary,
but some similar integrands were studied
by \cite{cafl:moro:owen:1997} who found that
those integrands were nearly additive.
QMC points like those of 
Sobol' are typically stratified in their one dimensional
margins. That is, for all $j\in1{:}s$
every interval $[a/n,(a+1)/n)$
for $a\in\ints_n$ has exactly one of the $x_{ij}$.
The additive parts of those integrands could be
integrated very accurately and the remainder contributed
only a tiny error.

We can quantify `nearly additive' using the ANOVA decomposition
\citep{fish:mack:1923}.
It can be extended to $f\in L_2[0,1]^s$  \citep{sobo:1969}
or to  any square integrable function of $s$ independent inputs
as in \cite{hoef:1948} and \cite{efro:stei:1981}.  We write
\begin{align}\label{eq:anova}
f(\bsx) = \sum_{u\subseteq 1:s}f_u(\bsx)
\end{align}
where the effect $f_u(\bsx)$ only depends on $\bsx$  through $\bsx_u$
for $u\subseteq1{:}s$. 

An ANOVA decomposition attributes
variance to $v$ instead of $u$ when $v\subsetneq u$. It does so
via a recursion where
$$
f_u(\bsx) = \int_{[0,1]^{s-|u|}} \biggl( f(\bsx) - \sum_{v\subsetneq u}f_v(\bsx) \biggr)\rd \bsx_{-u}
$$
starting with $f_\emptyset(\bsx)=\mu$ for all $\bsx$.
The ANOVA effects $f_u$ have $\var(f_u(\bsx))=:\sigma^2_u$ and by
orthogonality of the $f_u$ we have  $\sum_{u\subseteq1:s}\sigma_u^2=\sigma^2$.
\cite{cafl:moro:owen:1997} estimate that for one of those integrands
the main effects satisfy
$\sum_{j=1}^{360}\sigma^2_{\{j\}}\approx 0.9996\sigma^2$. A more
difficult integrand was then included that is only about $94.1$\% additive.
They say an integrand has effective dimension $r$ in the superposition
sense if $\sum_{|u|\le r}\sigma^2_u\ge 0.99\sigma^2$
and it has effective dimension $r$ in the truncation sense if
$\sum_{u\subseteq1:r}\sigma^2_u\ge 0.99\sigma^2$.
Favorable cases for (R)QMC have low effective dimension
along with enough regularity in $f_u$ for small $|u|$ to benefit
from low discrepancy sampling.

The error  $\hat\mu-\mu$ is a sum of errors in integrating
$2^s-1$ ANOVA effects $f_u$ for $u\ne\emptyset$.
In favorable settings the errors for small $|u|$ are
small because $\bsx_{i,u}\in [0,1]^{|u|}$ for $0\le i<n$ have low
discrepancy and the errors for large $|u|$ are small
because $\Vert f_u\Vert$ is negligible. In RQMC those $2^s-1$
errors are uncorrelated.
A similar decomposition argument of \cite{hlaw:1961}
uses what is now called an anchored decomposition instead
of the ANOVA.  
\cite{kuo:sloa:wasi:wozn:2010} describe such decompositions
in general.

The effective dimension is tricky to estimate.  A related notion is the
mean dimension 
$$
\nu(f) = \frac{\sum_{u\subseteq1:s}|u|\sigma^2_u}{\sum_{u\subseteq1:s}\sigma^2_u}
$$
available when the denominator, $\sigma^2$, is positive.
Using identities from \cite{jans:1999} and \cite{meandim} we can write the
numerator as the expected value of
\begin{align}\label{eq:4nu}
\frac12\sum_{j=1}^s \bigl(f(\bsx) - f(\bsx_{-j}{:}\tilde\bsx_j)\bigr)^2
\end{align}
where $\bsx_{-j}{:}\tilde\bsx_j$ is the point $\bsx$ with $x_j$ replaced by $\tilde x_j$,
and $\bsx,\tilde\bsx\simiid\dunif[0,1]^s$. Equation~\eqref{eq:4nu} supports sampling
based estimates of $\nu(f)$.


\section{Weighted spaces and tractability}\label{sec:weights}
Effective dimension and mean dimension pertain to a single integrand.
We can use weighted reproducing kernel Hilbert spaces to characterize
a class of integrands, with special interest in such classes that are
favorable for (R)QMC.  This work was initiated by \cite{sloa:wozn:1998}
building on \cite{hickdisc} and there is a good survey in \cite{dick:kuo:sloa:2013}
who include results on constructions to complement the existence
results described here.

The most cited works in this literature establish tractability results, 
where as described below, the effort to integrate over $[0,1]^s$
hardly depends on $s$. In those settings we choose $\bsx_i$
and study the error for a worst case $f$.  We see below how that
can be done using very strong assumptions on $f$.  Other settings
with $f$ chosen first followed by  random points have been
studied \citep{hick:wozn:2001} but they have received much less attention.
\cite{lecu:2009:fin} describes a setting where we pick points $\bsx_i$ and the
devil picks $f$ and by playing second we gain a factor of $O(n^{-1/2})$
which Bakhvalov also saw in his formulations.

We begin with weights $\gamma_u\ge0$, commonly all $\gamma_u>0$ and
then define the squared norm
$$
\Vert f\Vert_{\bsgamma}^2 =\sum_{u\subseteq1:s}\frac1{\gamma_u}
\int_{[0,1]^{|u|}} \biggl( \int_{[0,1]^{s-|u|}} \partial^u f(\bsx)\rd\bsx_{-u}\biggr)^2\rd\bsx_u
$$
where $\bsgamma$ contains all $\gamma_u$. 
We let $\cf_{\bsgamma} = \{f:[0,1]^s\to\real\mid \Vert f\Vert_{\bsgamma}<\infty\}$.
This is an `unanchored' space.  \cite{dick:kuo:sloa:2013} 
also describe a similar `anchored' space.

Larger weights $\gamma_u$ correspond to variable subsets
expected to be more important.  The norm penalizes functions heavily if, as quantified
by a mixed partial derivative,  they vary a lot for a variable set $u$ with small $\gamma_u$.
\cite{hick:1996:weights} used $\gamma_u = \gamma^{|u|}$ for some $\gamma>0$
and fixed $s$, now known as `order weights' because they depend only
on the  order $|u|$ of that subset. 
\cite{sloa:wozn:1998} use `product weights'
$\gamma_u=\prod_{j\in u}\gamma_j$ with $\gamma_j$ a decreasing function
of $j$.
Then sets with larger cardinalities or larger elements tend to be less important.
Product weights often take $\gamma_j = j^{-\eta}$ for
$\eta>0$, with larger $\eta$ imposing stronger penalties on interactions.

Product weights allow one to let $s\to\infty$ and find conditions
for tractability in which the cost to attain a given level of accuracy
can be made polynomial in $s$ or even bounded in $s$ (strong tractability).
\cite{dick:kuo:sloa:2013} describe many choices for the weights.  

First we define the ball
$$
\cf_{\bsgamma}(\rho) = \{f\in \cf_{\bsgamma}\mid \Vert f\Vert_{\bsgamma}\le \rho\}
$$
for $\rho>0$.
The worst case error for points $\bsx_i$ is taken to be
$$
\wce_{\bsgamma} =\wce_{\bsgamma}(\bsx_0,\dots,\bsx_{n-1}) = \sup_{f\in \cf_{\bsgamma}(1)}|\hat\mu(f)-\mu(f)|.
$$

For $n=0$, we could choose $\hat\mu=\hat\mu_0=0$
and our worst case error would be the `initial error'  $\ine_{\bsgamma} = \sup_{f\in\cf_{\bsgamma}(1)}|\mu(f)|$.  
Let $n_{\bsgamma,\varepsilon}$ be the smallest $n$ for which
there exist points with $\wce_{\bsgamma}/\ine_{\bsgamma}\le \varepsilon$.
The problem is tractable if $n_{\bsgamma,\varepsilon} \le Cs^q\varepsilon^{-p}$
for $C,p,q\in[0,\infty)$. If $q=0$ then the problem is strongly tractable
and $\varepsilon \le (C/n)^{1/p}$.

Strong tractability holds if $\sum_{j=1}^\infty \gamma_j<\infty$
which is satisfied for $\eta = 1+\delta$ for some $\delta>0$.
The exponent $p$ is
known to be in $[1,2]$ \citep{sloa:wozn:1998}
with $p=2$ corresponding to the MC rate. 
\cite{hick:wozn:2000} show that rates $O(n^{-1+\epsilon})$
(for $p$ near $1)$
can be attained if $\sum_{j=1}^{\infty}\gamma_j^{1/2}<\infty$.
This holds for product weights with $\eta = 2+\delta$.

If we have reduced the initial error by a factor of $\varepsilon$
then the error for $f/\Vert f\Vert_{\bsgamma}$ is at most $\varepsilon\times\ine_{\bsgamma}$.
From page 183 of \cite{dick:kuo:sloa:2013}, the unanchored space with $\gamma_j=j^{-\eta}$ has $\ine_{\bsgamma}=1$.
Therefore the error for integrating $f$ is at most
$\varepsilon\times\Vert f\Vert_{\bsgamma} \le\Vert f\Vert_{\bsgamma} (C/n)^{1/p}$ in
the strongly tractable setting.
It can be easy to check that $f\in \cf_{\bsgamma}$ but this
does not ensure small error for large $s$.
We also need $f$ to  be dominated by low order interactions among
the first few variables in order to avoid a large $\Vert f\Vert_{\bsgamma}$.
For example, $f=\prod_{j=1}^r(x_j-1/2)$ has $\Vert f\Vert_{\bsgamma}=(r!)^{\eta/2}$.
Then the bound only tells us that $|\hat\mu-\mu|\le (r!)^{\eta/2}(C/n)^{1/p}$
and to get this below $\varepsilon$ we need much larger $n$ for large $r$.

Instead of comparing the results to $\hat\mu=0$, we can
compare them to Monte Carlo.  To do that we may
choose $\rho$ large enough that $\max_{f\in\cf_{\bsgamma}(\rho)}\sigma^2(f)=1$.
Then MC sampling of $f\in\cf_{\bsgamma}(\rho)$ would have an RMSE of at most $1/\sqrt{n}$.
If $\gamma_u\le \gamma_{\{1\}}$
for $u\ne\emptyset$, then $\rho = \pi/\gamma_{\{1\}}^{1/2}$ 
\citep{effdimsobononper}
so $\rho=\pi$ for the product weights.
If we take $\eta=2$ which is almost enough to get $O(n^{-1+\epsilon})$
worst case error, then $\cf_{\bsgamma}(\rho)$ cannot contain any
integrand with $\sum_{|u|\ge3 }\sigma^2_u\ge0.01$
or with $\sum_{\lceil u\rceil \ge 10}\sigma^2_u\ge 0.01$ \citep{effdimsobononper}. 

\cite{bakh:1959} gets pessimistic results from a very large
function class. The tractability results using product weights
greatly reduce that pessimism but they make very optimistic assumptions
about the ANOVA decomposition of $f$.
The tractability theorems require smoothness of $f$
that MC does not require.  Those strong assumptions allow a worst
case analysis.

An average case error given fixed $\bsx_i$  is not necessarily more favorable.
The RMSE over a mean zero Gaussian process for $f$
with covariance equal to the inner product conforming to
$\Vert \cdot\Vert_{\bsgamma}$ matches the worst
case error over $\cf_{\bsgamma}(1)$.
A not very technical statement is in Proposition 6.1 of \cite{kana:etal:2018}
who cite \cite{ritt:2000}.

When we are most interested in the integral of just one
smooth function $f$, we may find that $f\in\cf_{\bsgamma}$
for all $\bsgamma$. 
Then it is hard to choose the $\bsgamma$ to work with.
In some settings there is a lot of knowledge about $f$ that
can be used to choose an appropriate $\bsgamma$.
\cite{kuo:nuye:2016} design weights
specifically for the solution of elliptic PDEs 
for phenomena such as the flow of water through a medium
with random permeability.
A median strategy is also effective
when the regularity and weights for $f$ are unknown
\citep{goda:suzu:mats:2024}.

\section{Lattices}\label{sec:lattices}

The other major branch of QMC methods is based on lattice sampling,
with point sets like the Fibonacci lattice in Figure~\ref{fig:examplepoints}.
There is only space to give a few of the key points.  There is an up to date
account in  \cite{dick:krit:pill:2022}. A less technical account suitable
for a first reading is in \cite{sloa:joe:1994}.

A rank one lattice has
\begin{align}\label{eq:r1lat}
x_{ij} = \Bigl\{\frac{a_{j}\times i}n\Bigr\} \in[0,1)\quad 0\le i<n
\end{align}
for a vector $\bsa\in\{1,2,\dots,n-1\}^s$. The components $a_j$ are 
carefully chosen integers relatively prime to $n$.  Usually $a_1=1$.
The lattices of \cite{koro:1960}  are a special case with
$\bsa = (1,a,a^2,\dots,a^{s-1})$ for $a\in\{1,2,\dots,n-1\}$.

The customary way to analyze this integration rule is to write
\begin{align*}
f(\bsx) &\simeq \sum_{\bsh\in\ints^s}\hat f(\bsh)\exp(2\pi\sqrt{-1}\bsx^\tran\bsh)\quad\text{for}\\
\hat f(\bsh) &= \int_{[0,1)^s}f(\bsx)\exp(-2\pi\sqrt{-1}\bsx^\tran\bsh)\rd\bsx.
\end{align*}
This Fourier decomposition holds in an $L_2$ sense for $f\in L_2[0,1)^s$.
It holds as an equality if $f$ has an absolutely convergent Fourier series.

For $\bsx_i$ from a rank one lattice
$$
\frac1n\sum_{i=0}^{n-1}\exp(2\pi\sqrt{-1}\bsx_i^\tran\bsh)
=\begin{cases}1, &\bsh^\tran\bsa=0\tmod n\\
0, &\text{else}
\end{cases}
$$
and then
$$
\hat\mu = \sum_{\bsh^\tran\bsa=0\tmod n}\hat f(\bsh).
$$
At the same time $\mu = \hat f(\bszero)$. Then we get an error contribution
of $\hat f(\bsh)$ from every nonzero $\bsh$ that satisfies $\bsh^\tran\bsa=0\tmod n$.  For smooth $f$,
the moduli $|\hat f(\bsh)|$ are bounded by a decreasing function of 
$\Vert\bsh\Vert$.
Then a good choice for $\bsa$ is one that does not satisfy $\bsa^\tran\bsh=0\tmod n$ for many values of $\bsh\ne\bszero$ 
with small $\Vert\bsh\Vert$.  There are numerous ways
to turn that notion into a precise figure of merit to optimize.  See \cite{lecu:mung:2016}
and \cite{lecu:mari:godi:puch:2022} for an up to date account of many of the choices
and the Latnet Builder software to choose~$\bsa$.

When $f:\real^s\to\real$ is one periodic (so $f(\bsx) = f(\bsx+\bsw)$
for $\bsw\in\ints^s$) with $r$ continuous derivatives then
the rate $O(n^{-r+\epsilon})$ can be attained for $\mu = \int_{[0,1)^s}f(\bsx)\rd\bsx$.  \cite{hick:2002} shows that
for $r=2$ and a non-periodic $f$, the rate $O(n^{-2+\epsilon})$ can
be attained by using $\tilde\bsx_i = B(\bsx_i)$ componentwise
for $B(z) = 1-|2z-1|$. That makes $f\circ B$
periodic and the lack of smoothness when any $x_j=1/2$ does not
change the rate.

The most common randomization
applied to lattices is the Cranley-Patterson rotation described
in Section~\ref{sec:rqmc}.
Randomly shifting the lattices does not generally improve their convergence rate.

The choice of whether to use lattices or digital methods seems
to be largely personal.  Neither dominates the other
in accuracy. Lattices require more tuning effort
to select the vector $\bsa$ than digital nets require.
\cite{cool:kuo:nuye:2006} show how to reduce the search
cost to $O(sn\log(n)^2)$ in a setting where both $s$ and $n$ can be very large.
However, once the vector $\bsa$ is chosen it
is quite easy to use.
For Sobol' points one has to choose the 
matrices $C_j$,
or equivalently, some quantities known as direction numbers.
The overwhelmingly popular choice there is to use the openly
published direction
numbers from \cite{joe:kuo:2008} instead of searching.
\cite{sobo:asot:krei:kuch:2011} describe some proprietary direction numbers.

\section{Rejection, MCMC and particles}\label{sec:nondevroye}

The most straightforward way to use (R)QMC is to use
a fixed transformation $\phi:[0,1]^s\to\real^d$ 
that turns $\bsx\sim\dunif[0,1]^s$
into $\phi(\bsx)\sim p$. Then we can average
the values of $f(\phi(\bsx_i))$.  The workflow is to
write the function $\phi$ and pass it $\bsx_i$
for a range of values $i$.  It is generally best for $\phi$ to invert
CDFs, for instance using $\Phi^{-1}(x_{ij})$ to generate a $\dnorm(0,1)$
random variable. 

To sample $\dnorm(\theta,\Sigma)$,
the usual approach in (R)QMC is to use $\theta + C\Phi^{-1}(\bsx)$
where $CC^\tran=\Sigma$ and $\Phi^{-1}$ is applied componentwise.
The accuracy of MC sampling is not affected by the choice of $C$
but in (R)QMC this choice can make a big difference.  Generating
Brownian motion at $d$ time points provides a good illustration
of this.  We can generate it one time step at a time, or we can generate it
via principal components \citep{acwo:broa:glas:1997} or we can 
use a Brownian bridge construction \citep{mosk:cafl:1996}
that generates the endpoint first before generating intermediate
points conditionally on the prior ones. These other algorithms
produce a rough skeleton of the Brownian path with their first
few variables and then refine it with the later ones.  For some
integrands that can make the first few variables extremely
important reducing the effective dimension.

\subsection{Acceptance-rejection}

Acceptance-rejection sampling with (R)QMC requires some extra
care and bookkeeping.  We need one or more uniform variables to generate
a proposal and another one to accept or reject it.  The number of
variables needed to accept a point is random and unbounded.
If we simply keep the unrejected points we generally
get fewer than $n$ of them. 
\cite{zhu:dick:2014} study the resulting star discrepancy. 

We can also keep making proposals using further dimensions of a given RQMC point
as \cite{hint:hofe:lemi:2022} do. That approach requires careful bookkeeping
to ensure that component $j$ of $\bsx_i$ is always used for the same
purpose, either variable generation or acceptance-rejection, and always
for the same attempt.

It is common for RQMC point sets to have many dimensions and
it is also common for acceptance-rejection methods to have an acceptance probability
very near to one \citep{horm:leyd:2004}.
In rare circumstances with a large number of rejections, we can
switch to plain MC sampling for subsequent proposals.
Padding out a high quality matrix with plain MC points produces
what is called a hybrid point set.  \cite{span:1995}
used that approach for particle transport.  When  some of the inputs are from MC we
would expect to get the ordinary MC convergence rate with good
odds of having a more favorable constant in the rate.



The acceptance-rejection decision introduces a discontinuity
that diminishes the effectiveness of (R)QMC.
See \cite{he:wang:2015} and \cite{yliu:2026}
for (R)QMC results when $f$ includes an indicator function.
Sometimes we can avoid acceptance-rejection by multiplying
the values $f(\bsx_i)$ by importance sampling weights.
That can remove the discontinuity we get from the acceptance rule.
In other settings we need unweighted draws from $p$ and
then weighting the draws is not sufficient.

\subsection{Markov chains}

(R)QMC can be used in Markov chain Monte Carlo algorithms.
Let the algorithm start with fixed or random $\bsz_0$
and then evolve as $\bsz_i=\phi(\bsz_{i-1},\bsx_i)$
where $\bsx_i\sim\dunif[0,1]^s$ for $i\ge1$.
We can write $\bsx_i = (u_{si-s+1},\dots,u_{si})$ for a
stream of values $u_i\in[0,1]$.
A usual MCMC algorithm takes $u_i\simiid\dunif[0,1]$.
We can replace this driving sequence by some more
equidistributed points.

The suitable driving sequences are those that are 
`completely uniformly distributed' or CUD.  This means that
if we form points $\bsv_i = (u_i,u_{i+1},\dots,u_{i+k-1})\in[0,1]^k$
then $D_n^*( \bsv_1,\dots,\bsv_n)\to0$ holds for all $k\in\natu$.
This is one definition of a random sequence in \cite{knut:1998:2:3}.
A CUD sequence gives a QMC version of MCMC.  Practically it
amounts to finding a small random number generator $u_1,\dots,u_N$
and taking $n=\lfloor N/s\rfloor$ steps to use essentially all of it.
For an RQMC version
we use a weakly CUD (WCUD) sequence, for which
$\Pr( D_n^*( \bsv_1,\dots,\bsv_n) > \epsilon)\to0$ holds for all $k\ge1$
and $\epsilon >0$.

Some consistency results for RQMC versions of MCMC appear in
\cite{owen:trib:2005} and \cite{chen:dick:owen:2011} emphasizing
discrete and continuous state spaces, respectively.
Empirical results in dissertations \cite{trib:2007} and \cite{chen:2011}
show strong variance reductions that grow with $n$ hinting at a
better than $O_p(n^{-1/2})$ convergence for the errors. The best
rates are for rejection-free algorithms like the Gibbs sampler and
for the posterior means of parameters in Bayesian computations.
\cite{chen:2011} establishes a better rate, but under quite strong
assumptions.

One limitation of those RQMC-based MCMC algorithms is that
error estimates based on replications only reflect variance not bias.  
That bias can be eliminated by a coupling argument of \cite{jaco:olea:atch:2020}.
Recent work by \cite{du:he:2025} uses WCUD sequences on coupled Gibbs samplers
to achieve an apparently better convergence rate for posterior means than with 
plain MCMC.

CUD sampling cannot turn a  non-ergodic sampler into an ergodic one.
If the sampler is slowly mixing then we can expect a (W)CUD driving
sequence to mix very nearly as well (or as badly) as is typical for an IID
sequence with less variation in how fast it mixes.

\subsection{Particles}

(R)QMC methods have been used in particle sampling.
At time $t\in\natu_0$ we have particles $\bsz_{0,t},\dots,\bsz_{n-1,t}\in\real^d$.
The points $\bsz_{i,0}$ are chosen from some distribution depending on context.
It may be that they are all the same, or they may be sampled IID from some
nontrivial distribution.
Then for $t\in\natu$, we let $\bsz_{i,t} = \phi_t(\bsz_{i,t-1},\bsx_{it})$
use $\bsx_{it}\in[0,1]^s$ to advance the $i$'th point.  The $i$'th particle
keeps moving until some stopping time $\tau_i\in\natu$.
The quantity of interest in \cite{lecu:leco:tuff:2008} is 
$$\mu = \e\biggl(\, \sum_{t=0}^\tau c_t(\bsz_{t})\biggr)$$
for cost functions $c_t$.

If we allow $\tau$ as large as $T$, then  an $n\times Ts$ matrix of variables
is enough to advance all of the particles. 
Instead of using $Ts$ uniform variables, the Array-QMC and Array-RQMC
algorithms reuse $s+1$ uniform variables.
\cite{leco:tuff:2004}
devise an Array-QMC solution for the case $d=1$. The idea is to match particles $z_{i,t-1}\in\real$ with
innovations $x_{it}$ so that the joint distribution $(z_{i,t-1},x_{it})$ resembles that of two independent
random variables. They do this with a QMC point set $\bsu_i\in[0,1)^2$ for $0\le i<n$.
The points $\bsu_i$ are constructed so that $u_{01}\le u_{11} \le \cdots \le u_{n-1,1}$.
The points $z_{i,t-1}$ are sorted yielding $z_{(0),t-1}\le z_{(1),t-1}\le\dots\le z_{(n-1),t-1}$. 
Then the $n$ particles are advanced via  $z_{i,t} = \phi(z_{(i),t-1},u_{i2})$.  The result is
that the update variables $u_{i2}$ are `locally uniformly distributed' over intervals of $z_{(i),t-1}$ in $\real$.
\cite{leco:tuff:2004} see much better results in numerical examples and bound the discrepancy
of $z_{i,t}$ for $0\le i<n$.

In Array-RQMC the QMC points $\bsx_{it}$ are replaced by a `fresh' randomization of some underlying QMC
points $\bsa_i$. When $\bsz_{it}$ are in $\real^d$ for $d>1$ a problem arises of how to order the points.
One very successful solution is to use a Hilbert space-filling curve
following \cite{gerb:chop:2015}.
That curve maps $[0,1]$ onto a rectangular region in $\real^d$ (containing all
of the $\bsz_{it}$) and it has a pseudo-inverse $h$ mapping that region
back to $[0,1]$.  We sort the points $\bsz_{it}$ in increasing order
of $h(\bsz_{it})$ getting $\bsz_{(i)t}$. Then QMC points $\bsa_i\in[0,1]^{s+1}$
are given a fresh randomization yielding points $\bsx_i\in[0,1]^{s+1}$
that are sorted into increasing order of $\bsx_{i1}$. Finally
$\bsz_{i,t+1} =\phi(\bsz_{(i),t},\bsx_i)$.

\cite{wos:arqmc:tr} see large gains using Array-RQMC to solve
some boundary value problems using the `walk on spheres' algorithm.
\cite{gerb:chop:2015} go beyond Array-RQMC to also handle
sequential quasi-Monte Carlo with importance reweighting
of the particles.

\subsection{Continuous time}

\cite{hofm:math:1997} use QMC instead of MC
to drive a one dimensional Langevin diffusion 
$\rd x_t  =-\alpha x_t\rd t + \beta(t)\rd W_t$ from $x_0=0$
with $\alpha>0$ and $\beta:[0,T]\to\real$ for Brownian motion $W_t$.
They note that uniformly distributed points are not enough and
use CUD points.

\cite{liu:2023}  uses WCUD points to drive a multivariate Langevin Monte Carlo
algorithm to estimate some posterior expectations.
When the potential function (negative log posterior)
has a Lipschitz gradient and is strongly convex and the step size $h$ is  small enough
she gets an integral error of $O(n^{-1+\epsilon}+h^{1/2})$ under a
BVHK condition on the integrand. She gets smaller errors for some
Bayesian posterior moments from WCUD points than from MC.

\section{Tradeoffs, limitations and gaps}\label{sec:tradeoffs}

The biggest limitation to the use of QMC and RQMC is in software,
despite the existence of good implementations of these algorithms.
A partial listing includes
SSJ \citep{lecu:buis:2005}, 
QMCPy \citep{qmcpy:2026}, the scipy.stats.qmc tools 
in SciPy version 1.7 \citep{roy:etal:2023}, QuasiMonteCarlo.jl \citep{quasimontecarlojl},
\cite{owen:rsobol}  and the
resources to build your own at \cite{nuye:magic:2017}.
They produce matrices $\cx\in[0,1]^{n\times s}$ whose $i$'th
row is $\bsx_i$ of an (R)QMC point set.  
QMCPy lets the user specify a desired error tolerance $\varepsilon$
conditionally on some assumptions about $f$.


\subsection{Software engineering}
What remains somewhat hard is
using that software in an application.  It is not necessarily hard
to code up RQMC for one application.
It is just that the alternative of using MC is almost effortless; there
we can access all our random quantities without consciously
considering whether they correspond to a given row $i$ or column~$j$.

For algorithms based on MC,
the user can call on random inputs (points or vectors or processes) from 
a vast array of implemented distributions. In (R)QMC the user must
manage the translation from quasi-uniform vectors $\bsx_i\in[0,1]^s$ to their 
desired quantities $\bsz_i$ ordinarily by inverting the CDF.
Inversion isn't always an enormous task because inverse CDFs are available
for many `named' distributions (Gaussian, Poisson, Gamma, Binomial, etc.).
But it does represent some friction compared to  plain MC.
Other transformations such as Box-Muller (advocated by \cite{okte:gonc:2011})
can also be chosen.   QMCPy incorporates some widely used transformations.

A general purpose drop-in replacement of
RQMC for MC in a language such as R or Python would make it easy
for people to compare RQMC to MC on their problems but producing
such an environment seems like
a large software engineering problem.
Similarly, the examples using WCUD points in MCMC have all been bespoke
computations for a specific model on a specific data set and there
is no code that lets a user flip a switch between MC and RQMC.

\subsection{Some subtleties}

As mentioned in the introduction, thinning (R)QMC points
or using burn-in or taking $n$ to be other than the
designed value can be very disadvantageous.
See \cite{firstsobol} for details and examples.

It is also awkward that  most QMC algorithms start off
with $\bsx_0=\bszero$.  Having a point at the origin is problematic
because $f$ could have a singularity there as often happens when
we transform uniform variables to Gaussian ones.
A poor practice is to skip
over the first Sobol' point and take the next $n=2^m$ points
in a Sobol' sequence.  That generally makes the sampled points
fail to be a $(t,m,s)$-net with a consequent loss of accuracy.
A far better practice is to use RQMC on the original $n$ points.
If one did want to skip points, then the best choice for
Sobol' sequences is to skip the first $2^M$ points for some $M\ge m$.
\cite{brat:fox:nied:1992} mention this.
The points in use are then a $(t,m,s)$-net, but they would not
retain the benefits of randomization.

Some (R)QMC methods generate errors that are $o(1/n)$.
Our experiences with MC don't prepare us for this.
Adding, removing or changing even one input point generally changes
$\hat\mu$ by an amount proportional to $1/n$, hence larger than the error.
In such cases $n=2^m+1$ can have a worse convergence
rate than $n=2^m$ has.

Halton points are not strongly affected by using sample sizes that
are round numbers like powers of ten.  The reason for this is
that for higher dimensions they have very few if any specially
good values of $n$.
They are also less affected by burn-in because any consecutive
$n$ points have a low discrepancy property. Thinning them
to every $k$'th point can make them fail to be asymptotically
uniform.

\subsection{Bayesian numerical analysis}

Bayesian numerical analysis is a counterpart to the frequentist
approach of getting estimates and confidence intervals described here.
It can use either QMC or MC points but it has some limitations
briefly described here.

One limitation is that the cost of the matrix
algebra typically grows proportionally to $n^3$.  For some
exceptions see \cite{jaga:hick:2019}. A second limitation is that uncertainty
quantification based on a Gaussian process model is
difficult.  In a frequentist approach we can study the 
uncertainty in $(1/n)\sum_{i=0}^{n-1}f(\bsx_i)$ by considering
what might have happened if the random $\bsx_i$
had come out differently.  A Bayesian approach can consider
what might have happened if $f$, modeled as a random 
function, had been different.
Modern pseudo-random number generators
have been very thoroughly tested by methods such as the Big Crush
of \cite{lecu:sima:2007}. They are one of the closest things we have to a `true model'
in mathematical science. There is no
counterpart to the Big Crush to validate a prior distribution on integrands $f$.
This makes the uncertainty quantification subjective.
For the related problem of interpolating from measured $f(\bsx_i)$ to
$f(\bsx)$ at a new point $\bsx$, Gaussian process predictions
are very effective.  This calls to mind some advice of 
\cite{box:1980} to use Bayesian methods in estimation
and sampling methods for model criticism.



\subsection{Best convergence rates}\label{sec:bestrates}

The most optimistic convergence rates proved for (R)QMC  
are seldom if ever seen for practically useful sample sizes
and integrands without a small effective dimension.
The higher order digital nets examples in \cite{dick:2011}  
have $s=1$ or $2$ and $d\le 3$.
The median of means computations for scrambled nets 
behave as predicted for smooth one dimensional examples.
\cite{superpolyone} show better than $O(n^{-3})$
squared error for the 6 dimensional OTL circuit function
but that function can be seen to have a low mean dimension. 
The periodizing change of variable used to get arbitrarily
good convergence rates for some lattice rules gives the
resulting integrands a norm that grows exponentially with $s$
\citep{kuo:sloa:wozn:2007}.
The quality parameter $t_u$ for projections of digital nets
cannot exceed $t$ and can be strictly smaller \citep{schm:2001}.
We need $n\ge b^{t_u+1}$ for the digital net property to say
anything more than that all $n$ points $\bsx_{i,u}$ are in $[0,1]^{|u|}$.
As a result we do not expect digital nets, scrambled or otherwise,
to  be much better than MC on high dimensional interactions.

\subsection{Workflow}

Our theoretical understanding of RQMC is primarily about the rate at
which the mean squared error decreases as $n\to\infty$ through
a series of good sample sizes $n$.  
The theorem conditions are typically classes $\cf$ of integrands of some
given regularity while the conclusions
are generally about worst (or average) cases over that class. 
A related approach, common in QMC has
results about how large $n$ must be to obtain $|\hat\mu_n-\mu|\le\epsilon$.

Now let's put ourselves in the position of an investigator who has just one
integrand to consider, perhaps similar to one in Table~\ref{tab:variance-all}
but more expensive to evaluate. The accuracy that this investigator will get
depends on their $f$ but not on any other $\tilde f\in\cf$.    Typically
$f\in\cf_j$ holds for a large collection of classes $j=1,\dots,J$. Some 
may be more optimistic and others less so. 
Scaling the integrand by some constant $c>0$ scales the integral, the estimate
and the error by $c$.  The scale can be a standard deviation, measure of total
variation or a Hilbert space norm and is generally harder to learn about than
$\mu$ itself is.  Some of the most precise theorems pertain to $\Vert f\Vert=1$
but the investigator will not ordinarily know the scale.

Investigators facing a new problem can run several different
RQMC evaluations and the results often come back very quickly.
When they are very focused on a single integrand $f$, their $\cf =\{f\}$
and they have to choose a solution approach comparing their $f$
to the $\cf$ for which results are available, informed by 
results they know of for some other $f$ similar to the one they're working on.
Based on some explorations of point sets and randomizations
they can learn which approaches seem to work best and then
use a larger sample size.  Our theorems do not describe 
such an interactive process that one naturally does.

This sort of workflow is not yet formally described
for RQMC.  By contrast there is some well developed scholarship on
Bayesian workflows involving MCMC algorithms and
how to use them in specific contexts \citep{baye:flow:2020}.

\subsection{Benchmarks}

The machine learning and artificial intelligence literatures make
substantial use of benchmark problems, and they can
be more influential than theorems. (R)QMC has a small set
of integrands that are commonly used such as an Asian
option finance problem and an isotropic function from \cite{keis:1996}.

Benchmarks are worked examples that complement
theorems. We can view a theorem as an infinite
set of examples.  The theorem statement may even include
the very function we care to integrate. However the conclusion
might only apply asymptotically leaving us to wonder 
about finite $n$ or about our norm on $f$.
We might similarly worry whether the integrand  represents a worst or average case
outcome that might fail to fit our integrand.  So numerically
worked examples are helpful.  They can be qualitatively similar
to our next integrand $f$ and they provide information for finite $n$.

There are a few larger collections, but RQMC could use more.
\cite{genz:1984} gives a collection of integrand families of
different difficulty levels. 
A small collection of finance integrands is in \cite{lecu:2009:fin}.
Ridge functions that let one vary the dimensionality while
holding other things fixed are in \cite{hoyt:owen:2020}.

Benchmarks eventually get overused and saturated.  The AI and
ML literatures respond to that by creating new benchmarks.

\subsection{Main application areas}

There are some application areas where (R)QMC has had
an outsized role.  Ever since \cite{pask:trau:1995} had their success
with high dimensional problems in financial valuation, (R)QMC
sampling has been widely used to value options. 
See the survey by \cite{lecu:2009:fin}.
Many of those integrands have a very low effective dimension.

QMC and RQMC both play a large role in graphical rendering.
They are included in the Oscar winning book \cite{phar:etal:2023} 
and the rendering tool `Blender'
(\url{https://developer.blender.org/docs}).
For a survey emphasizing (R)QMC see \cite{kell:2013}.
Graphical rendering requires some extremely sophisticated
algorithms that have to be very fast.  Compared to their
other tasks replacing MC by RQMC is a very minor difficulty.

\cite{kuo:schw:sloa:2012} use QMC to solve PDEs in random environments
where the $j$'th variable is the coefficient of a $j$'th eigenfunction in
a Karhunen-Lo\`eve expansion. The importance of the eigenfunctions
generally decreases with $j$.  They invented the `product and order dependent'
(POD) weights of the form $\gamma_{u} = \Gamma_{|u|}\prod_{j\in u}\gamma_j$.

(R)QMC methods are seeing use in variational Bayes.
See \cite{buch:wenz:mand:2018} and \cite{qmc4qn4vb}.
(R)QMC algorithms are embedded in
software to compute  multivariate normal and $t$
probabilities \citep{genz:bret:2009} as used in the
R package {\tt mvtnorm},  
Bayesian optimization \citep{bala:etal:2020} for uses
such as tuning hyperparameters for machine learning models,
the Cuba \citep{hahn:2005} code for integration problems in physics.


\subsection{Influence of examples}

Sometimes single or isolated examples are influential.
The best known is \cite{pask:trau:1995}
showing that some high dimensional integrals are quite amenable
to QMC, which motivated work on effective dimension
and weighted function spaces.  That scrambling can lower errors was seen
empirically in \cite{rtms} before being proved.
Sobol's point sets were devised for bounded
integrands but when colleagues used them successfully
on unbounded ones he showed why it worked \citep{sobo:1973}.
The recent work on median methods was strongly affected
by empirical results.

\section{Conclusions}\label{sec:conc}

RQMC is a compelling alternative to plain MC.  
It has very strong theoretical support, often showing
an improved rate of convergence over MC. It seldom comes out worse than MC.
The cost is some mild programming hassle.
RQMC generally requires some empirical investigation to see if the 
theoretically expected benefits materialize in a given application.
The greatest gains come for integrands that are dominated by
their main effects and low order interactions.
It can be difficult to predict when that will happen.  
The usual way to discover that an integrand is RQMC-friendly
is to run a few RQMC replicates and see how small
their variance is.  Much of the RQMC literature is about
obtaining better convergence rates or finite sample variance
reductions of 100-fold or more.  For computationally expensive
problems, five-fold reductions may already be very welcome.

\section*{Materials and Methods}

The example computations in Section~\ref{sec:examples} were done using SciPy
for the RQMC computations.  The code itself was just over 400 lines
of Python written by Claude. The author checked the code 
and it is available on request.  That code also produced the LaTeX table
in Table~\ref{tab:variance-all}.

\section*{Acknowledgments}

This paper includes results developed over many years.
My contributions depend greatly on financial support from the NSF,
the efforts of my co-authors, especially PhD advisees and,
inspirations from people at the MCQMC series of conferences.  
That series was founded
by Harald Niederreiter and organized by a steering committee
chaired first by Stefan Heinrich and then by Alexander Keller.

\bibliographystyle{imsart-nameyear} 
\bibliography{qmc}       

\begin{thebibliography}{160}

\bibitem[\protect\citeauthoryear{Acworth, Broadie and
  Glasserman}{1997}]{acwo:broa:glas:1997}
\begin{binproceedings}[author]
\bauthor{\bsnm{Acworth},~\bfnm{P.}\binits{P.}},
  \bauthor{\bsnm{Broadie},~\bfnm{M.}\binits{M.}} \AND
  \bauthor{\bsnm{Glasserman},~\bfnm{P.}\binits{P.}}
(\byear{1997}).
\btitle{A comparison of some {Monte Carlo} techniques for option pricing}.
In \bbooktitle{{Monte Carlo} and quasi-{Monte Carlo} methods '96}
(\beditor{\bfnm{H.}\binits{H.}~\bsnm{Niederreiter}},
  \beditor{\bfnm{P.}\binits{P.}~\bsnm{Hellekalek}},
  \beditor{\bfnm{G.}\binits{G.}~\bsnm{Larcher}} \AND
  \beditor{\bfnm{P.}\binits{P.}~\bsnm{Zinterhof}}, eds.)
\bpages{1--18}.
\bpublisher{Springer}.
\end{binproceedings}
\endbibitem

\bibitem[\protect\citeauthoryear{Andral}{2026}]{andr:2026}
\begin{binproceedings}[author]
\bauthor{\bsnm{Andral},~\bfnm{Charly}\binits{C.}}
(\byear{2026}).
\btitle{Combining normalizing flows and {quasi-Monte Carlo}}.
In \bbooktitle{International Conference on Monte Carlo and Quasi-Monte Carlo
  Methods in Scientific Computing}
(\beditor{\bfnm{Christiane}\binits{C.}~\bsnm{Lemieux}} \AND
  \beditor{\bfnm{Ben}\binits{B.}~\bsnm{Feng}}, eds.)
\bpages{151--168}.
\bpublisher{Springer}, \baddress{Cham, Switzerland}.
\end{binproceedings}
\endbibitem

\bibitem[\protect\citeauthoryear{Athey and Imbens}{2017}]{athe:imbe:2017}
\begin{bincollection}[author]
\bauthor{\bsnm{Athey},~\bfnm{Susan}\binits{S.}} \AND
  \bauthor{\bsnm{Imbens},~\bfnm{Guido~W}\binits{G.~W.}}
(\byear{2017}).
\btitle{The econometrics of randomized experiments}.
In \bbooktitle{Handbook of economic field experiments},
(\beditor{\bfnm{A.~V.}\binits{A.~V.}~\bsnm{Banerjee}} \AND
  \beditor{\bfnm{E.}\binits{E.}~\bsnm{Duflo}}, eds.)
\bvolume{1}
\bpages{73--140}.
\bpublisher{Elsevier}.
\end{bincollection}
\endbibitem

\bibitem[\protect\citeauthoryear{Bakhvalov}{1959}]{bakh:1959}
\begin{barticle}[author]
\bauthor{\bsnm{Bakhvalov},~\bfnm{N.~S.}\binits{N.~S.}}
(\byear{1959}).
\btitle{On approximate calculation of multiple integrals}.
\bjournal{Vestnik Moskovskogo Universiteta, Seriya Matematiki, Mehaniki,
  Astronomi, Fiziki, Himii}
\bvolume{4}
\bpages{3--18}.
\bnote{(In Russian)}.
\end{barticle}
\endbibitem

\bibitem[\protect\citeauthoryear{Balandat et~al.}{2020}]{bala:etal:2020}
\begin{barticle}[author]
\bauthor{\bsnm{Balandat},~\bfnm{Maximilian}\binits{M.}},
  \bauthor{\bsnm{Karrer},~\bfnm{Brian}\binits{B.}},
  \bauthor{\bsnm{Jiang},~\bfnm{Daniel}\binits{D.}},
  \bauthor{\bsnm{Daulton},~\bfnm{Samuel}\binits{S.}},
  \bauthor{\bsnm{Letham},~\bfnm{Benjamin}\binits{B.}},
  \bauthor{\bsnm{Wilson},~\bfnm{Andrew~Gordon}\binits{A.~G.}} \AND
  \bauthor{\bsnm{Bakshy},~\bfnm{Eytan}\binits{E.}}
(\byear{2020}).
\btitle{{BoTorch}: A framework for efficient {Monte-Carlo Bayesian}
  optimization}.
\bjournal{Advances in Neural Information Processing Systems (NeurIPS)}.
\end{barticle}
\endbibitem

\bibitem[\protect\citeauthoryear{Basu and Owen}{2017}]{basu:owen:2015}
\begin{barticle}[author]
\bauthor{\bsnm{Basu},~\bfnm{K.}\binits{K.}} \AND
  \bauthor{\bsnm{Owen},~\bfnm{A.~B.}\binits{A.~B.}}
(\byear{2017}).
\btitle{Scrambled geometric net integration over general product spaces}.
\bjournal{Foundations of Computational Mathematics}
\bvolume{17}
\bpages{467--496}.
\bnote{Published online 2015}.
\end{barticle}
\endbibitem

\bibitem[\protect\citeauthoryear{Bose}{1938}]{bose:1938}
\begin{barticle}[author]
\bauthor{\bsnm{Bose},~\bfnm{R.}\binits{R.}}
(\byear{1938}).
\btitle{On the Application of the Theory of Galois Fields to the Problem of
  Construction of Hyper-Graeco-Latin Squares}.
\bjournal{Sankhya}
\bvolume{3}
\bpages{323--338}.
\end{barticle}
\endbibitem

\bibitem[\protect\citeauthoryear{Box}{1980}]{box:1980}
\begin{barticle}[author]
\bauthor{\bsnm{Box},~\bfnm{George E.~P.}\binits{G.~E.~P.}}
(\byear{1980}).
\btitle{Sampling and {Bayes'} inference in scientific modelling and
  robustness}.
\bjournal{Journal of the Royal Statistical Society, Series A}
\bvolume{143}
\bpages{383--404}.
\end{barticle}
\endbibitem

\bibitem[\protect\citeauthoryear{Braaten and Weller}{1979}]{braa:well:1979}
\begin{barticle}[author]
\bauthor{\bsnm{Braaten},~\bfnm{Eric}\binits{E.}} \AND
  \bauthor{\bsnm{Weller},~\bfnm{George}\binits{G.}}
(\byear{1979}).
\btitle{An improved low-discrepancy sequence for multidimensional {quasi-Monte
  Carlo} integration}.
\bjournal{Journal of Computational Physics}
\bvolume{33}
\bpages{249--258}.
\end{barticle}
\endbibitem

\bibitem[\protect\citeauthoryear{Bratley, Fox and
  Niederreiter}{1992}]{brat:fox:nied:1992}
\begin{barticle}[author]
\bauthor{\bsnm{Bratley},~\bfnm{Paul}\binits{P.}},
  \bauthor{\bsnm{Fox},~\bfnm{Bennett~L}\binits{B.~L.}} \AND
  \bauthor{\bsnm{Niederreiter},~\bfnm{Harald}\binits{H.}}
(\byear{1992}).
\btitle{Implementation and tests of low-discrepancy sequences}.
\bjournal{ACM Transactions on Modeling and Computer Simulation (TOMACS)}
\bvolume{2}
\bpages{195--213}.
\end{barticle}
\endbibitem

\bibitem[\protect\citeauthoryear{Brauchart
  et~al.}{2014}]{brau:saff:sloa:wome:2014}
\begin{barticle}[author]
\bauthor{\bsnm{Brauchart},~\bfnm{Johann}\binits{J.}},
  \bauthor{\bsnm{Saff},~\bfnm{E}\binits{E.}},
  \bauthor{\bsnm{Sloan},~\bfnm{I}\binits{I.}} \AND
  \bauthor{\bsnm{Womersley},~\bfnm{R}\binits{R.}}
(\byear{2014}).
\btitle{QMC designs: optimal order quasi {Monte Carlo} integration schemes on
  the sphere}.
\bjournal{Mathematics of computation}
\bvolume{83}
\bpages{2821--2851}.
\end{barticle}
\endbibitem

\bibitem[\protect\citeauthoryear{Buchholz, Wenzel and
  Mandt}{2018}]{buch:wenz:mand:2018}
\begin{binproceedings}[author]
\bauthor{\bsnm{Buchholz},~\bfnm{A.}\binits{A.}},
  \bauthor{\bsnm{Wenzel},~\bfnm{F.}\binits{F.}} \AND
  \bauthor{\bsnm{Mandt},~\bfnm{S.}\binits{S.}}
(\byear{2018}).
\btitle{Quasi-{M}onte {C}arlo variational inference}.
In \bbooktitle{International Conference on Machine Learning}
\bpages{668--677}.
\bpublisher{PMLR}.
\end{binproceedings}
\endbibitem

\bibitem[\protect\citeauthoryear{Burley}{2020}]{burl:2020}
\begin{barticle}[author]
\bauthor{\bsnm{Burley},~\bfnm{Brent}\binits{B.}}
(\byear{2020}).
\btitle{Practical hash-based {Owen} scrambling}.
\bjournal{Journal of Computer Graphics Techniques}
\bvolume{9}
\bpages{1--20}.
\end{barticle}
\endbibitem

\bibitem[\protect\citeauthoryear{Caflisch, Morokoff and
  Owen}{1997}]{cafl:moro:owen:1997}
\begin{barticle}[author]
\bauthor{\bsnm{Caflisch},~\bfnm{R.~E.}\binits{R.~E.}},
  \bauthor{\bsnm{Morokoff},~\bfnm{W.}\binits{W.}} \AND
  \bauthor{\bsnm{Owen},~\bfnm{A.~B.}\binits{A.~B.}}
(\byear{1997}).
\btitle{Valuation of Mortgage Backed Securities using {Brownian} Bridges to
  Reduce Effective Dimension}.
\bjournal{Journal of Computational Finance}
\bvolume{1}
\bpages{27--46}.
\end{barticle}
\endbibitem

\bibitem[\protect\citeauthoryear{Chelson}{1976}]{chel:1976}
\begin{bphdthesis}[author]
\bauthor{\bsnm{Chelson},~\bfnm{P.}\binits{P.}}
(\byear{1976}).
\btitle{Quasi-random techniques for {Monte Carlo} methods},
\btype{PhD thesis},
\bpublisher{The Claremont Graduate School}.
\end{bphdthesis}
\endbibitem

\bibitem[\protect\citeauthoryear{Chen}{2011}]{chen:2011}
\begin{bphdthesis}[author]
\bauthor{\bsnm{Chen},~\bfnm{S.}\binits{S.}}
(\byear{2011}).
\btitle{Consistency and convergence rate of {Markov chain quasi-Monte Carlo}
  with examples},
\btype{PhD thesis},
\bpublisher{Stanford University}.
\end{bphdthesis}
\endbibitem

\bibitem[\protect\citeauthoryear{Chen, Dick and
  Owen}{2011}]{chen:dick:owen:2011}
\begin{barticle}[author]
\bauthor{\bsnm{Chen},~\bfnm{S.}\binits{S.}},
  \bauthor{\bsnm{Dick},~\bfnm{J.}\binits{J.}} \AND
  \bauthor{\bsnm{Owen},~\bfnm{A.~B.}\binits{A.~B.}}
(\byear{2011}).
\btitle{Consistency of {Markov chain quasi-Monte Carlo} on continuous state
  spaces}.
\bjournal{The Annals of Statistics}
\bvolume{39}
\bpages{673--701}.
\end{barticle}
\endbibitem

\bibitem[\protect\citeauthoryear{Chen, Srivastav and
  Travaglini}{2014}]{chen:sriv:trav:2014}
\begin{bbook}[author]
\beditor{\bsnm{Chen},~\bfnm{W.}\binits{W.}},
  \beditor{\bsnm{Srivastav},~\bfnm{A.}\binits{A.}} \AND
  \beditor{\bsnm{Travaglini},~\bfnm{G.}\binits{G.}}, eds.
(\byear{2014}).
\btitle{A panorama of discrepancy theory}.
\bpublisher{Springer}, \baddress{Cham, Switzerland}.
\end{bbook}
\endbibitem

\bibitem[\protect\citeauthoryear{Cl{\'e}ment
  et~al.}{2025}]{clem:doer:klam:paqu:2025}
\begin{barticle}[author]
\bauthor{\bsnm{Cl{\'e}ment},~\bfnm{Fran{\c{c}}ois}\binits{F.}},
  \bauthor{\bsnm{Doerr},~\bfnm{Carola}\binits{C.}},
  \bauthor{\bsnm{Klamroth},~\bfnm{Kathrin}\binits{K.}} \AND
  \bauthor{\bsnm{Paquete},~\bfnm{Lu{\'\i}s}\binits{L.}}
(\byear{2025}).
\btitle{Constructing optimal star discrepancy sets}.
\bjournal{Proceedings of the American Mathematical Society, Series B}
\bvolume{12}
\bpages{78--90}.
\end{barticle}
\endbibitem

\bibitem[\protect\citeauthoryear{Cockayne
  et~al.}{2019}]{cock:oate:sull:giro:2019}
\begin{barticle}[author]
\bauthor{\bsnm{Cockayne},~\bfnm{Jon}\binits{J.}},
  \bauthor{\bsnm{Oates},~\bfnm{Chris~J}\binits{C.~J.}},
  \bauthor{\bsnm{Sullivan},~\bfnm{Timothy~John}\binits{T.~J.}} \AND
  \bauthor{\bsnm{Girolami},~\bfnm{Mark}\binits{M.}}
(\byear{2019}).
\btitle{Bayesian probabilistic numerical methods}.
\bjournal{SIAM Review}
\bvolume{61}
\bpages{756--789}.
\end{barticle}
\endbibitem

\bibitem[\protect\citeauthoryear{{SciML
  Contributors}}{2019}]{quasimontecarlojl}
\begin{bmisc}[author]
\bauthor{\bsnm{{SciML Contributors}}}
(\byear{2019}).
\btitle{{QuasiMonteCarlo.jl}: Quasi-Monte Carlo Sampling for Julia}.
\bhowpublished{\url{https://docs.sciml.ai/QuasiMonteCarlo/stable/}}.
\bnote{Julia package}.
\end{bmisc}
\endbibitem

\bibitem[\protect\citeauthoryear{Cools, Kuo and
  Nuyens}{2006}]{cool:kuo:nuye:2006}
\begin{barticle}[author]
\bauthor{\bsnm{Cools},~\bfnm{R.}\binits{R.}},
  \bauthor{\bsnm{Kuo},~\bfnm{F.}\binits{F.}} \AND
  \bauthor{\bsnm{Nuyens},~\bfnm{D.}\binits{D.}}
(\byear{2006}).
\btitle{Constructing embedded lattice rules for multivariate integration}.
\bjournal{SIAM Journal on Scientific Computing}
\bvolume{28}
\bpages{2162--2188}.
\end{barticle}
\endbibitem

\bibitem[\protect\citeauthoryear{Cranley and Patterson}{1976}]{cran:patt:1976}
\begin{barticle}[author]
\bauthor{\bsnm{Cranley},~\bfnm{R.}\binits{R.}} \AND
  \bauthor{\bsnm{Patterson},~\bfnm{T.~N.~L.}\binits{T.~N.~L.}}
(\byear{1976}).
\btitle{Randomization of number theoretic methods for multiple integration}.
\bjournal{SIAM Journal of Numerical Analysis}
\bvolume{13}
\bpages{904--914}.
\end{barticle}
\endbibitem

\bibitem[\protect\citeauthoryear{Davis and Rabinowitz}{1984}]{davrab}
\begin{bbook}[author]
\bauthor{\bsnm{Davis},~\bfnm{P.~J.}\binits{P.~J.}} \AND
  \bauthor{\bsnm{Rabinowitz},~\bfnm{P.}\binits{P.}}
(\byear{1984}).
\btitle{Methods of Numerical Integration},
\bedition{2nd} ed.
\bpublisher{Academic Press}, \baddress{San Diego}.
\end{bbook}
\endbibitem

\bibitem[\protect\citeauthoryear{Devroye}{1986}]{devr:1986}
\begin{bbook}[author]
\bauthor{\bsnm{Devroye},~\bfnm{L.}\binits{L.}}
(\byear{1986}).
\btitle{Non-uniform Random Variate Generation}.
\bpublisher{Springer}, \baddress{New York}.
\end{bbook}
\endbibitem

\bibitem[\protect\citeauthoryear{Diaconis}{1988}]{diac:1988}
\begin{binproceedings}[author]
\bauthor{\bsnm{Diaconis},~\bfnm{P.}\binits{P.}}
(\byear{1988}).
\btitle{Bayesian Numerical Analysis}.
In \bbooktitle{Statistical Decision Theory and Related Topics IV, in two
  volumes}
\bvolume{1}
\bpages{163--176}.
\end{binproceedings}
\endbibitem

\bibitem[\protect\citeauthoryear{Dick}{2011}]{dick:2011}
\begin{barticle}[author]
\bauthor{\bsnm{Dick},~\bfnm{J.}\binits{J.}}
(\byear{2011}).
\btitle{Higher order scrambled digital nets achieve the optimal rate of the
  root mean square error for smooth integrands}.
\bjournal{The Annals of Statistics}
\bvolume{39}
\bpages{1372--1398}.
\end{barticle}
\endbibitem

\bibitem[\protect\citeauthoryear{Dick, Kritzer and
  Pillichshammer}{2022}]{dick:krit:pill:2022}
\begin{bbook}[author]
\bauthor{\bsnm{Dick},~\bfnm{J.}\binits{J.}},
  \bauthor{\bsnm{Kritzer},~\bfnm{P.}\binits{P.}} \AND
  \bauthor{\bsnm{Pillichshammer},~\bfnm{F.}\binits{F.}}
(\byear{2022}).
\btitle{Lattice Rules: Numerical Integration, Approximation, and Discrepancy}.
\bpublisher{Springer Nature}.
\end{bbook}
\endbibitem

\bibitem[\protect\citeauthoryear{Dick, Kuo and
  Sloan}{2013}]{dick:kuo:sloa:2013}
\begin{barticle}[author]
\bauthor{\bsnm{Dick},~\bfnm{J.}\binits{J.}},
  \bauthor{\bsnm{Kuo},~\bfnm{F.~Y.}\binits{F.~Y.}} \AND
  \bauthor{\bsnm{Sloan},~\bfnm{I.~H.}\binits{I.~H.}}
(\byear{2013}).
\btitle{High-dimensional integration: the {quasi-Monte Carlo} way}.
\bjournal{Acta Numerica}
\bvolume{22}
\bpages{133--288}.
\end{barticle}
\endbibitem

\bibitem[\protect\citeauthoryear{Dick and
  Pillichshammer}{2010}]{dick:pill:2010}
\begin{bbook}[author]
\bauthor{\bsnm{Dick},~\bfnm{J.}\binits{J.}} \AND
  \bauthor{\bsnm{Pillichshammer},~\bfnm{F.}\binits{F.}}
(\byear{2010}).
\btitle{Digital sequences, discrepancy and quasi-{Monte Carlo} integration}.
\bpublisher{Cambridge University Press}, \baddress{Cambridge}.
\end{bbook}
\endbibitem

\bibitem[\protect\citeauthoryear{Dobkin, Eppstein and
  Mitchell}{1996}]{dobk:epps:mitc:1996}
\begin{barticle}[author]
\bauthor{\bsnm{Dobkin},~\bfnm{D.~P.}\binits{D.~P.}},
  \bauthor{\bsnm{Eppstein},~\bfnm{D.}\binits{D.}} \AND
  \bauthor{\bsnm{Mitchell},~\bfnm{D.~P.}\binits{D.~P.}}
(\byear{1996}).
\btitle{Computing the discrepancy with applications to supersampling patterns}.
\bjournal{ACM Transactions on graphics}
\bvolume{15}
\bpages{354--376}.
\end{barticle}
\endbibitem

\bibitem[\protect\citeauthoryear{Dong et~al.}{2024}]{dong:etal:2024}
\begin{barticle}[author]
\bauthor{\bsnm{Dong},~\bfnm{Gracia~Yunruo}\binits{G.~Y.}},
  \bauthor{\bsnm{Hintz},~\bfnm{Erik}\binits{E.}},
  \bauthor{\bsnm{Hofert},~\bfnm{Marius}\binits{M.}} \AND
  \bauthor{\bsnm{Lemieux},~\bfnm{Christiane}\binits{C.}}
(\byear{2024}).
\btitle{Randomized {quasi-Monte Carlo} methods on triangles: extensible
  lattices and sequences}.
\bjournal{Methodology and Computing in Applied Probability}
\bvolume{26}
\bpages{15}.
\end{barticle}
\endbibitem

\bibitem[\protect\citeauthoryear{Du and He}{2025}]{du:he:2025}
\begin{barticle}[author]
\bauthor{\bsnm{Du},~\bfnm{Jiarui}\binits{J.}} \AND
  \bauthor{\bsnm{He},~\bfnm{Zhijian}\binits{Z.}}
(\byear{2025}).
\btitle{Unbiased {Markov chain quasi-Monte Carlo for Gibbs} samplers}.
\bjournal{SIAM/ASA Journal on Uncertainty Quantification}
\bvolume{13}
\bpages{1174--1199}.
\end{barticle}
\endbibitem

\bibitem[\protect\citeauthoryear{Efron}{1969}]{efro:1969}
\begin{barticle}[author]
\bauthor{\bsnm{Efron},~\bfnm{Bradley}\binits{B.}}
(\byear{1969}).
\btitle{Student's $t$-test under symmetry conditions}.
\bjournal{Journal of the American Statistical Association}
\bvolume{64}
\bpages{1278--1302}.
\end{barticle}
\endbibitem

\bibitem[\protect\citeauthoryear{Efron and Stein}{1981}]{efro:stei:1981}
\begin{barticle}[author]
\bauthor{\bsnm{Efron},~\bfnm{B.}\binits{B.}} \AND
  \bauthor{\bsnm{Stein},~\bfnm{C.}\binits{C.}}
(\byear{1981}).
\btitle{The Jackknife Estimate of Variance}.
\bjournal{Annals of Statistics}
\bvolume{9}
\bpages{586--596}.
\end{barticle}
\endbibitem

\bibitem[\protect\citeauthoryear{Faure}{1982}]{faur:1982}
\begin{barticle}[author]
\bauthor{\bsnm{Faure},~\bfnm{H.}\binits{H.}}
(\byear{1982}).
\btitle{Discr\'epance de Suites Associ\'ees \`a un syst\`eme de Num\'eration
  (en Dimension $s$)}.
\bjournal{Acta Arithmetica}
\bvolume{41}
\bpages{337--351}.
\end{barticle}
\endbibitem

\bibitem[\protect\citeauthoryear{Faure}{1992}]{faur:1992}
\begin{barticle}[author]
\bauthor{\bsnm{Faure},~\bfnm{H.}\binits{H.}}
(\byear{1992}).
\btitle{Good permutations for extreme discrepancy}.
\bjournal{Journal of Number Theory}
\bvolume{42}
\bpages{47--56}.
\end{barticle}
\endbibitem

\bibitem[\protect\citeauthoryear{Faure and Lemieux}{2009}]{faur:lemi:2009}
\begin{barticle}[author]
\bauthor{\bsnm{Faure},~\bfnm{Henri}\binits{H.}} \AND
  \bauthor{\bsnm{Lemieux},~\bfnm{Christiane}\binits{C.}}
(\byear{2009}).
\btitle{Generalized {Halton} sequences in 2008: A comparative study}.
\bjournal{ACM Transactions on Modeling and Computer Simulation (TOMACS)}
\bvolume{19}
\bpages{15:1--15:31}.
\end{barticle}
\endbibitem

\bibitem[\protect\citeauthoryear{Faure and Lemieux}{2017}]{faur:lemi:2017}
\begin{barticle}[author]
\bauthor{\bsnm{Faure},~\bfnm{Henri}\binits{H.}} \AND
  \bauthor{\bsnm{Lemieux},~\bfnm{Christiane}\binits{C.}}
(\byear{2017}).
\btitle{A review of discrepancy bounds for $(t, s)$ and $(t, e, s)$-sequences
  with numerical comparisons}.
\bjournal{Mathematics and Computers in Simulation}
\bvolume{135}
\bpages{63--71}.
\end{barticle}
\endbibitem

\bibitem[\protect\citeauthoryear{Fisher}{1926}]{fish:1926}
\begin{barticle}[author]
\bauthor{\bsnm{Fisher},~\bfnm{R.~A.}\binits{R.~A.}}
(\byear{1926}).
\btitle{The arrangement of field experiments}.
\bjournal{Journal of the Ministry of Agriculture}
\bvolume{33}
\bpages{503--515}.
\end{barticle}
\endbibitem

\bibitem[\protect\citeauthoryear{Fisher and Mackenzie}{1923}]{fish:mack:1923}
\begin{barticle}[author]
\bauthor{\bsnm{Fisher},~\bfnm{R.~A.}\binits{R.~A.}} \AND
  \bauthor{\bsnm{Mackenzie},~\bfnm{W.~A.}\binits{W.~A.}}
(\byear{1923}).
\btitle{The manurial response of different potato varieties}.
\bjournal{Journal of Agricultural Science}
\bvolume{13}
\bpages{311--320}.
\end{barticle}
\endbibitem

\bibitem[\protect\citeauthoryear{Friedel and Keller}{2002}]{frie:kell:2002}
\begin{binproceedings}[author]
\bauthor{\bsnm{Friedel},~\bfnm{Ilja}\binits{I.}} \AND
  \bauthor{\bsnm{Keller},~\bfnm{Alexander}\binits{A.}}
(\byear{2002}).
\btitle{Fast generation of randomized low-discrepancy point sets}.
In \bbooktitle{Monte Carlo and Quasi-Monte Carlo Methods 2000}
(\beditor{\bfnm{K.~T.}\binits{K.~T.}~\bsnm{Fang}},
  \beditor{\bfnm{F.~J.}\binits{F.~J.}~\bsnm{Hickernell}} \AND
  \beditor{\bfnm{H.}\binits{H.}~\bsnm{Niederreiter}}, eds.)
\bpages{257--273}.
\bpublisher{Springer}.
\end{binproceedings}
\endbibitem

\bibitem[\protect\citeauthoryear{Gelman et~al.}{2020}]{baye:flow:2020}
\begin{btechreport}[author]
\bauthor{\bsnm{Gelman},~\bfnm{Andrew}\binits{A.}},
  \bauthor{\bsnm{Vehtari},~\bfnm{Aki}\binits{A.}},
  \bauthor{\bsnm{Simpson},~\bfnm{Daniel}\binits{D.}},
  \bauthor{\bsnm{Margossian},~\bfnm{Charles~C}\binits{C.~C.}},
  \bauthor{\bsnm{Carpenter},~\bfnm{Bob}\binits{B.}},
  \bauthor{\bsnm{Yao},~\bfnm{Yuling}\binits{Y.}},
  \bauthor{\bsnm{Kennedy},~\bfnm{Lauren}\binits{L.}},
  \bauthor{\bsnm{Gabry},~\bfnm{Jonah}\binits{J.}},
  \bauthor{\bsnm{B{\"u}rkner},~\bfnm{Paul-Christian}\binits{P.-C.}} \AND
  \bauthor{\bsnm{Modr{\'a}k},~\bfnm{Martin}\binits{M.}}
(\byear{2020}).
\btitle{Bayesian workflow}
\btype{Technical Report},
\bpublisher{arXiv:2011.01808}.
\end{btechreport}
\endbibitem

\bibitem[\protect\citeauthoryear{Genz}{1984}]{genz:1984}
\begin{binproceedings}[author]
\bauthor{\bsnm{Genz},~\bfnm{A.}\binits{A.}}
(\byear{1984}).
\btitle{Testing multidimensional integration routines}.
In \bbooktitle{Proceedings of international conference on Tools, Methods and
  Languages for Scientific and Engineering Computation}
(\beditor{\bfnm{B.}\binits{B.}~\bsnm{Ford}},
  \beditor{\bfnm{J.~C.}\binits{J.~C.}~\bsnm{Rault}} \AND
  \beditor{\bfnm{F.}\binits{F.}~\bsnm{Thomasset}}, eds.)
\bpages{81--94}.
\bpublisher{Elsevier North-Holland, Inc.}
\end{binproceedings}
\endbibitem

\bibitem[\protect\citeauthoryear{Genz and Bretz}{2009}]{genz:bret:2009}
\begin{bbook}[author]
\bauthor{\bsnm{Genz},~\bfnm{Alan}\binits{A.}} \AND
  \bauthor{\bsnm{Bretz},~\bfnm{Frank}\binits{F.}}
(\byear{2009}).
\btitle{Computation of multivariate normal and $t$ probabilities}.
\bpublisher{Springer-Verlag}, \baddress{Berlin}.
\end{bbook}
\endbibitem

\bibitem[\protect\citeauthoryear{Gerber and Chopin}{2015}]{gerb:chop:2015}
\begin{barticle}[author]
\bauthor{\bsnm{Gerber},~\bfnm{M.}\binits{M.}} \AND
  \bauthor{\bsnm{Chopin},~\bfnm{N.}\binits{N.}}
(\byear{2015}).
\btitle{Sequential {quasi-Monte Carlo}}.
\bjournal{Journal of the Royal Statistical Society, Series B}
\bvolume{77}
\bpages{509--579}.
\end{barticle}
\endbibitem

\bibitem[\protect\citeauthoryear{Giles}{2008}]{gile:2008}
\begin{barticle}[author]
\bauthor{\bsnm{Giles},~\bfnm{M.~B.}\binits{M.~B.}}
(\byear{2008}).
\btitle{Multilevel {Monte Carlo} path simulation}.
\bjournal{Operations Research}
\bvolume{56}
\bpages{607--617}.
\end{barticle}
\endbibitem

\bibitem[\protect\citeauthoryear{Giles and Waterhouse}{2009}]{gile:wate:2009}
\begin{barticle}[author]
\bauthor{\bsnm{Giles},~\bfnm{Michael~B}\binits{M.~B.}} \AND
  \bauthor{\bsnm{Waterhouse},~\bfnm{Benjamin~J}\binits{B.~J.}}
(\byear{2009}).
\btitle{Multilevel {quasi-Monte Carlo} path simulation}.
\bjournal{Advanced Financial Modelling, Radon Series on Computational and
  Applied Mathematics}
\bvolume{8}
\bpages{165--181}.
\end{barticle}
\endbibitem

\bibitem[\protect\citeauthoryear{Gnewuch, Srivastav and
  Winzen}{2009}]{gnew:sriv:winz:2009}
\begin{barticle}[author]
\bauthor{\bsnm{Gnewuch},~\bfnm{M.}\binits{M.}},
  \bauthor{\bsnm{Srivastav},~\bfnm{A.}\binits{A.}} \AND
  \bauthor{\bsnm{Winzen},~\bfnm{C.}\binits{C.}}
(\byear{2009}).
\btitle{Finding optimal volume subintervals with $k$ points and calculating the
  star discrepancy are {NP}-hard problems}.
\bjournal{Journal of Complexity}
\bvolume{25}
\bpages{115--127}.
\end{barticle}
\endbibitem

\bibitem[\protect\citeauthoryear{Goda and L'Ecuyer}{2022}]{goda:lecu:2022}
\begin{barticle}[author]
\bauthor{\bsnm{Goda},~\bfnm{T.}\binits{T.}} \AND
  \bauthor{\bsnm{L'Ecuyer},~\bfnm{P.}\binits{P.}}
(\byear{2022}).
\btitle{Construction-free median {quasi-Monte Carlo} rules for function spaces
  with unspecified smoothness and general weights}.
\bjournal{SIAM Journal on Scientific Computing}
\bvolume{44}
\bpages{A2765--A2788}.
\end{barticle}
\endbibitem

\bibitem[\protect\citeauthoryear{Goda, Suzuki and
  Matsumoto}{2024}]{goda:suzu:mats:2024}
\begin{barticle}[author]
\bauthor{\bsnm{Goda},~\bfnm{Takashi}\binits{T.}},
  \bauthor{\bsnm{Suzuki},~\bfnm{Kosuke}\binits{K.}} \AND
  \bauthor{\bsnm{Matsumoto},~\bfnm{Makoto}\binits{M.}}
(\byear{2024}).
\btitle{A universal median {quasi-Monte Carlo} integration}.
\bjournal{SIAM Journal on Numerical Analysis}
\bvolume{62}
\bpages{533--566}.
\end{barticle}
\endbibitem

\bibitem[\protect\citeauthoryear{Griewank
  et~al.}{2018}]{grie:kuo:leov:sloa:2018}
\begin{barticle}[author]
\bauthor{\bsnm{Griewank},~\bfnm{A.}\binits{A.}},
  \bauthor{\bsnm{Kuo},~\bfnm{F.~Y.}\binits{F.~Y.}},
  \bauthor{\bsnm{Le{\"o}vey},~\bfnm{H.}\binits{H.}} \AND
  \bauthor{\bsnm{Sloan},~\bfnm{I.~H.}\binits{I.~H.}}
(\byear{2018}).
\btitle{High dimensional integration of kinks and jumps—{Smoothing} by
  preintegration}.
\bjournal{Journal of Computational and Applied Mathematics}
\bvolume{344}
\bpages{259--274}.
\end{barticle}
\endbibitem

\bibitem[\protect\citeauthoryear{Hahn}{2005}]{hahn:2005}
\begin{barticle}[author]
\bauthor{\bsnm{Hahn},~\bfnm{Thomas}\binits{T.}}
(\byear{2005}).
\btitle{{Cuba}—a library for multidimensional numerical integration}.
\bjournal{Computer Physics Communications}
\bvolume{168}
\bpages{78--95}.
\end{barticle}
\endbibitem

\bibitem[\protect\citeauthoryear{Hall}{1988}]{hall:1988}
\begin{barticle}[author]
\bauthor{\bsnm{Hall},~\bfnm{P.}\binits{P.}}
(\byear{1988}).
\btitle{Theoretical comparisons of bootstrap confidence intervals}.
\bjournal{The Annals of Statistics}
\bvolume{16}
\bpages{927--953}.
\end{barticle}
\endbibitem

\bibitem[\protect\citeauthoryear{Halton}{1960}]{halt:1960}
\begin{barticle}[author]
\bauthor{\bsnm{Halton},~\bfnm{J.~H.}\binits{J.~H.}}
(\byear{1960}).
\btitle{On the Efficiency of Certain Quasi-Random Sequences of Points in
  Evaluating Multi-Dimensional Integrals}.
\bjournal{Numerische Mathematik}
\bvolume{2}
\bpages{84--90}.
\end{barticle}
\endbibitem

\bibitem[\protect\citeauthoryear{Hammersley}{1960}]{hamm:1960}
\begin{barticle}[author]
\bauthor{\bsnm{Hammersley},~\bfnm{J.~M.}\binits{J.~M.}}
(\byear{1960}).
\btitle{{Monte Carlo} methods for solving multivariable problems}.
\bjournal{Annals of the New York Academy of Sciences}
\bvolume{86}
\bpages{844--874}.
\end{barticle}
\endbibitem

\bibitem[\protect\citeauthoryear{He and Wang}{2015}]{he:wang:2015}
\begin{barticle}[author]
\bauthor{\bsnm{He},~\bfnm{Z.}\binits{Z.}} \AND
  \bauthor{\bsnm{Wang},~\bfnm{X.}\binits{X.}}
(\byear{2015}).
\btitle{On the Convergence Rate of Randomized Quasi--{Monte Carlo} for
  Discontinuous Functions}.
\bjournal{SIAM Journal on Numerical Analysis}
\bvolume{53}
\bpages{2488--2503}.
\end{barticle}
\endbibitem

\bibitem[\protect\citeauthoryear{Hedayat, Sloane and
  Stufken}{1999}]{heda:sloa:stuf:1999}
\begin{bbook}[author]
\bauthor{\bsnm{Hedayat},~\bfnm{A.~S.}\binits{A.~S.}},
  \bauthor{\bsnm{Sloane},~\bfnm{N.~J.~A.}\binits{N.~J.~A.}} \AND
  \bauthor{\bsnm{Stufken},~\bfnm{J.}\binits{J.}}
(\byear{1999}).
\btitle{Orthogonal Arrays: Theory and Applications}.
\bpublisher{Springer}, \baddress{New York}.
\end{bbook}
\endbibitem

\bibitem[\protect\citeauthoryear{Heinrich}{1998}]{hein:1998}
\begin{barticle}[author]
\bauthor{\bsnm{Heinrich},~\bfnm{S.}\binits{S.}}
(\byear{1998}).
\btitle{{Monte Carlo} complexity of global solution of integral equations}.
\bjournal{Journal of Complexity}
\bvolume{14}
\bpages{151--175}.
\end{barticle}
\endbibitem

\bibitem[\protect\citeauthoryear{Heinrich}{2001}]{hein:2001}
\begin{bincollection}[author]
\bauthor{\bsnm{Heinrich},~\bfnm{S.}\binits{S.}}
(\byear{2001}).
\btitle{Multilevel {Monte Carlo} methods}.
In \bbooktitle{Large-Scale Scientific Computing},
(\beditor{\bfnm{S.}\binits{S.}~\bsnm{Margenov}},
  \beditor{\bfnm{J.}\binits{J.}~\bsnm{Wasniewski}} \AND
  \beditor{\bfnm{Y.}\binits{Y.}~\bsnm{Plamen}}, eds.).
\bseries{Lecture Notes in Computer Science}
\bvolume{2179}
\bpages{58--67}.
\bpublisher{Springer-Verlag}, \baddress{Heidelberg}.
\end{bincollection}
\endbibitem

\bibitem[\protect\citeauthoryear{Hickernell}{1996}]{hick:1996:weights}
\begin{barticle}[author]
\bauthor{\bsnm{Hickernell},~\bfnm{Fred~J}\binits{F.~J.}}
(\byear{1996}).
\btitle{Quadrature error bounds with applications to lattice rules}.
\bjournal{SIAM Journal on Numerical Analysis}
\bvolume{33}
\bpages{1995--2016}.
\end{barticle}
\endbibitem

\bibitem[\protect\citeauthoryear{Hickernell}{1998}]{hickdisc}
\begin{barticle}[author]
\bauthor{\bsnm{Hickernell},~\bfnm{F.~J.}\binits{F.~J.}}
(\byear{1998}).
\btitle{A generalized discrepancy and quadrature error bound}.
\bjournal{Mathematics of Computation}
\bvolume{67}
\bpages{299--322}.
\end{barticle}
\endbibitem

\bibitem[\protect\citeauthoryear{Hickernell}{2002}]{hick:2002}
\begin{binproceedings}[author]
\bauthor{\bsnm{Hickernell},~\bfnm{F.~J.}\binits{F.~J.}}
(\byear{2002}).
\btitle{Obtaining $O(N^{-2+\epsilon})$ convergence for lattice quadrature
  rules}.
In \bbooktitle{Monte Carlo and Quasi-Monte Carlo Methods in Scientific
  Computing}
(\beditor{\bfnm{K.~T.}\binits{K.~T.}~\bsnm{Fang}},
  \beditor{\bfnm{F.~J.}\binits{F.~J.}~\bsnm{Hickernell}} \AND
  \beditor{\bfnm{H.}\binits{H.}~\bsnm{Niederreiter}}, eds.)
\bpages{434--445}.
\bpublisher{Springer-Verlag}, \baddress{New York}.
\end{binproceedings}
\endbibitem

\bibitem[\protect\citeauthoryear{Hickernell, Lemieux and
  Owen}{2005}]{hick:lemi:owen:2005}
\begin{barticle}[author]
\bauthor{\bsnm{Hickernell},~\bfnm{F.~J.}\binits{F.~J.}},
  \bauthor{\bsnm{Lemieux},~\bfnm{C.}\binits{C.}} \AND
  \bauthor{\bsnm{Owen},~\bfnm{A.~B.}\binits{A.~B.}}
(\byear{2005}).
\btitle{Control variates for quasi-{Monte Carlo} (with discussion)}.
\bjournal{Statistical Science}
\bvolume{20}
\bpages{1--31}.
\end{barticle}
\endbibitem

\bibitem[\protect\citeauthoryear{Hickernell and
  Wo{\'z}niakowski}{2000}]{hick:wozn:2000}
\begin{barticle}[author]
\bauthor{\bsnm{Hickernell},~\bfnm{F.~J.}\binits{F.~J.}} \AND
  \bauthor{\bsnm{Wo{\'z}niakowski},~\bfnm{H.}\binits{H.}}
(\byear{2000}).
\btitle{Integration and approximation in arbitrary dimensions}.
\bjournal{Advances in Computational Mathematics}
\bvolume{12}
\bpages{25--58}.
\end{barticle}
\endbibitem

\bibitem[\protect\citeauthoryear{Hickernell and
  Wo{\'z}niakowski}{2001}]{hick:wozn:2001}
\begin{barticle}[author]
\bauthor{\bsnm{Hickernell},~\bfnm{Fred~J}\binits{F.~J.}} \AND
  \bauthor{\bsnm{Wo{\'z}niakowski},~\bfnm{Henryk}\binits{H.}}
(\byear{2001}).
\btitle{The price of pessimism for multidimensional quadrature}.
\bjournal{Journal of Complexity}
\bvolume{17}
\bpages{625--659}.
\end{barticle}
\endbibitem

\bibitem[\protect\citeauthoryear{Hintz, Hofert and
  Lemieux}{2022}]{hint:hofe:lemi:2022}
\begin{bincollection}[author]
\bauthor{\bsnm{Hintz},~\bfnm{Erik}\binits{E.}},
  \bauthor{\bsnm{Hofert},~\bfnm{Marius}\binits{M.}} \AND
  \bauthor{\bsnm{Lemieux},~\bfnm{Christiane}\binits{C.}}
(\byear{2022}).
\btitle{Quasi-random sampling with black box or acceptance-rejection inputs}.
In \bbooktitle{Advances in Modeling and Simulation: Festschrift for Pierre
  L'Ecuyer}
\bpages{261--281}.
\bpublisher{Springer}.
\end{bincollection}
\endbibitem

\bibitem[\protect\citeauthoryear{Hlawka}{1961}]{hlaw:1961}
\begin{barticle}[author]
\bauthor{\bsnm{Hlawka},~\bfnm{E.}\binits{E.}}
(\byear{1961}).
\btitle{Funktionen von beschr\"ankter Variation in der Theorie der
  Gleichverteilung}.
\bjournal{Annali di Matematica Pura Applicata}
\bvolume{54}
\bpages{324--334}.
\end{barticle}
\endbibitem

\bibitem[\protect\citeauthoryear{Ho and Owen}{2026}]{wos:arqmc:tr}
\begin{btechreport}[author]
\bauthor{\bsnm{Ho},~\bfnm{Valerie~NP}\binits{V.~N.}} \AND
  \bauthor{\bsnm{Owen},~\bfnm{Art~B}\binits{A.~B.}}
(\byear{2026}).
\btitle{Walk on spheres and {Array-RQMC}}
\btype{Technical Report},
\bpublisher{arXiv:2605.12844}.
\end{btechreport}
\endbibitem

\bibitem[\protect\citeauthoryear{Hoeffding}{1948}]{hoef:1948}
\begin{barticle}[author]
\bauthor{\bsnm{Hoeffding},~\bfnm{W.}\binits{W.}}
(\byear{1948}).
\btitle{A class of statistics with asymptotically normal distribution}.
\bjournal{Annals of Mathematical Statistics}
\bvolume{19}
\bpages{293--325}.
\end{barticle}
\endbibitem

\bibitem[\protect\citeauthoryear{Hofmann and Math{\'e}}{1997}]{hofm:math:1997}
\begin{barticle}[author]
\bauthor{\bsnm{Hofmann},~\bfnm{Norbert}\binits{N.}} \AND
  \bauthor{\bsnm{Math{\'e}},~\bfnm{Peter}\binits{P.}}
(\byear{1997}).
\btitle{On {quasi-Monte Carlo} simulation of stochastic differential
  equations}.
\bjournal{Mathematics of computation}
\bvolume{66}
\bpages{573--589}.
\end{barticle}
\endbibitem

\bibitem[\protect\citeauthoryear{H\"ormann, Leydold and
  Derflinger}{2004}]{horm:leyd:2004}
\begin{bbook}[author]
\bauthor{\bsnm{H\"ormann},~\bfnm{W.}\binits{W.}},
  \bauthor{\bsnm{Leydold},~\bfnm{J.}\binits{J.}} \AND
  \bauthor{\bsnm{Derflinger},~\bfnm{G.}\binits{G.}}
(\byear{2004}).
\btitle{Automatic nonuniform random variate generation}.
\bpublisher{Springer}, \baddress{Berlin}.
\end{bbook}
\endbibitem

\bibitem[\protect\citeauthoryear{Hoyt and Owen}{2020}]{hoyt:owen:2020}
\begin{barticle}[author]
\bauthor{\bsnm{Hoyt},~\bfnm{C.}\binits{C.}} \AND
  \bauthor{\bsnm{Owen},~\bfnm{A.~B.}\binits{A.~B.}}
(\byear{2020}).
\btitle{Mean dimension of ridge functions}.
\bjournal{SIAM Journal on Numerical Analysis}
\bvolume{58}
\bpages{1195--1216}.
\end{barticle}
\endbibitem

\bibitem[\protect\citeauthoryear{Jacob, O’Leary and
  Atchad{\'e}}{2020}]{jaco:olea:atch:2020}
\begin{barticle}[author]
\bauthor{\bsnm{Jacob},~\bfnm{Pierre~E}\binits{P.~E.}},
  \bauthor{\bsnm{O’Leary},~\bfnm{John}\binits{J.}} \AND
  \bauthor{\bsnm{Atchad{\'e}},~\bfnm{Yves~F}\binits{Y.~F.}}
(\byear{2020}).
\btitle{Unbiased Markov chain Monte Carlo methods with couplings}.
\bjournal{Journal of the Royal Statistical Society Series B}
\bvolume{82}
\bpages{543--600}.
\end{barticle}
\endbibitem

\bibitem[\protect\citeauthoryear{Jagadeeswaran and
  Hickernell}{2019}]{jaga:hick:2019}
\begin{barticle}[author]
\bauthor{\bsnm{Jagadeeswaran},~\bfnm{R.}\binits{R.}} \AND
  \bauthor{\bsnm{Hickernell},~\bfnm{F.~J.}\binits{F.~J.}}
(\byear{2019}).
\btitle{Fast automatic Bayesian cubature using lattice sampling}.
\bjournal{Statistics and Computing}
\bvolume{29}
\bpages{1215--1229}.
\end{barticle}
\endbibitem

\bibitem[\protect\citeauthoryear{Jain et~al.}{2026}]{jain:hick:owen:soro:2026}
\begin{barticle}[author]
\bauthor{\bsnm{Jain},~\bfnm{Aadit}\binits{A.}},
  \bauthor{\bsnm{Hickernell},~\bfnm{Fred~J}\binits{F.~J.}},
  \bauthor{\bsnm{Owen},~\bfnm{Art~B}\binits{A.~B.}} \AND
  \bauthor{\bsnm{Sorokin},~\bfnm{Aleksei~G}\binits{A.~G.}}
(\byear{2026}).
\btitle{Empirical {Bernstein} and betting confidence intervals for randomized
  {quasi-Monte Carlo}}.
\bjournal{Information and Inference}
\bvolume{15}
\bpages{iaag003}.
\end{barticle}
\endbibitem

\bibitem[\protect\citeauthoryear{Jansen}{1999}]{jans:1999}
\begin{barticle}[author]
\bauthor{\bsnm{Jansen},~\bfnm{M.~J.~W.}\binits{M.~J.~W.}}
(\byear{1999}).
\btitle{Analysis of variance designs for model output}.
\bjournal{Computer Physics Communications}
\bvolume{117}
\bpages{35--43}.
\end{barticle}
\endbibitem

\bibitem[\protect\citeauthoryear{Joe and Kuo}{2008}]{joe:kuo:2008}
\begin{barticle}[author]
\bauthor{\bsnm{Joe},~\bfnm{S.}\binits{S.}} \AND
  \bauthor{\bsnm{Kuo},~\bfnm{F.~Y.}\binits{F.~Y.}}
(\byear{2008}).
\btitle{Constructing {Sobol'} sequences with better two-dimensional
  projections}.
\bjournal{SIAM Journal on Scientific Computing}
\bvolume{30}
\bpages{2635--2654}.
\end{barticle}
\endbibitem

\bibitem[\protect\citeauthoryear{Kanagawa et~al.}{2018}]{kana:etal:2018}
\begin{btechreport}[author]
\bauthor{\bsnm{Kanagawa},~\bfnm{Motonobu}\binits{M.}},
  \bauthor{\bsnm{Hennig},~\bfnm{Philipp}\binits{P.}},
  \bauthor{\bsnm{Sejdinovic},~\bfnm{Dino}\binits{D.}} \AND
  \bauthor{\bsnm{Sriperumbudur},~\bfnm{Bharath~K}\binits{B.~K.}}
(\byear{2018}).
\btitle{Gaussian processes and kernel methods: A review on connections and
  equivalences}
\btype{Technical Report},
\bpublisher{arXiv:1807.02582}.
\end{btechreport}
\endbibitem

\bibitem[\protect\citeauthoryear{Keister}{1996}]{keis:1996}
\begin{barticle}[author]
\bauthor{\bsnm{Keister},~\bfnm{B.~D.}\binits{B.~D.}}
(\byear{1996}).
\btitle{Multidimensional quadrature algorithms}.
\bjournal{Computers in Physics}
\bvolume{10}
\bpages{119--122}.
\end{barticle}
\endbibitem

\bibitem[\protect\citeauthoryear{Keller}{2013}]{kell:2013}
\begin{bincollection}[author]
\bauthor{\bsnm{Keller},~\bfnm{Alexander}\binits{A.}}
(\byear{2013}).
\btitle{{Quasi-Monte Carlo} image synthesis in a nutshell}.
In \bbooktitle{Monte Carlo and Quasi-Monte Carlo Methods 2012}
(\beditor{\bfnm{Josef}\binits{J.}~\bsnm{Dick}},
  \beditor{\bfnm{Frances~Y.}\binits{F.~Y.}~\bsnm{Kuo}},
  \beditor{\bfnm{Gareth~W.}\binits{G.~W.}~\bsnm{Peters}} \AND
  \beditor{\bfnm{Ian~H.}\binits{I.~H.}~\bsnm{Sloan}}, eds.)
\bpages{213--249}.
\bpublisher{Springer}.
\end{bincollection}
\endbibitem

\bibitem[\protect\citeauthoryear{Kempthorne}{1947}]{kemp:1947}
\begin{barticle}[author]
\bauthor{\bsnm{Kempthorne},~\bfnm{O}\binits{O.}}
(\byear{1947}).
\btitle{A simple approach to confounding and fractional replication in
  factorial experiments}.
\bjournal{Biometrika}
\bvolume{34}
\bpages{255--272}.
\end{barticle}
\endbibitem

\bibitem[\protect\citeauthoryear{Knuth}{1998}]{knut:1998:2:3}
\begin{bbook}[author]
\bauthor{\bsnm{Knuth},~\bfnm{D.~E.}\binits{D.~E.}}
(\byear{1998}).
\btitle{The Art of Computer Programming, Volume 2: Seminumerical Algorithms},
\bedition{{3rd}} ed.
\bpublisher{Addison-Wesley}, \baddress{Reading MA}.
\end{bbook}
\endbibitem

\bibitem[\protect\citeauthoryear{Koksma}{1942/1943}]{koks:1942}
\begin{barticle}[author]
\bauthor{\bsnm{Koksma},~\bfnm{J.~F.}\binits{J.~F.}}
(\byear{1942/1943}).
\btitle{Een algemeene stelling uit de theorie der gelijkmatige verdeeling
  modulo 1}.
\bjournal{Mathematica B (Zutphen)}
\bvolume{11}
\bpages{7--11}.
\end{barticle}
\endbibitem

\bibitem[\protect\citeauthoryear{Korobov}{1960}]{koro:1960}
\begin{binproceedings}[author]
\bauthor{\bsnm{Korobov},~\bfnm{Nikolai~Mikhailovich}\binits{N.~M.}}
(\byear{1960}).
\btitle{Properties and calculation of optimal coefficients}.
In \bbooktitle{Doklady Akademii Nauk}
\bvolume{132}
\bpages{1009--1012}.
\bpublisher{Russian Academy of Sciences}.
\end{binproceedings}
\endbibitem

\bibitem[\protect\citeauthoryear{Kuo and Nuyens}{2016}]{kuo:nuye:2016}
\begin{barticle}[author]
\bauthor{\bsnm{Kuo},~\bfnm{F.~Y.}\binits{F.~Y.}} \AND
  \bauthor{\bsnm{Nuyens},~\bfnm{D.}\binits{D.}}
(\byear{2016}).
\btitle{Application of {quasi-Monte Carlo} methods to elliptic {PDEs} with
  random diffusion coefficients: a survey of analysis and implementation}.
\bjournal{Foundations of Computational Mathematics}
\bvolume{16}
\bpages{1631--1696}.
\end{barticle}
\endbibitem

\bibitem[\protect\citeauthoryear{Kuo, Schwab and
  Sloan}{2012}]{kuo:schw:sloa:2012}
\begin{barticle}[author]
\bauthor{\bsnm{Kuo},~\bfnm{Frances~Y}\binits{F.~Y.}},
  \bauthor{\bsnm{Schwab},~\bfnm{Christoph}\binits{C.}} \AND
  \bauthor{\bsnm{Sloan},~\bfnm{Ian~H}\binits{I.~H.}}
(\byear{2012}).
\btitle{{Quasi-Monte Carlo} finite element methods for a class of elliptic
  partial differential equations with random coefficients}.
\bjournal{SIAM Journal on Numerical Analysis}
\bvolume{50}
\bpages{3351--3374}.
\end{barticle}
\endbibitem

\bibitem[\protect\citeauthoryear{Kuo, Sloan and
  Wo{\'z}niakowski}{2007}]{kuo:sloa:wozn:2007}
\begin{barticle}[author]
\bauthor{\bsnm{Kuo},~\bfnm{Frances~Y}\binits{F.~Y.}},
  \bauthor{\bsnm{Sloan},~\bfnm{Ian~H}\binits{I.~H.}} \AND
  \bauthor{\bsnm{Wo{\'z}niakowski},~\bfnm{Henryk}\binits{H.}}
(\byear{2007}).
\btitle{Periodization strategy may fail in high dimensions}.
\bjournal{Numerical Algorithms}
\bvolume{46}
\bpages{369--391}.
\end{barticle}
\endbibitem

\bibitem[\protect\citeauthoryear{Kuo et~al.}{2010}]{kuo:sloa:wasi:wozn:2010}
\begin{barticle}[author]
\bauthor{\bsnm{Kuo},~\bfnm{F.}\binits{F.}},
  \bauthor{\bsnm{Sloan},~\bfnm{I.}\binits{I.}},
  \bauthor{\bsnm{Wasilkowski},~\bfnm{G.}\binits{G.}} \AND
  \bauthor{\bsnm{Wo{\'z}niakowski},~\bfnm{H.}\binits{H.}}
(\byear{2010}).
\btitle{On decompositions of multivariate functions}.
\bjournal{Mathematics of computation}
\bvolume{79}
\bpages{953--966}.
\end{barticle}
\endbibitem

\bibitem[\protect\citeauthoryear{L{\'e}cot and Tuffin}{2004}]{leco:tuff:2004}
\begin{bincollection}[author]
\bauthor{\bsnm{L{\'e}cot},~\bfnm{C.}\binits{C.}} \AND
  \bauthor{\bsnm{Tuffin},~\bfnm{B.}\binits{B.}}
(\byear{2004}).
\btitle{{Quasi-Monte Carlo} methods for estimating transient measures of
  discrete time {Markov} chains}.
In \bbooktitle{Monte Carlo and Quasi-Monte Carlo Methods 2002}
\bpages{329--343}.
\bpublisher{Springer}.
\end{bincollection}
\endbibitem

\bibitem[\protect\citeauthoryear{L'Ecuyer}{2009}]{lecu:2009:fin}
\begin{barticle}[author]
\bauthor{\bsnm{L'Ecuyer},~\bfnm{Pierre}\binits{P.}}
(\byear{2009}).
\btitle{{Quasi-Monte Carlo} methods with applications in finance}.
\bjournal{Finance and Stochastics}
\bvolume{13}
\bpages{307--349}.
\end{barticle}
\endbibitem

\bibitem[\protect\citeauthoryear{L'Ecuyer}{2023}]{lecu:rqmc:wsc23:data}
\begin{bmisc}[author]
\bauthor{\bsnm{L'Ecuyer},~\bfnm{Pierre}\binits{P.}}
(\byear{2023}).
\btitle{Randomized Quasi-Monte Carlo Data for Small Function Examples}.
\bhowpublished{\url{https://github.com/pierrelecuyer/rqmc-samples-wsc23}}.
\bnote{Data and Java code accompanying the paper ``Confidence Intervals for
  Randomized Quasi-Monte Carlo Estimators.'' Accessed July 29, 2026}.
\end{bmisc}
\endbibitem

\bibitem[\protect\citeauthoryear{L'Ecuyer and Buist}{2005}]{lecu:buis:2005}
\begin{binproceedings}[author]
\bauthor{\bsnm{L'Ecuyer},~\bfnm{Pierre}\binits{P.}} \AND
  \bauthor{\bsnm{Buist},~\bfnm{Eric}\binits{E.}}
(\byear{2005}).
\btitle{Simulation in {Java} with {SSJ}}.
In \bbooktitle{Proceedings of the Winter Simulation Conference, 2005}
\bpages{10--pp}.
\bpublisher{IEEE}.
\end{binproceedings}
\endbibitem

\bibitem[\protect\citeauthoryear{L'Ecuyer, L{\'e}cot and
  Tuffin}{2008}]{lecu:leco:tuff:2008}
\begin{barticle}[author]
\bauthor{\bsnm{L'Ecuyer},~\bfnm{Pierre}\binits{P.}},
  \bauthor{\bsnm{L{\'e}cot},~\bfnm{Christian}\binits{C.}} \AND
  \bauthor{\bsnm{Tuffin},~\bfnm{Bruno}\binits{B.}}
(\byear{2008}).
\btitle{A randomized quasi-{M}onte {C}arlo simulation method for {M}arkov
  chains}.
\bjournal{Operations Research}
\bvolume{56}
\bpages{958--975}.
\end{barticle}
\endbibitem

\bibitem[\protect\citeauthoryear{L'Ecuyer and Lemieux}{1999}]{lecu:lemi:1999}
\begin{binproceedings}[author]
\bauthor{\bsnm{L'Ecuyer},~\bfnm{Pierre}\binits{P.}} \AND
  \bauthor{\bsnm{Lemieux},~\bfnm{Christiane}\binits{C.}}
(\byear{1999}).
\btitle{{Quasi-Monte Carlo} via linear shift-register sequences}.
In \bbooktitle{Proceedings of the 31st Winter Simulation Conference}
\bpages{632--639}.
\end{binproceedings}
\endbibitem

\bibitem[\protect\citeauthoryear{L'Ecuyer and Munger}{2016}]{lecu:mung:2016}
\begin{barticle}[author]
\bauthor{\bsnm{L'Ecuyer},~\bfnm{P.}\binits{P.}} \AND
  \bauthor{\bsnm{Munger},~\bfnm{D.}\binits{D.}}
(\byear{2016}).
\btitle{Algorithm 958: {Lattice Builder}: A General Software Tool for
  Constructing Rank-1 Lattice Rules}.
\bjournal{ACM Transactions on Mathematical Software}
\bvolume{42}
\bpages{Article 15}.
\end{barticle}
\endbibitem

\bibitem[\protect\citeauthoryear{L'Ecuyer and Simard}{2007}]{lecu:sima:2007}
\begin{barticle}[author]
\bauthor{\bsnm{L'Ecuyer},~\bfnm{P.}\binits{P.}} \AND
  \bauthor{\bsnm{Simard},~\bfnm{R.}\binits{R.}}
(\byear{2007}).
\btitle{{TestU01}: a {C} library for empirical testing of random number
  generators}.
\bjournal{{ACM} transactions on mathematical software}
\bvolume{33}
\bpages{article 22}.
\end{barticle}
\endbibitem

\bibitem[\protect\citeauthoryear{L'Ecuyer
  et~al.}{2022}]{lecu:mari:godi:puch:2022}
\begin{binproceedings}[author]
\bauthor{\bsnm{L'Ecuyer},~\bfnm{Pierre}\binits{P.}},
  \bauthor{\bsnm{Marion},~\bfnm{Pierre}\binits{P.}},
  \bauthor{\bsnm{Godin},~\bfnm{Maxime}\binits{M.}} \AND
  \bauthor{\bsnm{Puchhammer},~\bfnm{Florian}\binits{F.}}
(\byear{2022}).
\btitle{A tool for custom construction of {QMC} and {RQMC} point sets}.
In \bbooktitle{Monte Carlo and Quasi-Monte Carlo Methods. MCQMC 2020}
(\beditor{\bfnm{Alexander}\binits{A.}~\bsnm{Keller}}, ed.)
\bpages{51--70}.
\bpublisher{Springer}.
\end{binproceedings}
\endbibitem

\bibitem[\protect\citeauthoryear{L'Ecuyer et~al.}{2023}]{ci4rqmc}
\begin{binproceedings}[author]
\bauthor{\bsnm{L'Ecuyer},~\bfnm{Pierre}\binits{P.}},
  \bauthor{\bsnm{Nakayama},~\bfnm{Marvin~K}\binits{M.~K.}},
  \bauthor{\bsnm{Owen},~\bfnm{Art~B}\binits{A.~B.}} \AND
  \bauthor{\bsnm{Tuffin},~\bfnm{Bruno}\binits{B.}}
(\byear{2023}).
\btitle{Confidence intervals for randomized quasi-{Monte Carlo} estimators}.
In \bbooktitle{2023 Winter Simulation Conference (WSC)}
(\beditor{\bfnm{C.~G.}\binits{C.~G.}~\bsnm{Corlu}},
  \beditor{\bfnm{S.~R.}\binits{S.~R.}~\bsnm{Hunter}},
  \beditor{\bfnm{H.}\binits{H.}~\bsnm{Lam}},
  \beditor{\bfnm{B.~S.}\binits{B.~S.}~\bsnm{Onggo}},
  \beditor{\bfnm{J.}\binits{J.}~\bsnm{Shortle}} \AND
  \beditor{\bfnm{B.}\binits{B.}~\bsnm{Biller}}, eds.)
\bpages{445--456}.
\bpublisher{IEEE}.
\end{binproceedings}
\endbibitem

\bibitem[\protect\citeauthoryear{Liu}{2023}]{liu:2023}
\begin{barticle}[author]
\bauthor{\bsnm{Liu},~\bfnm{Sifan}\binits{S.}}
(\byear{2023}).
\btitle{{Langevin} quasi-{Monte Carlo}}.
\bjournal{Advances in Neural Information Processing Systems}
\bvolume{36}.
\end{barticle}
\endbibitem

\bibitem[\protect\citeauthoryear{Liu}{2026a}]{liu:2026}
\begin{barticle}[author]
\bauthor{\bsnm{Liu},~\bfnm{Sifan}\binits{S.}}
(\byear{2026}a).
\btitle{Transport {quasi-Monte Carlo}}.
\bjournal{SIAM Journal on Scientific Computing}
\bvolume{48}
\bpages{C658--C683}.
\end{barticle}
\endbibitem

\bibitem[\protect\citeauthoryear{Liu}{2026b}]{yliu:2026}
\begin{barticle}[author]
\bauthor{\bsnm{Liu},~\bfnm{Yang}\binits{Y.}}
(\byear{2026}b).
\btitle{Randomized {quasi-Monte Carlo} and {Owen's} boundary growth condition:
  a spectral analysis}.
\bjournal{IMA Journal of Numerical Analysis}
\bvolume{46}
\bpages{1060--1097}.
\end{barticle}
\endbibitem

\bibitem[\protect\citeauthoryear{Liu and Owen}{2006}]{meandim}
\begin{barticle}[author]
\bauthor{\bsnm{Liu},~\bfnm{R.}\binits{R.}} \AND
  \bauthor{\bsnm{Owen},~\bfnm{A.~B.}\binits{A.~B.}}
(\byear{2006}).
\btitle{Estimating mean dimensionality of Analysis of Variance Decompositions}.
\bjournal{Journal of the American Statistical Association}
\bvolume{101}
\bpages{712--721}.
\end{barticle}
\endbibitem

\bibitem[\protect\citeauthoryear{Liu and Owen}{2021}]{qmc4qn4vb}
\begin{barticle}[author]
\bauthor{\bsnm{Liu},~\bfnm{Sifan}\binits{S.}} \AND
  \bauthor{\bsnm{Owen},~\bfnm{Art~B}\binits{A.~B.}}
(\byear{2021}).
\btitle{Quasi-{Monte Carlo quasi-Newton} in variational {Bayes}}.
\bjournal{Journal of Machine Learning Research}
\bvolume{22}
\bpages{1--23}.
\end{barticle}
\endbibitem

\bibitem[\protect\citeauthoryear{Liu and Owen}{2023}]{liu:preintegration}
\begin{barticle}[author]
\bauthor{\bsnm{Liu},~\bfnm{Sifan}\binits{S.}} \AND
  \bauthor{\bsnm{Owen},~\bfnm{Art~B}\binits{A.~B.}}
(\byear{2023}).
\btitle{Preintegration via active subspace}.
\bjournal{SIAM Journal on Numerical Analysis}
\bvolume{61}
\bpages{495--514}.
\end{barticle}
\endbibitem

\bibitem[\protect\citeauthoryear{Loh}{2003}]{loh:2003}
\begin{barticle}[author]
\bauthor{\bsnm{Loh},~\bfnm{W.~L.}\binits{W.~L.}}
(\byear{2003}).
\btitle{On the asymptotic distribution of scrambled net quadrature}.
\bjournal{Annals of Statistics}
\bvolume{31}
\bpages{1282--1324}.
\end{barticle}
\endbibitem

\bibitem[\protect\citeauthoryear{Matou\v{s}ek}{1998}]{mato:1998:2}
\begin{barticle}[author]
\bauthor{\bsnm{Matou\v{s}ek},~\bfnm{J.}\binits{J.}}
(\byear{1998}).
\btitle{On the {L$^2$}--discrepancy for anchored boxes}.
\bjournal{Journal of Complexity}
\bvolume{14}
\bpages{527--556}.
\end{barticle}
\endbibitem

\bibitem[\protect\citeauthoryear{Moskowitz and Caflisch}{1996}]{mosk:cafl:1996}
\begin{barticle}[author]
\bauthor{\bsnm{Moskowitz},~\bfnm{B.}\binits{B.}} \AND
  \bauthor{\bsnm{Caflisch},~\bfnm{R.~E.}\binits{R.~E.}}
(\byear{1996}).
\btitle{Smoothness and dimension reduction in {quasi-Monte Carlo} methods}.
\bjournal{Mathematical and Computer Modelling}
\bvolume{23}
\bpages{37--54}.
\end{barticle}
\endbibitem

\bibitem[\protect\citeauthoryear{Nakayama and Tuffin}{2024}]{naka:tuff:2024}
\begin{barticle}[author]
\bauthor{\bsnm{Nakayama},~\bfnm{Marvin~K}\binits{M.~K.}} \AND
  \bauthor{\bsnm{Tuffin},~\bfnm{Bruno}\binits{B.}}
(\byear{2024}).
\btitle{Sufficient conditions for central limit theorems and confidence
  intervals for randomized {quasi-Monte Carlo methods}}.
\bjournal{ACM Transactions on Modeling and Computer Simulation}
\bvolume{34}
\bpages{1--38}.
\end{barticle}
\endbibitem

\bibitem[\protect\citeauthoryear{Niederreiter}{1978}]{nied:1978}
\begin{barticle}[author]
\bauthor{\bsnm{Niederreiter},~\bfnm{H.}\binits{H.}}
(\byear{1978}).
\btitle{Quasi-{Monte Carlo} methods and pseudo-random numbers}.
\bjournal{Bulletin of the American Mathematical Society}
\bvolume{84}
\bpages{957--1041}.
\end{barticle}
\endbibitem

\bibitem[\protect\citeauthoryear{Niederreiter}{1992}]{nied:1992}
\begin{bbook}[author]
\bauthor{\bsnm{Niederreiter},~\bfnm{H.}\binits{H.}}
(\byear{1992}).
\btitle{Random Number Generation and Quasi-{Monte Carlo} Methods}.
\bpublisher{SIAM}, \baddress{Philadelphia, PA}.
\end{bbook}
\endbibitem

\bibitem[\protect\citeauthoryear{Novak and
  Wo{\'z}niakowski}{2010}]{nova:wozn:2010}
\begin{bbook}[author]
\bauthor{\bsnm{Novak},~\bfnm{Erich}\binits{E.}} \AND
  \bauthor{\bsnm{Wo{\'z}niakowski},~\bfnm{Henryk}\binits{H.}}
(\byear{2010}).
\btitle{Tractability of Multivariate Problems, Volume {II}: Standard
  Information for Functionals}.
\bseries{EMS Tracts in Mathematics}
\bvolume{12}.
\bpublisher{European Mathematical Society}, \baddress{Z{\"u}rich}.
\bdoi{10.4171/084}
\end{bbook}
\endbibitem

\bibitem[\protect\citeauthoryear{Nuyens}{2017}]{nuye:magic:2017}
\begin{bmisc}[author]
\bauthor{\bsnm{Nuyens},~\bfnm{Dirk}\binits{D.}}
(\byear{2017}).
\btitle{The ``Magic Point Shop'' of {QMC} Point Generators and Generating
  Vectors}.
\bhowpublished{\url{https://people.cs.kuleuven.be/~dirk.nuyens/qmc-generators/}}.
\bnote{Software and generating vectors, version 14 December 2017; accessed 24
  July 2026}.
\end{bmisc}
\endbibitem

\bibitem[\protect\citeauthoryear{{\"O}kten and
  G{\"o}nc{\"u}}{2011}]{okte:gonc:2011}
\begin{barticle}[author]
\bauthor{\bsnm{{\"O}kten},~\bfnm{Giray}\binits{G.}} \AND
  \bauthor{\bsnm{G{\"o}nc{\"u}},~\bfnm{Ahmet}\binits{A.}}
(\byear{2011}).
\btitle{Generating low-discrepancy sequences from the normal distribution:
  {Box--Muller} or inverse transform?}
\bjournal{Mathematical and Computer Modelling}
\bvolume{53}
\bpages{1268--1281}.
\end{barticle}
\endbibitem

\bibitem[\protect\citeauthoryear{Ouyang, Wang and He}{2024}]{ouya:wang:he:2024}
\begin{barticle}[author]
\bauthor{\bsnm{Ouyang},~\bfnm{Du}\binits{D.}},
  \bauthor{\bsnm{Wang},~\bfnm{Xiaoqun}\binits{X.}} \AND
  \bauthor{\bsnm{He},~\bfnm{Zhijian}\binits{Z.}}
(\byear{2024}).
\btitle{Achieving high convergence rates by {quasi-Monte Carlo} and importance
  sampling for unbounded integrands}.
\bjournal{SIAM Journal on Numerical Analysis}
\bvolume{62}
\bpages{2393--2414}.
\end{barticle}
\endbibitem

\bibitem[\protect\citeauthoryear{Owen}{1995}]{rtms}
\begin{binproceedings}[author]
\bauthor{\bsnm{Owen},~\bfnm{A.~B.}\binits{A.~B.}}
(\byear{1995}).
\btitle{Randomly Permuted $(t,m,s)$-Nets and $(t,s)$-Sequences}.
In \bbooktitle{Monte Carlo and Quasi-Monte Carlo Methods in Scientific
  Computing}
(\beditor{\bfnm{H.}\binits{H.}~\bsnm{Niederreiter}} \AND
  \beditor{\bfnm{P.~J.~S.}\binits{P.~J.~S.}~\bsnm{Shiue}}, eds.)
\bpages{299--317}.
\bpublisher{Springer-Verlag}, \baddress{New York}.
\end{binproceedings}
\endbibitem

\bibitem[\protect\citeauthoryear{Owen}{1997a}]{snetvar}
\begin{barticle}[author]
\bauthor{\bsnm{Owen},~\bfnm{A.~B.}\binits{A.~B.}}
(\byear{1997}a).
\btitle{{Monte Carlo} Variance of Scrambled Net Quadrature}.
\bjournal{SIAM Journal on Numerical Analysis}
\bvolume{34}
\bpages{1884--1910}.
\end{barticle}
\endbibitem

\bibitem[\protect\citeauthoryear{Owen}{1997b}]{smoovar}
\begin{barticle}[author]
\bauthor{\bsnm{Owen},~\bfnm{A.~B.}\binits{A.~B.}}
(\byear{1997}b).
\btitle{Scrambled Net Variance for Integrals of Smooth Functions}.
\bjournal{Annals of Statistics}
\bvolume{25}
\bpages{1541--1562}.
\end{barticle}
\endbibitem

\bibitem[\protect\citeauthoryear{Owen}{1998}]{snxs}
\begin{barticle}[author]
\bauthor{\bsnm{Owen},~\bfnm{A.~B.}\binits{A.~B.}}
(\byear{1998}).
\btitle{Scrambling {S}obol' and {N}iederreiter-{X}ing points}.
\bjournal{Journal of Complexity}
\bvolume{14}
\bpages{466-489}.
\end{barticle}
\endbibitem

\bibitem[\protect\citeauthoryear{Owen}{2005}]{variation}
\begin{binproceedings}[author]
\bauthor{\bsnm{Owen},~\bfnm{A.~B.}\binits{A.~B.}}
(\byear{2005}).
\btitle{Multidimensional variation for quasi-{Monte Carlo}}.
In \bbooktitle{International Conference on Statistics in honour of Professor
  Kai-Tai Fang's 65th birthday}
(\beditor{\bfnm{J.}\binits{J.}~\bsnm{Fan}} \AND
  \beditor{\bfnm{G.}\binits{G.}~\bsnm{Li}}, eds.).
\end{binproceedings}
\endbibitem

\bibitem[\protect\citeauthoryear{Owen}{2008}]{localanti}
\begin{barticle}[author]
\bauthor{\bsnm{Owen},~\bfnm{A.~B.}\binits{A.~B.}}
(\byear{2008}).
\btitle{Local antithetic sampling with scrambled nets}.
\bjournal{Annals of Statistics}
\bvolume{36}
\bpages{2319--2343}.
\end{barticle}
\endbibitem

\bibitem[\protect\citeauthoryear{Owen}{2019}]{effdimsobononper}
\begin{barticle}[author]
\bauthor{\bsnm{Owen},~\bfnm{A.~B.}\binits{A.~B.}}
(\byear{2019}).
\btitle{Effective dimension of weighted {pre-Sobolev} spaces with dominating
  mixed partial derivatives}.
\bjournal{SIAM Journal on Numerical Analysis}
\bvolume{57}
\bpages{547--562}.
\end{barticle}
\endbibitem

\bibitem[\protect\citeauthoryear{Owen}{2020}]{firstsobol}
\begin{binproceedings}[author]
\bauthor{\bsnm{Owen},~\bfnm{Art~B}\binits{A.~B.}}
(\byear{2020}).
\btitle{On dropping the first {Sobol'} point}.
In \bbooktitle{International conference on Monte Carlo and quasi-Monte Carlo
  methods in scientific computing, {MCQMC} 2020}
(\beditor{\bfnm{A.}\binits{A.}~\bsnm{Keller}}, ed.)
\bpages{71--86}.
\bpublisher{Springer}.
\end{binproceedings}
\endbibitem

\bibitem[\protect\citeauthoryear{Owen}{2026}]{owen:rsobol}
\begin{bmisc}[author]
\bauthor{\bsnm{Owen},~\bfnm{Art~B.}\binits{A.~B.}}
(\byear{2026}).
\btitle{Fast scrambled {Sobol'} points in {R}}.
\bhowpublished{\url{https://artowen.su.domains/code/}}.
\end{bmisc}
\endbibitem

\bibitem[\protect\citeauthoryear{Owen and Pan}{2022}]{thelogs}
\begin{bincollection}[author]
\bauthor{\bsnm{Owen},~\bfnm{Art~B}\binits{A.~B.}} \AND
  \bauthor{\bsnm{Pan},~\bfnm{Zexin}\binits{Z.}}
(\byear{2022}).
\btitle{Where are the logs?}
In \bbooktitle{Advances in Modeling and Simulation: Festschrift for Pierre
  L'Ecuyer}
(\beditor{\bfnm{Z.}\binits{Z.}~\bsnm{Botev}},
  \beditor{\bfnm{A.}\binits{A.}~\bsnm{Keller}} \AND
  \beditor{\bfnm{B.}\binits{B.}~\bsnm{Tuffin}}, eds.)
\bpages{381--400}.
\bpublisher{Springer}.
\end{bincollection}
\endbibitem

\bibitem[\protect\citeauthoryear{Owen and Pan}{2024}]{scrambledhalton}
\begin{barticle}[author]
\bauthor{\bsnm{Owen},~\bfnm{Art~B}\binits{A.~B.}} \AND
  \bauthor{\bsnm{Pan},~\bfnm{Zexin}\binits{Z.}}
(\byear{2024}).
\btitle{Gain coefficients for scrambled {Halton} points}.
\bjournal{SIAM Journal on Numerical Analysis}
\bvolume{62}
\bpages{1021--1038}.
\end{barticle}
\endbibitem

\bibitem[\protect\citeauthoryear{Owen and Rudolf}{2020}]{owen:rudo:2020}
\begin{barticle}[author]
\bauthor{\bsnm{Owen},~\bfnm{A.~B.}\binits{A.~B.}} \AND
  \bauthor{\bsnm{Rudolf},~\bfnm{D.}\binits{D.}}
(\byear{2020}).
\btitle{A strong law of large numbers for scrambled net integration}.
\bjournal{SIAM Review}
\bvolume{63}
\bpages{360--372}.
\end{barticle}
\endbibitem

\bibitem[\protect\citeauthoryear{Owen and Tribble}{2005}]{owen:trib:2005}
\begin{barticle}[author]
\bauthor{\bsnm{Owen},~\bfnm{A.~B.}\binits{A.~B.}} \AND
  \bauthor{\bsnm{Tribble},~\bfnm{S.~D.}\binits{S.~D.}}
(\byear{2005}).
\btitle{A quasi-{Monte Carlo} {Metropolis} algorithm}.
\bjournal{Proceedings of the National Academy of Sciences}
\bvolume{102}
\bpages{8844--8849}.
\end{barticle}
\endbibitem

\bibitem[\protect\citeauthoryear{Pan}{2026}]{pan:2026:automatic}
\begin{barticle}[author]
\bauthor{\bsnm{Pan},~\bfnm{Zexin}\binits{Z.}}
(\byear{2026}).
\btitle{Automatic optimal-rate convergence of randomized nets using
  median-of-means}.
\bjournal{Mathematics of Computation}
\bvolume{95}
\bpages{1415--1446}.
\end{barticle}
\endbibitem

\bibitem[\protect\citeauthoryear{Pan and Owen}{2023a}]{superpolyone}
\begin{barticle}[author]
\bauthor{\bsnm{Pan},~\bfnm{Z.}\binits{Z.}} \AND
  \bauthor{\bsnm{Owen},~\bfnm{A.~B.}\binits{A.~B.}}
(\byear{2023}a).
\btitle{Super-polynomial accuracy of one dimensional randomized nets using the
  median-of-means}.
\bjournal{Mathematics of Computation}
\bvolume{92}
\bpages{805--837}.
\end{barticle}
\endbibitem

\bibitem[\protect\citeauthoryear{Pan and Owen}{2023b}]{nonzerogain}
\begin{barticle}[author]
\bauthor{\bsnm{Pan},~\bfnm{Zexin}\binits{Z.}} \AND
  \bauthor{\bsnm{Owen},~\bfnm{Art~B}\binits{A.~B.}}
(\byear{2023}b).
\btitle{The nonzero gain coefficients of {Sobol's} sequences are always powers
  of two}.
\bjournal{Journal of Complexity}
\bvolume{75}
\bpages{101700}.
\end{barticle}
\endbibitem

\bibitem[\protect\citeauthoryear{Pan and Owen}{2024}]{superpolymulti}
\begin{barticle}[author]
\bauthor{\bsnm{Pan},~\bfnm{Zexin}\binits{Z.}} \AND
  \bauthor{\bsnm{Owen},~\bfnm{Art~B}\binits{A.~B.}}
(\byear{2024}).
\btitle{Super-polynomial accuracy of multidimensional randomized nets using the
  median-of-means}.
\bjournal{Mathematics of Computation}
\bvolume{93}
\bpages{2265--2289}.
\end{barticle}
\endbibitem

\bibitem[\protect\citeauthoryear{Pan and Owen}{2025}]{skewrqmc}
\begin{barticle}[author]
\bauthor{\bsnm{Pan},~\bfnm{Zexin}\binits{Z.}} \AND
  \bauthor{\bsnm{Owen},~\bfnm{Art~B}\binits{A.~B.}}
(\byear{2025}).
\btitle{Skewness of a randomized {quasi-Monte Carlo} estimate}.
\bjournal{Journal of Complexity}
\bvolume{90}
\bpages{101956}.
\end{barticle}
\endbibitem

\bibitem[\protect\citeauthoryear{Paskov and Traub}{1995}]{pask:trau:1995}
\begin{barticle}[author]
\bauthor{\bsnm{Paskov},~\bfnm{S.}\binits{S.}} \AND
  \bauthor{\bsnm{Traub},~\bfnm{J.}\binits{J.}}
(\byear{1995}).
\btitle{Faster Valuation of Financial Derivatives}.
\bjournal{The Journal of Portfolio Management}
\bvolume{22}
\bpages{113--120}.
\end{barticle}
\endbibitem

\bibitem[\protect\citeauthoryear{Pharr, Jakob and
  Humphreys}{2023}]{phar:etal:2023}
\begin{bbook}[author]
\bauthor{\bsnm{Pharr},~\bfnm{Matt}\binits{M.}},
  \bauthor{\bsnm{Jakob},~\bfnm{Wenzel}\binits{W.}} \AND
  \bauthor{\bsnm{Humphreys},~\bfnm{Greg}\binits{G.}}
(\byear{2023}).
\btitle{Physically based rendering: {From} theory to implementation}.
\bpublisher{MIT Press}.
\end{bbook}
\endbibitem

\bibitem[\protect\citeauthoryear{Rezende and Mohamed}{2015}]{reze:moha:2015}
\begin{binproceedings}[author]
\bauthor{\bsnm{Rezende},~\bfnm{Danilo}\binits{D.}} \AND
  \bauthor{\bsnm{Mohamed},~\bfnm{Shakir}\binits{S.}}
(\byear{2015}).
\btitle{Variational inference with normalizing flows}.
In \bbooktitle{International conference on machine learning}
\bpages{1530--1538}.
\bpublisher{PMLR}.
\end{binproceedings}
\endbibitem

\bibitem[\protect\citeauthoryear{Ritter}{2000}]{ritt:2000}
\begin{bbook}[author]
\bauthor{\bsnm{Ritter},~\bfnm{Klaus}\binits{K.}}
(\byear{2000}).
\btitle{Average-case analysis of numerical problems}
\bvolume{1733}.
\bpublisher{Springer-Verlag}, \baddress{Berlin}.
\end{bbook}
\endbibitem

\bibitem[\protect\citeauthoryear{Roth}{1954}]{roth:1954}
\begin{barticle}[author]
\bauthor{\bsnm{Roth},~\bfnm{K.~F.}\binits{K.~F.}}
(\byear{1954}).
\btitle{On irregularities of distribution}.
\bjournal{Mathematica}
\bvolume{1}
\bpages{73--79}.
\end{barticle}
\endbibitem

\bibitem[\protect\citeauthoryear{Roy et~al.}{2023}]{roy:etal:2023}
\begin{barticle}[author]
\bauthor{\bsnm{Roy},~\bfnm{Pamphile~T}\binits{P.~T.}},
  \bauthor{\bsnm{Owen},~\bfnm{Art~B}\binits{A.~B.}},
  \bauthor{\bsnm{Balandat},~\bfnm{Maximilian}\binits{M.}} \AND
  \bauthor{\bsnm{Haberland},~\bfnm{Matt}\binits{M.}}
(\byear{2023}).
\btitle{{Quasi-Monte Carlo} methods in {Python}}.
\bjournal{Journal of Open Source Software}
\bvolume{8}
\bpages{5309}.
\end{barticle}
\endbibitem

\bibitem[\protect\citeauthoryear{Rusch
  et~al.}{2024}]{rusc:kirk:bron:lemu:rus:2024}
\begin{barticle}[author]
\bauthor{\bsnm{Rusch},~\bfnm{T~Konstantin}\binits{T.~K.}},
  \bauthor{\bsnm{Kirk},~\bfnm{Nathan}\binits{N.}},
  \bauthor{\bsnm{Bronstein},~\bfnm{Michael~M}\binits{M.~M.}},
  \bauthor{\bsnm{Lemieux},~\bfnm{Christiane}\binits{C.}} \AND
  \bauthor{\bsnm{Rus},~\bfnm{Daniela}\binits{D.}}
(\byear{2024}).
\btitle{Message-Passing {Monte Carlo}: Generating low-discrepancy point sets
  via graph neural networks}.
\bjournal{Proceedings of the National Academy of Sciences}
\bvolume{121}
\bpages{e2409913121}.
\end{barticle}
\endbibitem

\bibitem[\protect\citeauthoryear{Schmid}{1999}]{schm:1999}
\begin{binproceedings}[author]
\bauthor{\bsnm{Schmid},~\bfnm{W.~Ch.}\binits{W.~C.}}
(\byear{1999}).
\btitle{The exact quality parameter of nets derived from {Sobol'} and
  {Niederreiter} sequences}.
In \bbooktitle{Recent advances in numerical methods and applications}
(\beditor{\bfnm{O.}\binits{O.}~\bsnm{Illiev}},
  \beditor{\bfnm{M.~S.}\binits{M.~S.}~\bsnm{Kaschiev}},
  \beditor{\bfnm{S.~D.}\binits{S.~D.}~\bsnm{Margenov}},
  \beditor{\bfnm{Bl.~H.}\binits{B.~H.}~\bsnm{Sendov}} \AND
  \beditor{\bfnm{P.~S.}\binits{P.~S.}~\bsnm{Vassilevski}}, eds.)
\bpages{287--295}.
\bpublisher{World Scientific}, \baddress{Singapore}.
\end{binproceedings}
\endbibitem

\bibitem[\protect\citeauthoryear{Schmid}{2001}]{schm:2001}
\begin{barticle}[author]
\bauthor{\bsnm{Schmid},~\bfnm{W.~Ch.}\binits{W.~C.}}
(\byear{2001}).
\btitle{Projections of digital nets and sequences}.
\bjournal{Mathematics and computers in simulation}
\bvolume{55}
\bpages{239--247}.
\end{barticle}
\endbibitem

\bibitem[\protect\citeauthoryear{Sch\"urer and Schmid}{2009}]{schu:schm:2009}
\begin{binproceedings}[author]
\bauthor{\bsnm{Sch\"urer},~\bfnm{R.}\binits{R.}} \AND
  \bauthor{\bsnm{Schmid},~\bfnm{W.~C.}\binits{W.~C.}}
(\byear{2009}).
\btitle{{MinT}--new features and new results}.
In \bbooktitle{Monte Carlo and Quasi-Monte Carlo Methods 2008}
(\beditor{\bfnm{P.}\binits{P.}~\bsnm{L'Ecuyer}} \AND
  \beditor{\bfnm{A.~B.}\binits{A.~B.}~\bsnm{Owen}}, eds.)
\bpages{501--512}.
\bpublisher{Springer-Verlag}, \baddress{Berlin}.
\end{binproceedings}
\endbibitem

\bibitem[\protect\citeauthoryear{Sloan and Joe}{1994}]{sloa:joe:1994}
\begin{bbook}[author]
\bauthor{\bsnm{Sloan},~\bfnm{I.~H.}\binits{I.~H.}} \AND
  \bauthor{\bsnm{Joe},~\bfnm{S.}\binits{S.}}
(\byear{1994}).
\btitle{Lattice Methods for Multiple Integration}.
\bpublisher{Oxford Science Publications}, \baddress{Oxford}.
\end{bbook}
\endbibitem

\bibitem[\protect\citeauthoryear{Sloan and
  Wo{\'z}niakowski}{1998}]{sloa:wozn:1998}
\begin{barticle}[author]
\bauthor{\bsnm{Sloan},~\bfnm{Ian~H.}\binits{I.~H.}} \AND
  \bauthor{\bsnm{Wo{\'z}niakowski},~\bfnm{Henryk}\binits{H.}}
(\byear{1998}).
\btitle{When are quasi-{Monte Carlo} algorithms efficient for high dimensional
  integration?}
\bjournal{Journal of Complexity}
\bvolume{14}
\bpages{1--33}.
\end{barticle}
\endbibitem

\bibitem[\protect\citeauthoryear{Sobol'}{1967}]{sobo:1967:tran}
\begin{barticle}[author]
\bauthor{\bsnm{Sobol'},~\bfnm{I.~M.}\binits{I.~M.}}
(\byear{1967}).
\btitle{The Distribution of Points in a Cube and the Accurate Evaluation of
  Integrals}.
\bjournal{USSR Computational Mathematics and Mathematical Physics}
\bvolume{7}
\bpages{86--112}.
\end{barticle}
\endbibitem

\bibitem[\protect\citeauthoryear{Sobol'}{1969}]{sobo:1969}
\begin{bbook}[author]
\bauthor{\bsnm{Sobol'},~\bfnm{I.~M.}\binits{I.~M.}}
(\byear{1969}).
\btitle{Multidimensional Quadrature Formulas and {H}aar Functions}.
\bpublisher{Nauka}, \baddress{Moscow}.
\bnote{(In Russian)}.
\end{bbook}
\endbibitem

\bibitem[\protect\citeauthoryear{Sobol'}{1973}]{sobo:1973}
\begin{barticle}[author]
\bauthor{\bsnm{Sobol'},~\bfnm{I.~M.}\binits{I.~M.}}
(\byear{1973}).
\btitle{Calculation of improper integrals using uniformly distributed
  sequences}.
\bjournal{Soviet Math Dokl}
\bvolume{14}
\bpages{734--738}.
\end{barticle}
\endbibitem

\bibitem[\protect\citeauthoryear{Sobol'
  et~al.}{2011}]{sobo:asot:krei:kuch:2011}
\begin{barticle}[author]
\bauthor{\bsnm{Sobol'},~\bfnm{I.~M.}\binits{I.~M.}},
  \bauthor{\bsnm{Asotsky},~\bfnm{D.}\binits{D.}},
  \bauthor{\bsnm{Kreinin},~\bfnm{A.}\binits{A.}} \AND
  \bauthor{\bsnm{Kucherenko},~\bfnm{S.}\binits{S.}}
(\byear{2011}).
\btitle{Construction and comparison of high-dimensional {Sobol'} generators}.
\bjournal{Wilmott magazine}
\bvolume{2011}
\bpages{64--79}.
\end{barticle}
\endbibitem

\bibitem[\protect\citeauthoryear{Sorokin et~al.}{2026}]{qmcpy:2026}
\begin{barticle}[author]
\bauthor{\bsnm{Sorokin},~\bfnm{Aleksei~G}\binits{A.~G.}},
  \bauthor{\bsnm{Hickernell},~\bfnm{Fred~J}\binits{F.~J.}},
  \bauthor{\bsnm{Choi},~\bfnm{Sou-Cheng~T}\binits{S.-C.~T.}},
  \bauthor{\bsnm{Rathinavel},~\bfnm{Jagadeeswaran}\binits{J.}},
  \bauthor{\bsnm{Robbe},~\bfnm{Pieterjan}\binits{P.}} \AND
  \bauthor{\bsnm{Jain},~\bfnm{Aadit}\binits{A.}}
(\byear{2026}).
\btitle{QMCPy: A {Python} Framework for {(Quasi-) Monte Carlo} Algorithms}.
\bjournal{Journal of Open Source Software}
\bvolume{11}
\bpages{9705}.
\end{barticle}
\endbibitem

\bibitem[\protect\citeauthoryear{Spanier}{1995}]{span:1995}
\begin{binproceedings}[author]
\bauthor{\bsnm{Spanier},~\bfnm{J.}\binits{J.}}
(\byear{1995}).
\btitle{{Quasi-{Monte Carlo} Methods for Particle Transport Problems}}.
In \bbooktitle{Monte Carlo and Quasi-Monte Carlo Methods in Scientific
  Computing}
(\beditor{\bfnm{H.}\binits{H.}~\bsnm{Niederreiter}} \AND
  \beditor{\bfnm{P.~Jau-Shyong}\binits{P.~J.-S.}~\bsnm{Shiue}}, eds.)
\bpages{121--148}.
\bpublisher{Springer-Verlag}, \baddress{New York}.
\end{binproceedings}
\endbibitem

\bibitem[\protect\citeauthoryear{Surjanovic and Bingham}{2013}]{surj:bing:2013}
\begin{bmisc}[author]
\bauthor{\bsnm{Surjanovic},~\bfnm{S.}\binits{S.}} \AND
  \bauthor{\bsnm{Bingham},~\bfnm{D.}\binits{D.}}
(\byear{2013}).
\btitle{Virtual library of simulation experiments: test functions and
  datasets}.
\bhowpublished{\url{https://www.sfu.ca/~ssurjano/}}.
\end{bmisc}
\endbibitem

\bibitem[\protect\citeauthoryear{Suzuki}{2026}]{suzu:2026}
\begin{barticle}[author]
\bauthor{\bsnm{Suzuki},~\bfnm{Kosuke}\binits{K.}}
(\byear{2026}).
\btitle{Coarse Scrambling for {Sobol'} and {Niederreiter} Sequences}.
\bjournal{SIAM Journal on Numerical Analysis}
\bvolume{64}
\bpages{1342--1363}.
\end{barticle}
\endbibitem

\bibitem[\protect\citeauthoryear{Tribble}{2007}]{trib:2007}
\begin{bphdthesis}[author]
\bauthor{\bsnm{Tribble},~\bfnm{S.~D.}\binits{S.~D.}}
(\byear{2007}).
\btitle{{Markov chain Monte Carlo} algorithms using completely uniformly
  distributed driving sequences},
\btype{PhD thesis},
\bpublisher{Stanford University}.
\end{bphdthesis}
\endbibitem

\bibitem[\protect\citeauthoryear{van~der Corput}{1935}]{vand:1935:I}
\begin{barticle}[author]
\bauthor{\bparticle{van~der} \bsnm{Corput},~\bfnm{J.~G.}\binits{J.~G.}}
(\byear{1935}).
\btitle{Verteilungsfunktionen {I}}.
\bjournal{Nederl. Akad. Wetensch. Proc.}
\bvolume{38}
\bpages{813--821}.
\end{barticle}
\endbibitem

\bibitem[\protect\citeauthoryear{Warnock}{1972}]{warn:1972}
\begin{bincollection}[author]
\bauthor{\bsnm{Warnock},~\bfnm{T.~T.}\binits{T.~T.}}
(\byear{1972}).
\btitle{Computational investigations of low discrepancy point sets}.
In \bbooktitle{Applications of number theory to numerical analysis}
(\beditor{\bfnm{S.~K.}\binits{S.~K.}~\bsnm{Zaremba}}, ed.)
\bpages{319--343}.
\bpublisher{Academic Press}, \baddress{New York}.
\end{bincollection}
\endbibitem

\bibitem[\protect\citeauthoryear{Waudby-Smith and
  Ramdas}{2024}]{waud:ramd:2024}
\begin{barticle}[author]
\bauthor{\bsnm{Waudby-Smith},~\bfnm{Ian}\binits{I.}} \AND
  \bauthor{\bsnm{Ramdas},~\bfnm{Aaditya}\binits{A.}}
(\byear{2024}).
\btitle{Estimating means of bounded random variables by betting}.
\bjournal{Journal of the Royal Statistical Society Series B}
\bvolume{86}
\bpages{1--27}.
\end{barticle}
\endbibitem

\bibitem[\protect\citeauthoryear{Weyl}{1916}]{weyl:1916}
\begin{barticle}[author]
\bauthor{\bsnm{Weyl},~\bfnm{H.}\binits{H.}}
(\byear{1916}).
\btitle{\"{Uber} die Gleichverteilung von Zahlen mod. Eins}.
\bjournal{Mathematische Annalen}
\bvolume{77}
\bpages{313--352}.
\end{barticle}
\endbibitem

\bibitem[\protect\citeauthoryear{Yue and Mao}{1999}]{yue:mao:1999}
\begin{barticle}[author]
\bauthor{\bsnm{Yue},~\bfnm{R.~X.}\binits{R.~X.}} \AND
  \bauthor{\bsnm{Mao},~\bfnm{S.~S.}\binits{S.~S.}}
(\byear{1999}).
\btitle{On the variance of quadrature over scrambled nets and sequences}.
\bjournal{Statistics \& probability letters}
\bvolume{44}
\bpages{267--280}.
\end{barticle}
\endbibitem

\bibitem[\protect\citeauthoryear{Zhu and Dick}{2014}]{zhu:dick:2014}
\begin{barticle}[author]
\bauthor{\bsnm{Zhu},~\bfnm{Houying}\binits{H.}} \AND
  \bauthor{\bsnm{Dick},~\bfnm{Josef}\binits{J.}}
(\byear{2014}).
\btitle{Discrepancy bounds for deterministic acceptance-rejection samplers}.
\bjournal{Electronic Journal of Statistics}
\bvolume{8}
\bpages{678--707}.
\end{barticle}
\endbibitem

\end{thebibliography}

\end{document}